\documentclass[12pt]{elsart}
\usepackage{epsfig,rotate}
\newcommand{\ds}{\displaystyle} 
\newcommand{\jfigps}[2]{\begin{figure}[tbp]
\centerline{\psfig{file=#1.ps,height=9cm}}
\caption{\small #2} \label{fig:#1} \end{figure}}
\newcommand{\jfigpsb}[2]{\begin{figure}[tbp]
\centerline{\psfig{file=#1.ps,height=12cm}} \vspace*{-1cm}
\caption{\small #2} \label{fig:#1} \end{figure}}

\begin{document}
\begin{frontmatter}
\title{PHYSICS OF STRANGE MATTER}
\author[gie]{Carsten Greiner}
\author[lbnl]{and J\"urgen Schaffner-Bielich}
\address[gie]{Institut f\"ur Theoretische Physik, Universit\"at Giessen,
35392 Giessen, Germany}
\address[lbnl]{Nuclear Science Division, Lawrence Berkeley National Laboratory,
Berkeley, CA 94720, USA}
\end{frontmatter}
\section{INTRODUCTION}
All known nuclei are made of the two nucleons, the proton and the neutron.
Besides those two lightest baryons there exist still a couple of other
stable (but weakly decaying) baryons, the hyperons. These were found for the first time
in cosmic ray experiments and were entitled as `strange' because of their
unusual long lifetimes of about $10^{-10}$ sec.
For their notation in the hadron zoo of elementary particles one introduced
the new quantum number `strangeness' (or `hypercharge').  In the quark picture
baryons are interpreted as being made out of three constituents, the quarks.
Correspondingly the proton (uud) and neutron (udd) are made out of the
two lightest quarks, the up and down quark. For the description of the hyperons
a third flavour, the strange quark, was demanded. The lightest hyperon, the
$\Lambda$-particle (uds), and the three $\Sigma $-particles (uus,uds,dds) contain
one strange quark, the two more heavy $\Xi$-particles (uss,dss) contains
two strange quark and the $\Omega $-particle (sss) is solely made out of three
strange quarks.

Up to now strangeness remains an
experimentally as theoretically rather largely unexplored
degree of freedom in strongly interacting
baryonic matter. This lack of investigation reflects
the experimental task in producing nuclei containing (weakly decaying)
strange baryons, which is conventionally limited by replacing one
neutron (or at maximum two) by a strange $\Lambda$-particle in scattering
experiments with pions or kaons.
There exist a broad knowledge about single hypernuclei,
i.e. nuclei, where one nucleon is substituted
by a $\Lambda $ (or $\Sigma $) by means of the exchange reaction
$\pi ^{+} + n \rightarrow \Lambda + K^+$ (the $K^+$ ($\bar{s}$u)
denotes the positively charged kaon, the lightest strange meson).
Over the last two decades a rich phenomenology has resulted for such hypernuclei.

However, there exist more or less no experimental insight how more than one
hyperon behave inside a nuclei (or nuclear matter). The technical
problem is to create within a tiny moment, smaller than the decay time
of a hyperon, enough hyperons and then to bring them together with nucleons
to form any potential multi hypernuclei.
By employing a relativistic shell model calculation,
which gives a rather excellent description of normal
nuclei and single $\Lambda $-hypernuclei, it was
found that such configurations might exist as (small or large) bound
multi hypernuclei (MEMO - metastable exotic multihypernuclear object).
The reasoning for such exotic bound nuclei and the resulting extension
of the periodic table to the strangeness degree of freedom
will be discussed in detail in section 3.

Hypermatter could also be realized in a completely different picture.
Indeed, this second and much more speculative possibility was raised
by physicists much earlier. The fundamental theory of strong interactions,
quantum chromodynamics, does not forbid the principle existence of `larger'
hadronic particles, so called multiquark states. Today only the
mesons and baryons are known in nature. However, there could exist states with
more than three quarks. Going further with this speculation one comes
to the conclusion that only multiquark states with nearly the same number
of up, down and strange quarks might exist as stable configurations.
According to very schematic model calculations such
states could exist as metastable or even absolutely stable objects,
being then more bound than $^{56}Fe$. Such a very speculative form
of hypermatter is called strange quark matter. We will start our
discussion with an introduction to the physics
of strange quark matter in section 2.

\begin{figure}[ht]
\vspace*{\fill}
\centerline{\psfig{figure=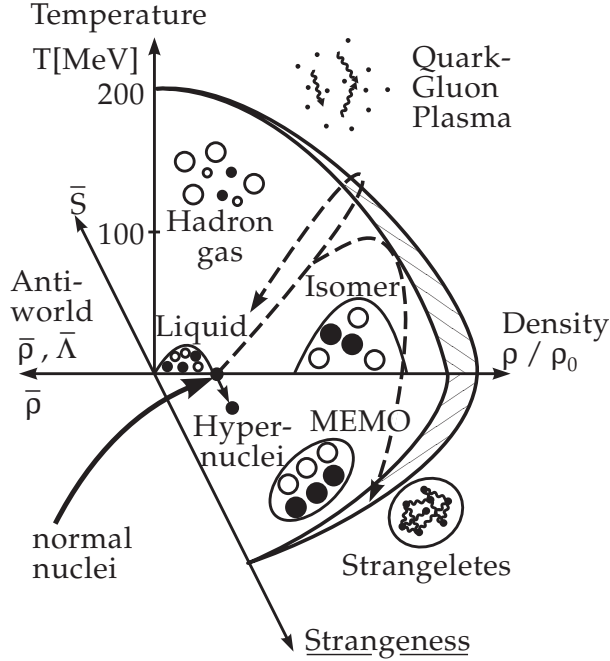,width=8cm}}
\caption{
Phase diagram of (hot) nuclear matter including the strangeness
degree of freedom -- MEMOs and possibly also strangelets establish
as stable multistrange configurations.
\label{phdia}}
\end{figure}

Central (ultra-)relativistic heavy ion collisions provide
the only source
for the formation of either multi-hypernuclear
(strange hadronic matter) objects, consisting of nucleons,
$\Lambda $'s and $\Xi $'s, or strangelets
(strange multiquark droplets).
To be more specific,
on the average the occurrence of
20 $\Lambda $'s, 10 $\Sigma $'s and 1 $\Xi $'s
per event
for $Au(11.7 \, AGeV)Au$ and of
60 $\Lambda $'s, 40 $\Sigma $'s and 5 $\Xi $'s per event
for $Pb(160 \, AGeV)Pb$ are expected to be created.
In Fig. \ref{phdia} we depict qualitatively what we want to elaborate
in section 4: Customarily the equation of state of hot and dense hadronic
matter (being created in a relativistic heavy ion collision) is
characterized by means of a phase diagram ($\rho_B \leftrightarrow T$), where
at some critical temperature and/or nonstrange baryon density eventually
a phase transition
to a deconfined quark gluon plasma (QGP) state does occur. However, the EOS to be passed through during
a heavy ion collision incorporates also a new degree of freedom, the
{\em net} strangeness (counting a surplus of strange over antistrange
quarks). Like the occurrence of bound nonstrange nuclear matter,
multihypernuclear matter, or small droplets
(MEMOs) of this new state,
may be revealed.
In addition, also the phase transition to the deconfined state is
affected by the possible conglomeration of the strangeness degree of freedom.
In particular, if the strangelet does exist in principle, it has to be regarded
as a cold, stable and bound manifestation of that phase being a remnant
or `ash' of the originally hot QGP-state.
We will close section 4 in critically emphasizing the detection
possibilities of such exotic states by their properties and lifetimes,
also in respect to the present experimental
undertaking at Brookhaven and at CERN.

In section 5 we finally want to give some insight on how the physics
of strange matter can affect the physical picture of dense neutron stars.
\clearpage
\section{STRANGE QUARK MATTER}

The proposal that hypothetical strange quark matter droplets (`strangelets')
at zero temperature and in $\beta$-equilibrium might be absolutely
stable has stimulated substantial activity. Such a scenario
would be of fundamental importance, as, for example, it could
explain a large fraction of the non-observable mass of the universe
in certain cosmological models, and could modify substantially
our understanding of the structure and stability of neutron stars
\cite{SMSc,SMProc} (see also section 5).

Some years ago we proposed that such strange quark matter droplets might
as well play an important role in ultrarelativistic heavy-ion
collisions \cite{PRL87,PRD88,PRD91}.
One important goal in ultrarelativistic heavy-ion physics is the observation of
a temporarily created quark gluon plasma.
An enhanced production of
strange particles from the QGP, especially an enhanced $K^+/\pi ^+$-ratio \cite{Koc86},
and the $J/\psi $-suppression \cite{Mat88} were proposed as possible signatures for
such a novel state of deconfined, strongly interacting matter.
It was subsequently shown, however, that such indications, as indeed were
experimentally found, can be still
understood as being due to the formation
of a very (energy-) dense and hot region of matter consisting of
confined hadrons or precursor effects of QGP formation.
Therefore, it seems that perhaps the only unambiguous way to detect
the transient existence of a QGP might be the
experimental observation of exotic remnants (`ashes' of the QGP),
like the formation of strange quark matter droplets.
To this interesting aspect we will turn back in section 4.

In this first section we now summarize the reasons for the speculation
of (cold) stable or metastable strange quark matter in the bulk.
In addition, being more relevant for heavy ion physics,
also a brief description of finite pieces of strange quark matter will be given
(see also section 4).

Strange quark matter or strangelets are thought to be confined
(bulk) objects containing a large number of delocalized
quarks $(u...u\, , \, d...d\, , \, s...s)$, so-called multiquark droplets.
Multiquark states consisting only of u- and d-quarks must have a mass
larger than ordinary nuclei, otherwise normal nuclei would
certainly be unstable, which, of course, is not the case.
However, the situation is different for
droplets of SQM, which would contain approximately the same
amount of u-, d- and s-quarks.
Bodmer was the first person to consider this new form of matter \cite{Bod71};
he proposed that strange multiquark clusters,
being much more compressed than ordinary
nuclei, may exist as long-lived exotic isomers of nuclear matter inside
neutron stars.

\begin{figure}[ht]
\vspace*{\fill}
\centerline{\psfig{figure=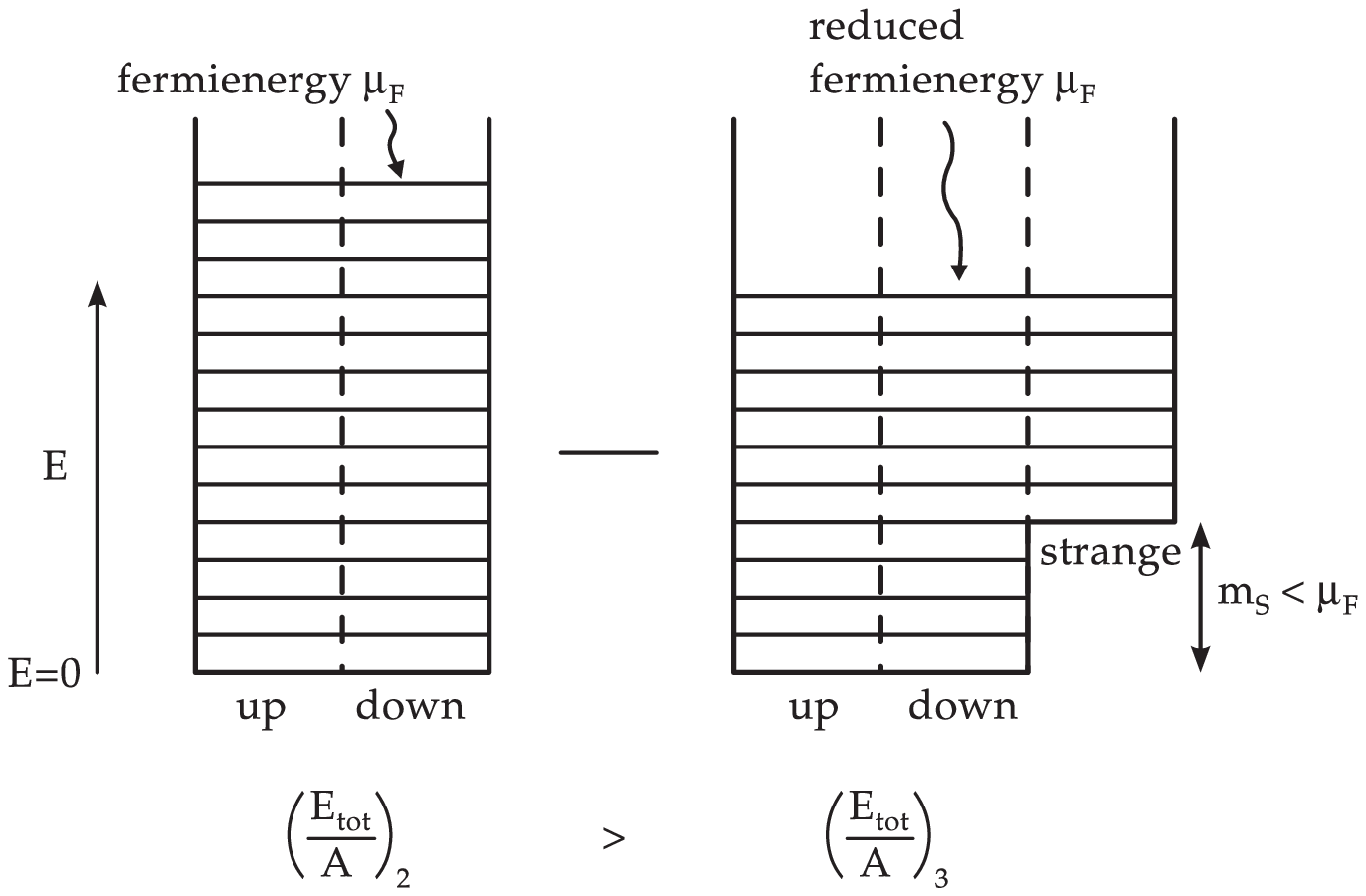,height=2.5in}}
\caption{
Schematic illustration of the energy levels inside
a multiquark bag with two or three flavours.
\label{scheme}
}
\end{figure}

Chin and Kerman \cite{Chi79} and independently
McLerran and Bjorken \cite{Bj79} postulated two reasons why such
huge hadronic states should be relatively stable:
\begin{enumerate}
\item The (weak) decay of a s-quark into a
d-quark would be suppressed or forbidden because
the lowest single particle states are occupied.
\item The strange quark mass is lower than the Fermi energy of the u-
or d-quark in such a quark droplet; the opening of a new flavour degree of
freedom tends to lower the Fermi energy and hence also the mass of the
strangelet (see Fig. \ref{scheme}).
\end{enumerate}
According to this picture, SQM should appear as a nearly neutral and massive
state because the number of strange quarks is nearly equal to the number
of massless up or down quarks and so the strange quarks neutralize that
hypothetical form of nuclear matter.

\begin{figure}[ht]
\vspace*{\fill}
\centerline{\rotate[r]{\psfig{figure=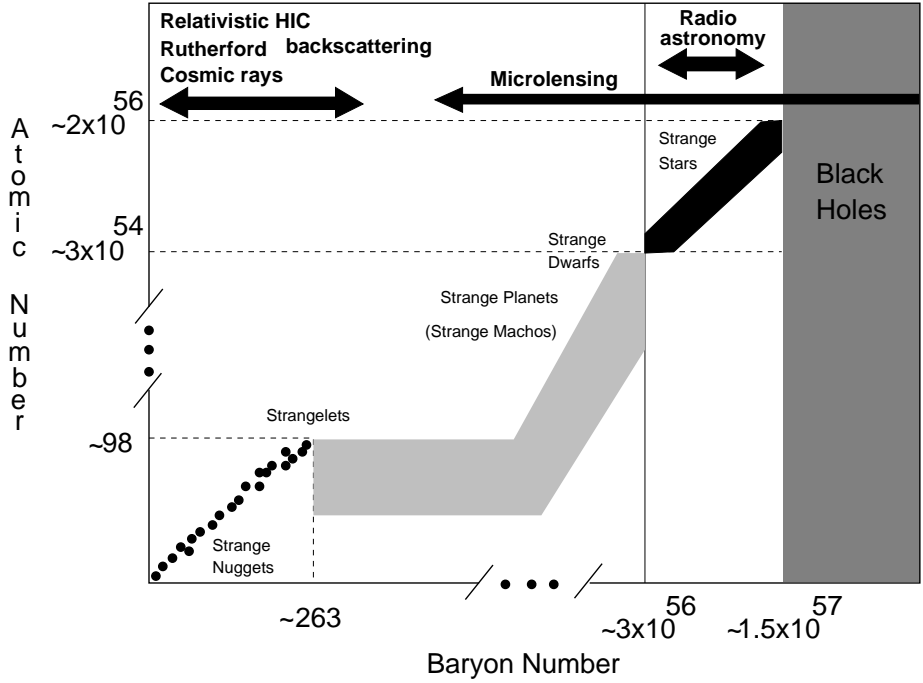,width=9cm}}}
\caption{
The possible places where to find stable strange quark matter.
If being absolutely stable SQM might exist in forms of small nuggets
(strangelets) being not (much) larger than normal nuclei. It might
also span as a charge neutral state the empty `nuclear desert' \cite{Ruj84}
within the range
from $A_B \sim 300$ up to sizes of neutron stars $A_B \sim 10^{56}$.
\label{weber}}
\end{figure}

It was then later Witten who realized and hence
raised the intriguing possibility that
strange quark matter might in principle also
be absolutely stable and may also provide an
explanation for {\em cold dark matter} in the universe \cite{Wit84}.
This would be the case
if the mass of a strangelet is smaller than the mass of the
corresponding ordinary nucleus with the same baryon number and hence be
the true groundstate of nuclear matter!
If being stable and nearly neutral, it could exist
at {\em all possible sizes} \cite{Ruj84}, as the small Coulomb energy is not sufficient for a
break up into smaller pieces \cite{Far84}. Such a speculation is illustrated
in Fig. \ref{weber} \cite{WeGle}.

Presently such a highly speculative scenario cannot be ruled out.
Normal nuclei could
only transform into these novel states by a
higher order weak decay;
the decay into one single $\Lambda $-particle would be energetically forbidden.
The time-scale for such collective decays is extremely large;
accordingly, normal nuclei would remain intact.

On the other hand, it is also conceivable that the mass per baryon
of a strange droplet is
lower than the mass of the strange $\Lambda $-baryon, but larger than the nucleon
mass. The droplet is then in a metastable state, it cannot decay
(strongly) into
$\Lambda $'s \cite{Chi79}.

For a very simple quantification
of these ideas one typically models quark matter
in {\em bulk} by a Fermi gas equation of state of interacting quarks
which to first
order in $\alpha_c = g^2/16\pi $ is given by \cite{Fr78}
\begin{eqnarray}
\lefteqn{\Omega_i(m_i,\mu_i) \,  =  } \nonumber \\[1em]
&&-\frac{1}{4\pi^2} \,  \left(
\mu_i (\mu_i^2-m_i^2)^{1/2}(\mu_i^2-\frac{5}{2}m_i^2) +
\frac{3}{2} m_i^4 \ln \frac{\mu_i + (\mu_i^2-m_i^2)^{1/2}}{m_i} \right.
\label{cg1}
\\[0.5em]
&&  \left. - \, \frac{8}{\pi} \alpha_c \left[
3\, \left(\mu_i (\mu_i^2-m_i^2)^{1/2} \, - \,
m_i^2 \ln \frac{\mu_i + (\mu_i^2-m_i^2)^{1/2}}{m_i} \, \right) ^2
\,  - 2 \, (\mu_i^2 - m_i^2) \right] \, \right) \,  . \nonumber
\end{eqnarray}
Here $m_i$ and $\mu_i$ denote the (current) mass and the chemical potential,
respectively, of the quark flavour i=u,d,s.
For the total potential the vacuum excitation energy $BV$
has to be added, which corresponds to the energy difference between the
`false', perturbative vacuum inside the `bag' and the true vacuum at the
outside, and which, in return, confines the quarks, i.e.
\begin{equation}
\label{cg2}
\Omega (\mu_q,\mu_s; m_s; \alpha_c) \, = \, \sum_{i=u,d,s} \,
\Omega_i (m_i,\mu_i)  \, + \, BV \, \, \, .
\end{equation}
$m_u \approx m_d \approx 0$ and $\mu_u = \mu_d$ has been implicitly
taken for an isospin symmetric situation. From this expression (\ref{cg2})
the energy per baryon in the groundstate can be
readily obtained by varying E/A with respect to the baryon density
to describe the system at zero pressure. To fix the amount of net strangeness
in the system the strangeness fraction $f_s$ is introduced
as the ratio of net strangeness and total baryon number, i.e.
\begin{equation}
\label{cg3}
f_s \, = \, \frac{n_{s} \, - \, n_{\bar{s}}}{\frac{1}{3}\,
[(n_{q}  - n_{\bar{q}}) \, + \, (n_{s}  - n_{\bar{s}})]}
\, \stackrel{T=0}{\equiv } \, N_s/A_B \, \, \, .
\end{equation}

\begin{figure}[ht]
\vspace*{9.5cm}
\caption{
Energy per baryon for strange quark matter at zero
temperature and zero pressure, as a function of the strangeness fraction
$f_{s}$. The dashed line
defines the corresponding mass of a hyperonic matter ground-state.
\label{SQMEA}
}
\end{figure}

The energy per baryon of quark matter in the groundstate thus depends solely
on the explicit strangeness fraction $f_s$
and on the three intrinsic parameter, the strange (current) mass
$m_s$, the coupling constant $\alpha _c$ and the bag constant $B^{1/4}$.
For most purposes to come a non-zero $\alpha _c$ can be `absorbed'
by a slight increase in the bag parameter.
In Fig. \ref{SQMEA}, E/A is depicted as a function of $f_s$ and vanishing coupling constant,
but for various choices of the bag constant and the strange quark mass.
The minimum in the energy is achieved for a strangeness fraction $f_s \sim
0.7$, if the more popular and smaller mass $m_s = 150 $ MeV is taken.
According to the formula $Z/A = (1-f_s)/2$ smaller pieces of
strange quark matter would then be slightly positive
at its most stable position.
A necessary condition for the stability of strange quark matter
against strong decay is that its energy per baryon must be
smaller than that of the corresponding hyperonic matter
(dashed-dotted curve),  which can be determined as
\begin{eqnarray}
\label{cg4}
m_{Hyp} &=& f_{s} m_{\Lambda} + (1-f_{s})m_N -\epsilon_B
\hspace{2cm}; \hspace{1cm}
0 \leq f_{s} \leq 1 \nonumber \\ & &
( f_{s} -1) m_{\Xi} + (2-f_{s}) m_{\Lambda} - \epsilon_B
\hspace{1cm}; \hspace{1cm}
1 \leq f_{s} \leq 2  \\ & &
( f_{s} -2) m_{\Omega} + (3-f_{s} ) m_{\Xi} - \epsilon_B
\hspace{1cm}; \hspace{1cm}
2 \leq f_{s} \leq 3 \nonumber
\end{eqnarray}
where $\Lambda$, $\Xi$ and $\Omega$ are the masses of the strange hyperons
and $\epsilon_{B}$ is the binding energy per nucleon, which
is, for simplicity, taken to be the
infinite nuclear matter parameter of 16 MeV.

If popular parameters within bag models, like the MIT bag model
employed here, are extrapolated
to strange quark droplets, both of the above considered pictures turn
out to be possible. However, metastable strangelets relax somewhat the
stringent conditions on the choice of parameters of the bag model
required by an absolutely stable state \cite{Far84}, namely small
Bag constants $B^{\frac{1}{4}} \stackrel{< }{\sim } 150$ MeV (which,
within the same model extrapolated to finite temperatures,
would give a too low critical transition temperatures $T_c \sim 100$ MeV
for the onset of deconfinement). For large pieces of absolutely stable
strange matter one has to include the Coulomb effect due to the still
tiny net positive charge \cite{Far84}. These have been neglected in the
present figure \ref{SQMEA}.
It turns out that then the
minimum of the energy per baryon number will be shifted smoothly to
$f_s \rightarrow 1$ instead of $f_s \approx 0.7$, so that the strange matter
piece still carries a very small positive charge.

On the other hand, from Fig. \ref{SQMEA} it follows that for bag parameters
B$^{1/4}$ lower than 190 MeV strange quark droplets can only decay
via weak interactions, i.e., they would be {\em metastable}.
The baryon density of a strangelet would be about $1.5-2$ of
normal nuclear matter density.
For larger B-values (large) strangelets are instable.

The idea of the existence of metastable strangelets was actually raised
earlier than the possibility of a new groundstate of normal nuclear matter.
Chin and Kerman \cite{Chi79} had employed the original favoured MIT bag parameters
of $B^{1/4}=145$ MeV, $g_c^2/4\pi = 2.2$
($\alpha _c = 0.55$) and a strange quark mass
of about $m_s = 279 $ MeV \cite{DeG75,Com1}.
The minimum they find is shifted to a higher strangeness content. In quark
matter one-gluon exchange is repulsive, if the quarks are massless and
relativistic, and attractive if the quarks are massive and behave already
nonrelativistic. Accordingly, metastable strange quark matter
in the ground state might as well be slightly negatively charged.
(This cannot be the case if it would exist in absolutely stable form, for
than, normal, positively charged nuclei would be attracted and absorbed
at the surface. Such catastrophic consequences have to be excluded.)
One should remark, however, that the linear expansion to obtain the expression
for E/A used in Ref. \cite{Chi79} is not valid, as the energy density
as well as the baryon density turn negative for $\alpha_c > \pi /8$, so that
a perturbative expansion is meaningless and signals the
breakdown of perturbation theory. This basically sheds some light how
{\em `seriously'} the calculations, the conclusions
and the speculations have to be taken!

Originally (strange) quark matter in bulk was thought to exist
only in the interior of neutron stars where the pressure is as
high that the neutron matter melts into its quark substructure
\cite{Col75,Ba76,Fr78}
(see also section 5).
In Witten's prospective scenario hot strange quark matter
nuggets could have condensed out of the
deconfined phase during the phase transition in the expanding and
cooling early universe \cite{Wit84}.
These would carry most of the tiny surplus in baryon number of the whole universe.
The baryon number would remain inside if the heat and entropy of the nuggets is
carried away mainly by neutrino emission instead of mesons and especially
baryons. If an absolutely stable groundstate would exist, i.e. if
{\em cold} strange quark matter is the true groundstate of nuclear matter,
the hot nuggets would further cool and instead of suffering a complete hadronization
they might settle into these new states and hence could resolve the
dark matter problem. Since then the idea of absolute stability has
stimulated a lot of work on potential consequences in astrophysics
\cite{SMProc,Astro}.

Nuclear collisions, however, do allow also
the detection and study of only short lived, metastable
strangelets, which will decay by flavour changing weak processes.
Most of the experiments are sensitive to strangelets with lifetimes
up to $10^{-8}-10^{-9}$ sec. In addition, the strangelets to be expected
will be small. Hence, in the last part of this subsection,
we already address the potential importance of {\em finite} size effects.
Later in section 4  we come back to this analysis in somewhat
more detail in order to try to pin down possible short-lived or long-lived
candidates for the present and future experimental undertakings.
Due to these effects and the wider range
of employed phenomenological parameters, smaller strangelets, if they exist, are more
likely to be metastable than being absolutely stable.

Small pieces of (strange) quark matter are described
within a spherical MIT-bag by filling up the bag
with exact single-particle Dirac-states \cite{Far84,Vas86,PRD88}.
These calculations have been done with $\alpha _c =0$.
The first MIT boundary condition,
which ensures confinement, is basically a Bogolyubov-type
condition for the scalar mass of the quark,
i.e. $M(r>R_{bag}) \rightarrow \infty$,
 and leads to the following equation for the single particle
energies:
\begin{eqnarray}
\label{cg5}
j_{l_{\kappa}}(pR)&=&-sign(\kappa)\frac{p}{E+m_{i}}
           j_{l_{-\kappa}}(pR)
\end{eqnarray}
with
\begin{eqnarray}
p&=&\frac{\omega^{i}_{\kappa,\alpha}}{R}  \nonumber  \\
E^{i}_{\kappa,\alpha}&=&((\frac{\omega^{i}_{\kappa,\alpha}}{R})^{2}
      +m_{i}^{2})^{\frac{1}{2}}
\end{eqnarray}
Here $m_{i}$ is the quark mass (i=u,d,s),
with $m_{u}=m_{d}=0$ MeV and $m_{s}=150$ MeV. R defines the radius
of the bag, $\kappa$ is the angular momentum quantum number and $\alpha$
labels the eigenvalues in this quantum state $\kappa$.
The total energy of the finite system is obtained by summing the
`lowest' energy eigenvalues and adding the phenomenological
bag energy $BV$ for the perturbative vacuum:
\begin{equation}
\label{cg6}
E_{total} \, = \,
\sum_{i=u,d,s} \,
\sum_{\kappa \, ; \, \alpha }^{E^{i}_{\kappa, \alpha} \leq \mu_i} \,
N_{\kappa } \,
E^{i}_{\kappa, \alpha} \, \, + \, \, B \, \left( \frac{4\pi }{3} R^3 \right)
\, \, \, ,
\end{equation}
where the degeneracy $N_{\kappa }= 6 |\kappa |$.
In addition, a center of momentum correction and
a Coulomb term $((N_{0}-2N_{s})/R)^{2}/60\alpha$ is included.
(In the original MIT model \cite{DeG75} a zero-point energy term
of the form $-Z_0/R$ is also included as a more or less phenomenological
parameter for fitting the light hadron spectra. Such an additional
attractive term may have some noticeable effect on very small strangelets.)
The volume V or the radius R, respectively, has to be chosen in a way that
the total pressure of the quarks inside the bag compensates the vacuum
pressure B, or, in other words,  $\partial E_{total} / \partial R =0$.

\begin{figure}[ht]
\vspace*{\fill}
\centerline{\psfig{figure=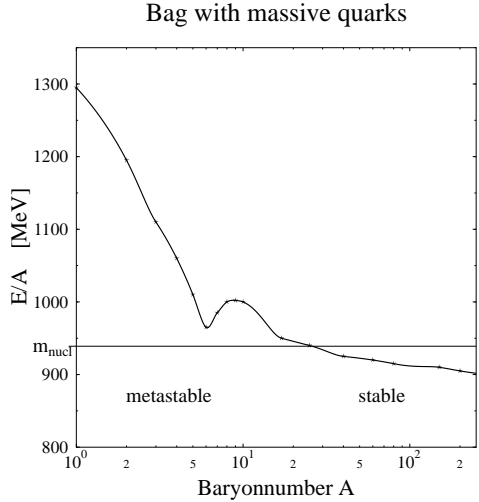,height=3.0in}}
\caption{
E/A versus A in a hadronic (MIT-)\-Bag model for noninteracting
quarks. A constant strangeness fraction $f_s$=1, i.e.
$N_u=N_d=N_s$, is assumed
($m_s=150$ MeV, $B^{\frac{1}{4}}= 145$ MeV).
\label{SQMEAfin}
}
\end{figure}

Fig. \ref{SQMEAfin}
shows the dependence of the energy per baryon
of a giant bag on the baryon number for
$B^{\frac{1}{4}}=145$ MeV and for a
(constant) strangeness fraction of
$f_{s}=1$, i.e. SU(3) symmetric and charge neutral matter.
One clearly sees a pronounced shell structure.
Especially the first magic number shows up nicely at a baryon number of 6, that is,
if all quarks completely fill the $1S_{1/2}$ ground state
(the so called `quark-alphas' \cite{Mi88}).
Generally speaking, the energy per baryon
for a smaller strangelet with a baryon number $N_{B}<50$
is about 50-100 MeV/N
larger as compared to the infinite quark matter calculations.
This can be interpreted in terms of a surface
correction \cite{Far84}.
If one varies the number of
up and down and strange quarks,
strangelets do have a large variety of nearly degenerate states.
This fact is interesting for the heavy ion experiments because
these nearly degenerate states will be stable against strong decay and
thus would all be detectable if their lifetime against weak decay is
long enough (see section 4).

We close this section with some cautionary remarks:
\begin{itemize}
\item
The existence of strange quark matter cannot rigorously
be predicted by theory, although some general and appealing arguments
for its stability do exist. A detailed understanding cannot be achieved
by employing more or less relatively simple, phenomenological models.
Ultimately only the experimentalists (in astrophysics or heavy ion physics)
may prove its existence or nonexistence.
\item
For smaller strangelets, as the above calculations suggest, finite size
effects and a profound shell structure might be quite important.
For very small strangelets the colour magnetic part of the gluon exchange
should be important and should further favour SU(3) symmetric
multiquark states to exist. In this respect
the prominent H-Dibaryon proposed by Jaffe \cite{Ja77} would be the
smallest strangelet.
\item
Although the idea of absolute stable pieces of strange quark matter is
intriguing in its consequences for astrophysics,
metastable strangelets only need to be smaller in mass then the corresponding
hyperonic multi-baryon configuration, which is much higher
than that of a normal nuclei. Because of this argument it might well be that
(small) strangelets, if they do exist, are metastable
and not absolutely stable (see section 4).
\vspace*{10mm}
\end{itemize}

\clearpage


\section{STRANGE HADRONIC MATTER}


Strange hadronic matter consists of baryons, i.e.\ nucleons and hyperons as
quasiparticles are the basic constituents. In this case, there exists some
knowledge about the properties of hyperons inside nuclei from hypernuclear
physics up to two units of strangeness.
They extend the chart of nuclides to a third dimension in addition to baryon
number and isospin. The extension to it will 
open new perspectives for nuclear physics.
This is what we are going to discuss in the following subsections.

First we will give a short introduction to hypernuclear physics.
Then we list all possible metastable combinations of nucleons and hyperons on
the basis of symmetry considerations 
which are stable against strong interactions.
If they are bound, 
they will have lifetimes of the order of the weak decay lifetime
of the hyperons of 10$^{-10}$ s. 

The extrapolation to systems with a large amount of strangeness will rely then
on several basic features: the knowledge that the $\Lambda$-nucleon and the
$\Lambda\Lambda$ interaction is attractive, the existence of metastable
combinations of nucleons and hyperons or hyperons alone, 
and the effect of the Pauli-blocking in the hyperon world.

The properties of strange hadronic matter are then discussed on the basis of
the 
relativistic mean field model, which is able to reproduce fairly well the
properties of nuclei and hypernuclei. 
Here we will focus on the properties of superheavy elements. As the hyperons
$\Sigma^-$ and $\Xi^-$ are negatively charged, a similar situation as for
strangelets occur and superheavy elements with $A>250$ and a low charge are
possible.


\subsection{Hypernuclei}


Our intention is not to give a detailed overview about hypernuclear
physics, as there exist excellent reviews in the literature
\cite{Chr89,Ban90}, 
but to give a short summary of the main features.

Hypernuclei are known since 1953 as Danysz and Pniewski observe the
first $\Lambda$-hypernuclei in a cosmic ray emulsion experiment \cite{Dan53}
(in some of the older publications hypernuclei are called
hyperfragments).
In the following decades one has found $\Lambda$-hypernuclei up to a maximum
mass number of $A=15$ which is due to the resolution
of emulsion experiments.
The lightest one is the hypernucleus $^3_\Lambda$H, a system of
one proton, one neutron and one $\Lambda$. Another interesting fact is
the existence of the hypernucleus $^9_\Lambda$Be: the presence of the
$\Lambda$ glues together and stabilizes
the nuclear core $^8$Be which normally decays to two $\alpha$-particles!
Other examples for the enhanced stability of hypernuclear systems compared
to their nuclear cores are the hypernuclei $^9_\Lambda$Li and $^9_\Lambda$B.
We will present even more impressive ones in the forthcoming
sections.

In the 70's programs started at the AGS \cite{Bon74}
and at CERN \cite{Bru76}
for detecting $\Lambda$-hypernuclei by using meson beams and spectroscopy
techniques.
Here one uses the reaction:
\begin{equation}
K^- + {\rm n} \longrightarrow \Lambda + \pi^- \quad .
\end{equation}
A special feature of this reaction is, that there exist a magic momentum
of the ingoing kaon where the $\Lambda$ is produced at rest
inside the nucleus facilitating the formation of a bound hypernuclear system.
Another generation of hypernuclear spectroscopy uses the reverse reaction:
\begin{equation}
\pi^+ + {\rm n} \longrightarrow \Lambda + K^+ .
\end{equation}
This reaction yields a lower resolution, but deep lying states can be
detected more easily \cite{Mot88}.
The study of these reactions started in the 80's at the AGS
\cite{Miln85,Pile91} and at the KEK in Japan \cite{Akei89}.

\jfigps{}{The chart of $\Lambda$ hypernuclei as well as $\Xi$, $\Sigma$
and double $\Lambda$ hypernuclei known so far (taken from 
\cite{Ban90}).}

The hypernuclear chart with the hypernuclear states is shown in
Fig.~\ref{fig:hypchart}. 
Experimentally one has seen hypernuclei up to a baryon number of $A=209$
(Tungsten) where one neutron is replaced by a $\Lambda$.
Moreover one is able to extract the single particle binding energies
from excited states (like 1f$_\Lambda$) to very deep lying states
1s$_\Lambda$ which opens the possibility to study deeply bound
probes inside the nucleus over a wide range of mass number.
This was an excellent test (and success) of the shell model \cite{Fesh90}.
An essential feature of the $\Lambda$-nucleon interaction is that
one has seen no spin-orbit splitting experimentally,
which leads to the statement that the $\Lambda$ behaves as a
spinless neutron \cite{Bru76,Bert79} (a theoretical discussion of this topic
can be found in \cite{Mil85}).
The potential depth of the $\Lambda$ in nuclear matter can be derived
from the experimental data to be
\begin{equation}
U_\Lambda^{(N)} = 27 - 30 \, {\rm MeV}
\end{equation}
which has to be compared with the corresponding nuclear depth of about 60 MeV.
Calculations using Hartree-Fock models \cite{Mil88} and Relativistic
Mean-Field (RMF) models \cite{Bou77,Bro77,Bog81,Rufa87,Mar89,Rufa90}
are able to reproduce very nicely
the observed trend of the single particle energy with the mass number.

\jfigps{}{The potentials and the single particle levels for the
hypernucleus $^{17}_\Lambda$O in the RMF model.}

Fig.~\ref{fig:hypspe} shows as an example a calculation for the hypernucleus
$^{17}_\Lambda$O. The potential for the $\Lambda$ is shallower than for the
nucleon but of a similar radius. 
The 1s state of the $\Lambda$ is a separate
state within this $\Lambda$ potential.
The surface thickness is bigger for the
$\Lambda$ density distribution than for the nucleons. This will cause
a $\Lambda$-halo as the $\Lambda$ is not as deeply bound as the nucleons.
The single particle energies for the 1p states of the $\Lambda$ 
are very close to each other, resembling the weak spin-orbit force of the
$\Lambda$-nucleon interaction. 

The lifetime of the hypernuclei has been also measured. They decay weakly
by emission of a pion (mesonic decay channel):
\begin{equation}
\Lambda \to {\rm p} + \pi^-  \qquad \mbox{and} \qquad
\Lambda \to {\rm n} + \pi^0
\end{equation}
like the free $\Lambda$ releasing 38 and 41 MeV, respectively.
Moreover a new reaction,
the nonmesonic decay, opens in the nuclear medium
\begin{equation}
\Lambda + {\rm p} \to {\rm n} + {\rm p} 
\qquad \mbox{and} \qquad \Lambda + {\rm n} \to {\rm n} + {\rm n}
\end{equation}
which releases an energy of about 177 MeV. The mesonic channel is dominant
for very light systems but it is negligible for heavier hypernuclei like 
the hypernucleus $^{12}_\Lambda$C \cite{Bar88}.
This results from the dominant absorption of the pion inside the nucleus
and of the larger phase space for the nonmesonic decay.
Finally, the lifetime of $\Lambda$-hypernuclei has been determined to be at the
order of $10^{-10}$ s even for
$A\approx 200$ \cite{Pol88}, which is close to the lifetime
of the $\Lambda$ in free space.

There exists also some scarce information about $\Sigma$- 
and $\Xi$-hypernuclei.
The experimental situation concerning the existence of bound
$\Sigma$-hypernuclear systems is still controversial. The strong process
$\Sigma+N\to \Lambda+N$ smears out possible peaks in the meson spectra.
Nevertheless, one has found surprisingly small structures in hypernuclear
spectra.
All these structures are lying
in the continuum.
Recent analysis of the data 
about $\Sigma^-$-atoms indicate that the isoscalar potential depth
of a $\Sigma$ in a bath of nucleons in its groundstate is repulsive
\cite{Gal94} which would prevent a formation of bound $\Sigma$ hypernuclear
states. The isovector potential can be attractive, as for example in light
systems like $4^{}$He$_{\Sigma^+}$ and might form a bound state.
Indications for such a bound state have been seen in \cite{Hay90}.

$\Xi$-hypernuclei have already been found shortly after the first observation
of a $\Lambda$-hypernucleus in emulsion experiments using kaon beams.
There exist seven emulsion events in the literature (for a review see
\cite{Dov83}) showing that the strong conversion process
$\Xi + N\to\Lambda+\Lambda$ still allows for bound systems.
The potential depth of a $\Xi$ in nuclear matter has been derived by
using Wood-Saxon potentials to be
\begin{equation}
   U_\Xi^{(N)} = 21 - 24  \mbox{ MeV}
\end{equation}
whereas relativistic mean field calculations get a little higher value
of $U_\Xi^{(N)} = 28$ MeV \cite{Sch94}.

There exist also three $\Lambda\Lambda$-hypernuclear events by caption of
a $\Xi^-$ and a detection of their mesonic decays
(for an excellent review see \cite{Dal89}).
In particular these are the hypernuclei
$_{\Lambda\Lambda}^6$He \cite{Pro66}, $_{\Lambda\Lambda}^{10}$Be
\cite{Dan63}, and $_{\Lambda\Lambda}^{13}$B \cite{Aoki91,Dov91}.
These data constitute our only (and poor)
knowledge about the hyperon-hyperon interactions.
One defines the following two quantities:
\begin{eqnarray}
B_{\Lambda\Lambda}(^A_{\Lambda\Lambda}Z) &=&
B_\Lambda(^A_{\Lambda\Lambda}Z) +
B_\Lambda(^{A-1}_{\Lambda}Z) \nonumber \\
\triangle B_{\Lambda\Lambda}(^A_{\Lambda\Lambda}Z) &=&
B_\Lambda(^A_{\Lambda\Lambda}Z) -
B_\Lambda(^{A-1}_{\Lambda}Z)
\end{eqnarray}
where $B_{\Lambda\Lambda}$ is the sum of the binding energies of the
two $\Lambda$'s and $\triangle B_{\Lambda\Lambda}$ is a measure of
the $\Lambda\Lambda$ interaction. The experimental values are listed
in table~\ref{tab:dophyp}.
The $\Lambda\Lambda$ interaction is attractive and relatively strong.
The value of $4-5$ MeV has to be contrasted with the corresponding values
of $6-7$ MeV for the NN interaction and
$2-3$ MeV for the $\Lambda$N interaction.

\begin{table}
\begin{center}
\begin{tabular}{|c|c|c|}
\hline
Hypernucleus & $B_{\Lambda\Lambda} [{\rm MeV}]$
 & $\Delta B_{\Lambda\Lambda} [{\rm MeV}]$ \\
\hline
$_{\Lambda\Lambda}^6$He    & $10.9\pm 0.6$ & $ 4.7\pm 0.6$ \cr
$_{\Lambda\Lambda}^{10}$Be & $17.7\pm 0.4$ & $ 4.3\pm 0.4$ \cr
$_{\Lambda\Lambda}^{13}$B  & $27.5\pm 0.7$ & $ 4.8\pm 0.7$ \cr
\hline
\end{tabular}
\end{center}
\caption{The two observables
$\Delta B_{\Lambda\Lambda}$ and $B_{\Lambda\Lambda}$
of the known double $\Lambda$-hypernuclei.}
\label{tab:dophyp}
\end{table}

Hence, the $\Lambda\Lambda$ interaction is much stronger than the $\Lambda$N
interaction and about 3/4 of the NN interaction.
This has to be taken into account for hyperon-rich systems.


\subsection{Classification of Strange Hadronic Matter}


The strong interaction conserves strangeness and charge and the baryon number.
Therefore combinations of nucleons and hyperons, which are stable against
weak interactions, so called metastable combinations, have to be classified
according to these quantum numbers.

\begin{table}
\begin{center}
\begin{tabular}{|c|c|c|c|c|c|}
\hline
$-S\backslash Z$& -2 & -1 & 0 & +1 & +2 \\
\hline
0& & &nn&np&pp\\
\hline
1& &$\Sigma^-$n&$\Lambda$ n&$\Lambda$ p&$\Sigma^+$p\\
\hline
2&$\Sigma^-\Sigma^-$&$\Xi^-$n&$\Lambda\Lambda$&$\Xi^0$p&$\Sigma^+\Sigma^+$\\
\hline
3&$\Xi^-\Sigma^-$&$\Xi^-\Lambda$&$\Xi^0\Lambda$&$\Xi^0\Sigma^+$& \\
\hline
4&$\Xi^-\Xi^-$&$\Xi^0\Xi^-$&$\Xi^0\Xi^0$& & \\
\hline
5&$\Xi^-\Omega^-$&$\Xi^0\Omega^-$& & & \\
\hline
6&$\Omega^-\Omega^-$& & & & \\
\hline
\end{tabular}
\end{center}
\caption{
The possible configurations of metastable partners of nucleons and hyperons
classified according to their total strangeness $S$ and charge number $Z$.}
\label{tab:dublets}
\end{table}

First, in a nuclear environment, strong reactions
in the medium among different baryon species can occur.
There exist two strong reactions,
which have a considerable lower mass difference (or Q-value) as the
others:
$\Xi^-+p\to \Lambda + \Lambda$ and
$\Xi^0+n\to \Lambda + \Lambda$, releasing 28 and 23 MeV energy,
respectively. 
Indeed these two reactions will play a major role in the discussion of
bound strange hadronic systems and we will keep them in mind.
All other reactions have a rather large Q-value of more than 50 MeV
which can not be overcome by binding energy differences 
(remember that the maximum
binding of a nucleon is about 60~MeV).
Nevertheless, the final products of all reactions constitutes the
combinations of two different species which are for a given charge and
strangeness the deepest lying state and which are therefore stable against
strong interactions. All possible metastable combinations of two different
species are listed in table \ref{tab:dublets}.
Next consider a composite system of an
arbitrary number of nucleons and hyperons. One can easily see, that the
stability against strong interactions does not depend on the number of
baryons, e.g.\ on the number of neutrons or $\Sigma^-$'s
for the combination $\{ {\rm n},\Sigma^- \}$. 
Therefore our considerations of metastability
are valid for an arbitrary number of respective baryons.
Note that we only study the stability against strong interactions here.
Whether the
combination is indeed bound or not will be discussed afterwards.

\begin{table}
\begin{center}
\begin{tabular}{|c|c|c|c|c|c|}
\hline
$-S\backslash Z$& -2 & -1 & 0 & +1 & +2 \\
\hline
1&&&$\qquad \qquad$ &$\Lambda$np&\\
\hline
2&&&&&\\
\hline
3&$\Xi^-\Sigma^-$n&$\Xi^-\Lambda$n& &$\Xi^0\Lambda$ p&$\Xi^0\Sigma^+$p\\
\hline
4&&&&&\\
\hline
5&&$\Xi^-\Xi^0\Lambda$& & &\\
\hline
6& & & & & \\
\hline
7&$\Omega^- \Xi^- \Xi^0$& & & & \\
\hline
\end{tabular}
\end{center}
\caption{
The possible configurations of metastable combinations of
three different baryon species
classified according to their total strangeness $S$ and charge number $Z$.}
\label{tab:triplets}
\end{table}

Next one can think of metastable combinations of three or even more different
species.
The former ones are listed in table \ref{tab:triplets}.
Of special interest is that systems consisting of 
$\{ \Lambda,\Xi^0,\Xi^- \}$-hyperons are metastable (they form purely
hyperonic matter to be discussed later).
More than three different baryon species can not form a metastable
combination in free space as one can see from this table.
For a metastable object of four different baryon species every
combination of three of them must be one of the metastable triplets given
in table \ref{tab:triplets}. As this can not be the case there does not exist
metastable combinations of more than three different baryons in free space.
The situation changes if one introduces interactions which will be presented
in the following section.


\subsection{Relativistic Mean-Field Model}


In the preceding sections we have discussed the input for a theoretical
description of strange hadronic matter. In the following we present
an extended relativistic mean field model which is able to reproduce
fairly well the experimental data of nuclei and hypernuclei and implements
all hyperons. For reviews about this model for finite nuclei see
\cite{Ser86,Rei89}.

The basic idea of this approach is the meson-exchange picture, where
baryons interact via the exchange of mesons, usually a scalar meson
$\sigma$ which parametrizes effectively the two-pion exchange \cite{Ser92},
the vector meson $\omega$ (hereafter the field $V_\mu $)
and the isovector meson $\rho$ (field $\vec{R}_\mu $) for isotopic trends.
One starts from the Lagrangian
\begin{equation}
{\bf L} = {\bf L}_{\rm Baryon} + {\bf L}_{\rm Meson} +
           {\bf L}_{\rm Coupling} + {\bf L}_{\rm Coulomb}
\end{equation}
with free terms for the baryon and meson fields
\begin{eqnarray}
{\bf L}_{\rm Baryon}&=&
\sum_{\rm B} \overline{\Psi}_B(i \gamma^\mu\partial_\mu - m_B)
                       \Psi_B \\ \cr
{\bf L}_{\rm Meson}&=&\frac{1}{2}\partial^\mu \sigma \partial_\mu \sigma
                     -  U(\sigma) -\frac{1}{4}G^{\mu\nu}G_{\mu\nu}
                       + \frac{1}{2}m_\omega^2 V^\mu V_\mu\\ \cr
                    & & -\frac{1}{4}\vec{B}^{\mu\nu}\vec{B}_{\mu\nu}
                        + \frac{1}{2}m_\rho^2 \vec{R}^\mu \vec{R}_\mu
\end{eqnarray}
where the sum runs over all baryons of the baryon octet
(p,n,$\Lambda,\Sigma^+,\Sigma^0,\Sigma^-,\Xi^0$, $\Xi^-$).
The term $U(\sigma)$ stands for the scalar selfinteraction
\begin{equation}
U(\sigma) = \frac{1}{2}m_\sigma^2 \sigma^2
+ \frac{b}{3}\sigma^3 + \frac{c}{4}\sigma^4
\end{equation}
which has been introduced by Boguta and Bodmer \cite{Bog77} to get a correct
compressibility of nuclear matter. Another stabilized functional form
has been given by Reinhard \cite{Rei88}.
The interaction is introduced by a minimal coupling of meson fields and
baryon bilinear forms
\begin{eqnarray}
{\bf L}_{\rm Coupling}&=&
    - \sum_{\rm B} g_{\sigma B}\overline{\Psi}_B\Psi_B\sigma \\ \cr
& & - \sum_{\rm B} g_{\omega B}\overline{\Psi}_B\gamma^\mu\Psi_B V_\mu \\ \cr
& & - \sum_{\rm B} g_{\rho B}\overline{\Psi}_B\gamma^\mu
                        \vec{\tau}_B\Psi_B\vec{R}_\mu
\quad .
\end{eqnarray}
For finite nuclei one takes also account of the coulomb force
\begin{equation}
{\bf L}_{\rm Coulomb} = -\frac{1}{4}F^{\mu\nu}F_{\mu\nu}
- \sum_{\rm B} q_B e \overline{\Psi}_B\gamma_\mu\Psi_B A^\mu
\end{equation}
where $q_B$ is the charge number of the baryon.
We have used here the following notation
\begin{eqnarray}
G^{\mu\nu} &=&\partial^\mu V^\nu-\partial^\nu V^\mu \cr
\vec{B}^{\mu\nu} &=&\partial^\mu \vec{R}^\nu-\partial^\nu \vec{R}^\mu
+ i  \vec{R}^\mu\times\vec{R}^\nu \cr
F^{\mu\nu} &=&\partial^\mu A^\nu-\partial^\nu A^\mu \quad .
\end{eqnarray}
The equations of motions can be derived in the standard way
\begin{equation}
\partial_\nu \frac{\partial {\bf L}}{\partial(\partial_\nu q_i)} -
 \frac{\partial {\bf L}}{\partial q_i} =0 \, , \quad
q_i=\sigma , V_\mu , R_\mu , A_\mu , \Psi_B, \overline{\Psi}_B
\end{equation}
and one gets Klein-Gordon and Proca equations for the
meson fields with source terms coming from the baryon fields
\begin{eqnarray}
\left(\partial_\mu\partial^\mu + U'(\sigma)\right) \, \sigma &=&
- \sum_B g_{\sigma B} \overline{\Psi}_B\Psi_B \label{eq:kgs} \\ \cr
\partial^\mu G_{\mu\nu} + m_\omega^2 V_\mu &=&
\sum_B g_{\omega B} \overline{\Psi}_B\gamma_\mu\Psi_B \\ \label{eq:procav} \cr
\partial^\mu \vec{R}_{\mu\nu} + m_\rho^2 \vec{R}_\mu &=&
\sum_B g_{\rho B} \overline{\Psi}_B \vec{\tau}_B \gamma_\mu\Psi_B
\label{eq:procar}
\end{eqnarray}
where
\begin{equation}
U'(\sigma) = \frac{\partial}{\partial\sigma} U(\sigma) \quad .
\end{equation}
The equation for the electromagnetic field reads
\begin{equation}
\partial^\mu F_{\mu\nu} + m_\omega^2 A_\mu =
q_B e \overline{\Psi}_B\gamma_\mu\Psi_B \quad .
\label{eq:coul}
\end{equation}
The Dirac equation for each baryon
\begin{equation}
\left\{ i \gamma^\mu \partial_\mu + m_B
   + g_{\sigma B} \sigma
   + g_{\omega B} \gamma^\mu V_\mu
   + g_{\rho B} \gamma^\mu \vec{\tau}_B \vec{R}_\mu
   + q_B e \gamma^\mu A_\mu \right\} \Psi_B = 0
\label{eq:dirac}
\end{equation}
contains now potential terms from the meson and coulomb fields.
The scalar field shifts the mass to the effective mass defined as
\begin{equation}
m^*_B = m_B + g_{\sigma B} \sigma \quad.
\end{equation}
Analogously, the vector fields shift the energy and momentum of the baryon.
The equations (\ref{eq:kgs})-(\ref{eq:coul}) and (\ref{eq:dirac}) constitute
a set of coupled differential equations which have to be simplified
in order to solve them.
In the following we will use the relativistic mean field (RMF) approximation,
where the meson fields (and the Coulomb field)
are replaced by their classical expectation values
\begin{equation}
\sigma \to \left< \sigma \right> \,, \quad
V_\mu \to \left< V_\mu \right> \,, \quad
\vec{R}_\mu \to \left< \vec{R}_\mu \right> \,, \quad
A_\mu \to \left< A_\mu \right>
\end{equation}
i.e.\ quantum fluctuations of these fields are neglected.
Furthermore the many body wavefunctions of the baryons are treated on the
Hartree-level.
They are approximated by a sum of single particle wave functions
\begin{equation}
\Psi_B = \sum_\alpha a_{\alpha ,B}^+ \psi_{\alpha ,B} \quad ,
\end{equation}
where the sum runs only over the occupied states.
The contribution coming from the Dirac-sea is therefore also neglected
(this is the no-sea approximation).
Note that in the so called relativistic Hartree-approximation
(RHA) the quantum fluctuations
coming from the Dirac-sea are implemented by renormalizing the Lagrangian
\cite{Ser92} which we will not discuss here.
It is important to know that all non-diagonal coupling terms vanish in the
relativistic mean field approximation or at the Hartree-level.
They contribute at the next order which takes into account the Fock-term
(this is the Dirac-Hartree-Fock approximation).
The coupling term to the pseudoscalar fields (pion and kaon fields)
and pseudovector fields vanishes in the RMF approximation due to the $\gamma_5$
matrix which mixes upper and lower components of the wave functions.
Also only the third component of the isovector field is present in the
RMF approximation.

For spherical and static systems (nuclei in its groundstate)
the fields do only depend on the radius
\begin{equation}
\sigma = \sigma (|\vec r|) \, , \quad
V_\mu = \delta_{\mu 0} V_0 (|\vec r|) \, , \quad
\vec{R}_\mu = \delta_{\mu 0} \delta_{i 0} R_{0,0} (|\vec r|) \, , \quad
A_\mu = \delta_{\mu 0} A_0 (|\vec r|)
\end{equation}
and the spatial components of the vector fields $V_i$, $\vec{R}_i$ and $A_i$
are zero.
Hence one gets the following radial equations for the meson and Coulomb fields
\begin{eqnarray}
\left(-\frac{1}{r}{d^2\over dr^2}\cdot r+U'(\sigma) \right)\sigma(r)&=&
-\sum_B g_{\sigma B} \rho_{s,B}(r) \label{eq:skg1} \\ \cr
\left(-\frac{1}{r}{d^2\over dr^2}\cdot r+m_\omega ^2\right)V_0(r)&=&
\sum_B g_{\omega B} \rho_{v,B}(r) \label{eq:skg2} \\ \cr
\left(-\frac{1}{r}{d^2\over dr^2}\cdot r+m_\rho ^2\right)R_{0,0}(r)&=&
\sum_B g_{\rho B} \tau_{0,B}\rho_{{\rm iso},B}(r) \label{eq:skg3} \\ \cr
-\frac{1}{r}{d^2\over dr^2}(r\cdot A_0(r))&=& \sum_B q_B e \rho_{v,B}(r)
\label{eq:scoul}
\end{eqnarray}
where the scalar, vector, and isovector densities are given by
the sum over the occupied single particle wave functions
\begin{eqnarray}
\rho_{s,B}(r) &=& \overline{\Psi}_B\Psi_B
  = \sum_{\alpha =1}^N \omega_{\alpha ,B}\varphi_{\alpha ,B}^+(r)
    \gamma_0\varphi_{\alpha ,B} (r) \\ \cr
\rho_{v,B}(r) &=& \overline{\Psi}_B \gamma_0 \Psi_B
  = \sum_{\alpha =1}^N \omega_{\alpha ,B}\varphi_{\alpha ,B}^+(r)
    \varphi_{\alpha ,B} (r) \\ \cr
\rho_{{\rm iso},B}(r) &=& \overline{\Psi}_B\gamma_0\tau_{0,B}\Psi_B
  = \sum_{\alpha =1}^N\omega_{\alpha ,B}\varphi_{\alpha ,B}^+(r)\tau_{0,B}
    \varphi_{\alpha ,B} (r)
\end{eqnarray}
with the occupation number $\omega_{\alpha ,B}$ for each baryon species
and $\tau_{0,B}$ stands for the isospin of the baryon.
The wavefunctions for the baryons can be separated into upper and lower
components by the ansatz
\begin{equation}
\varphi_{\alpha ,B}(r) = \left(\matrix{
{\ds iG_{\alpha ,B}(r)\over\ds r}& Y_{j_\alpha l_\alpha m_\alpha} \cr\cr
{\ds F_{\alpha ,B}(r) \over\ds r} {\ds \sigma \cdot r \over\ds r} &
Y_{j_\alpha l_\alpha m_\alpha} \cr } \right)
\end{equation}
with the spherical harmonics $Y_{jlm}$ and the normalization
\begin{equation}
\int\limits_0^\infty
\{ {\vert G_{\alpha ,B}(r)\vert}^2 + {\vert F_{\alpha ,B}(r)\vert}^2 \}dr=1
\quad .
\end{equation}
The Dirac equations can then be written in terms of these two
components
\begin{eqnarray}
\lefteqn{\varepsilon_{\alpha ,B} G_{\alpha ,B}(r) \, = \,
\left(-{d \over dr}+{\kappa_{\alpha ,B}\over r}\right)F_{\alpha ,B}(r)}
 \label{eq:sd1} \\
&+& \left(m_B+g_{\sigma B}\sigma (r) + g_{\omega B} V_0(r)
   +g_{\rho B}\tau_{0,B}R_{0,0}(r)
   +q_B e A_0(r) \right) G_{\alpha ,B}(r) \nonumber
\end{eqnarray}
and
\begin{eqnarray}
\lefteqn{\varepsilon_{\alpha ,B} F_{\alpha ,B}(r) \, = \,
\left(+{d \over dr}+{\kappa_{\alpha ,B}\over r}\right)G_{\alpha ,B}(r)}
\label{eq:sd2} \\
&-& \left(m_B+g_{\sigma B}\sigma (r) - g_{\omega B} V_0 (r)
   -g_{\rho B}\tau_{0,B}R_{0,0}(r)
   -q_B e A_0(r) \right) F_{\alpha ,B}(r)  \nonumber
\end{eqnarray}
with
\begin{equation}
\kappa_{\alpha ,B}=
\cases{-(j_\alpha +{1\over 2})\,,\quad j_\alpha =l_\alpha +\frac{1}{2} \cr\cr
       +(j_\alpha +{1\over 2})\,,\quad j_\alpha =l_\alpha -\frac{1}{2} }
\end{equation}
The densities can be rewritten to
\begin{eqnarray}
\rho_{s,B}(r)&=&{1\over 4\pi r^2}
\sum\limits_{\alpha=1}^N \omega_{\alpha ,B} (2j_\alpha+1)
\lbrack G_{\alpha ,B}(r)^2-F_{\alpha ,B}(r)^2\rbrack \label{eq:sdens} \\
\rho_{v,B}(r)&=&{1\over 4\pi r^2}
\sum\limits_{\alpha=1}^N \omega_{\alpha ,B} (2j_\alpha+1)
\lbrack G_{\alpha ,B}(r)^2+F_{\alpha ,B}(r)^2\rbrack \label{eq:vdens} \\
\rho_{{\rm iso},B}(r)&=&{1\over 4\pi r^2}
\sum\limits_{\alpha=1}^N \omega_{\alpha ,B} (2j_\alpha+1)
\tau_{0,B} \lbrack G_{\alpha ,B}(r)^2+F_{\alpha ,B}(r)^2\rbrack
\label{eq:isodens} \quad .
\end{eqnarray}
The occupation numbers are
\begin{equation}
\omega_\alpha = \cases{+1 \cr  -1\cr}
\end{equation}
for closed shells. Otherwise one chooses the occupation number in the
interval $0<\omega_{\alpha ,B} < 1$ according to a schematic pairing
\begin{equation}
\omega_{\alpha ,B} = {1\over 2}\Biggl\lbrack
1-{\varepsilon_{\alpha ,B}-\varepsilon_{{\rm Fermi},B} \over
\sqrt{\left(\varepsilon_{\alpha ,B} -\varepsilon_{{\rm Fermi},B}\right)^2
+\Delta^2}} \Biggr\rbrack
\end{equation}
analogous to Skyrme-Hartree-Fock calculations \cite{Blo76} where
\begin{equation}
\Delta={11.2 MeV \over \sqrt A} 
\end{equation}
is a constant gap derived from nuclear pairing properties.
The Fermi-energy $\varepsilon_{{\rm Fermi},B}$ is evaluated in such a way,
that the baryon number is conserved for each species.

The coupled equations (\ref{eq:skg1})-(\ref{eq:scoul}) and
(\ref{eq:sd1}),(\ref{eq:sd2})
together with the above expressions for the densities can be solved
iteratively using standard techniques (for details see \cite{Rufa88}).
The parameters of the model are fitted to the properties of spherical
nuclei \cite{Rei86,Rufa88} where we will use the stabilized parameter set PL-Z
\cite{Rei88} in the following.
The RMF model gives a description of nuclei as good as
the conventional Skyrme-Hartree-Fock calculations \cite{Rei89}.
Also the binding energy of hypernuclei can be nicely reproduced
(see \cite{Bou77,Bro77,Bog81}, for fits to recent hypernuclear data
see \cite{Mar89,Rufa90}).

\jfigps{}{
The single particle energy of several hypernuclei in
comparison with RMF calculations.}

Fig. \ref{fig:hypnuc} shows the measured single particle energy for
different hypernuclei in comparison with RMF calculations.
One sees nicely the different single particle levels of the $\Lambda$. Note
that the spin-orbit splitting for $\Lambda$ levels is negligible therefore the
shells can be labeled by s-shell, p-shell and so on. Bulk matter
($A\to \infty$) corresponds to the crossing of the energy levels with the
y-axis. All levels merge to the value of the $\Lambda$ potential depth in
nuclear matter of about 27 MeV.

It turns out that the two coupling constants of the $\Lambda$
($g_{\sigma\Lambda}$ and $g_{\omega\Lambda}$) are strongly correlated
because they are fixed by the potential depth of the $\Lambda$
\begin{equation}
U_\Lambda = - g_{\sigma\Lambda}\sigma - g_{\omega\Lambda} V_0
\end{equation}
in saturated nuclear matter \cite{Sch92}. Hence one can choose
for example SU(6)-symmetry for the vector coupling constants
(see table~\ref{tab:su6})
and fixes the scalar coupling constants to the potential depth of the
corresponding hyperon.

Nevertheless, this model, hereafter named model 1, is not able
to reproduce the observed strongly attractive $\Lambda\Lambda$ interaction
as can be seen in fig.~\ref{fig:pic4} irrespectively of the chosen
vector coupling constant. Model 1 does not give the necessary strong attraction
between the two $\Lambda$'s. An additional interaction has to be invoked
between hyperons. This can be done by using a SU(3) symmetric model
and introducing two additional meson fields, the scalar meson $f_0(975)$
(denoted as $\sigma^*$ in the following) and the vector meson
$\phi(1020)$ \cite{Sch93,Sch94}
which we will discuss in the following.


\subsection{Hyperon-Hyperon interactions}


The additional Lagrangian for the new mesons is straight forward to derive
\begin{equation}
{\bf L}' = {\bf L} + {\bf L}^{YY}_{\rm Meson}
  + {\bf L}^{YY}_{\rm Coupling}
\end{equation}
with
\begin{eqnarray}
{\bf L}^{YY}_{\rm Meson} & = &
   \frac{1}{2}\left(\partial_\nu\sigma^*\partial^\nu\sigma^*
     - m_{\sigma^*}^2{\sigma^*}^2\right)
 - \frac{1}{4} S_{\mu\nu}S^{\mu\nu} + \frac{1}{2} m_\phi^2\phi_\mu\phi^\mu
\\ \cr
{\bf L}^{YY}_{Coupling} & = &
 - \sum_B g_{\sigma^* B}{\overline \Psi}_B\Psi_B\sigma^*
 - \sum_B g_{\phi B}{\overline\Psi}_B\gamma_\mu\Psi_B \phi^\mu
\end{eqnarray}
and the notation
\begin{equation}
S_{\mu\nu} = \partial_\mu\phi_\nu - \partial_\nu\phi_\mu
\quad .
\end{equation}
The new meson fields give new potential terms for the Dirac equations
\begin{eqnarray}
\lefteqn{\varepsilon_{\alpha ,B} \varphi_{\alpha ,B}(r) \, = \,
\gamma_0\biggl\lbrace
-i \vec \gamma \cdot \vec \nabla
+ m_B + g_{\sigma B}\sigma (r) + g_{\sigma^* B}\sigma^* (r) }
\\ &&
+ g_{\omega B}\gamma^0 V_0(r) + g_{\phi B}\gamma^0\phi_0(r)
+ g_{\rho B}\gamma^0 R_{0,0}(r)
+ q_B e \gamma^0 A_0(r)
\biggr\rbrace \varphi_{\alpha ,B}(r) \nonumber
\end{eqnarray}
and two new meson field equations
\begin{eqnarray}
\left(-\Delta + m_{\sigma^*}^2\right)\sigma^* (r)&=&
-\sum_B g_{\sigma^* B} \rho_{s,B}(r)
 \\ \cr
\left(-\Delta + m_\phi ^2\right)\phi_0 (r)&=&
\sum_B g_{\phi B} \rho_{v,B}(r)
\end{eqnarray}
which are written here in the static RMF approximation.

\begin{table}
\begin{center}
\begin{tabular}{|c|c|c|c|c|}
\hline
$g_{MB}$ & Nucleon & $\Lambda$ & $\Sigma$ & $\Xi$ \cr
\hline
$\omega$ & 3 & 2 & 2 & 1 \cr
\hline
$\rho$ & 1 & 0 & 2 & 1 \cr
\hline
$\phi$ & 0 & $-\sqrt{2}$ & $-\sqrt{2}$ & $-2\sqrt{2}$ \cr
\hline
\end{tabular}
\end{center}
\caption{The vector coupling constants in SU(6)-symmetry relative
to $g_{NN\rho}$.}
\label{tab:su6}
\end{table}

The vector coupling constants to the $\phi$-field are given by SU(6)-symmetry
(see table~\ref{tab:su6}). 
How does one get these relations?

Here we discuss the coupling scheme for the baryons to mesons in SU(3)
symmetry. The baryons are members of an octet, the mesons of a nonet (a octet
plus a singlet) under
SU(3) symmetry. 
Coupling two octets (for example two baryons) results in 
\begin{equation}
8\otimes 8 = 1\oplus 8^1 \oplus 8^2 \oplus 10 \oplus 10^* \oplus 27
\quad .
\end{equation}
These terms have to be multiplied with the octet of the mesons now and the
result must be an overall singlet, so that the interaction (and the Lagrangian)
is invariant under
SU(3) symmetry. As there appears two octets in the above product, 
there are two possibilities to form an overall singlet in the
Lagrangian. They are usually called symmetric and antisymmetric coupling.
The latter one is given by 
\begin{equation}
{\bf L}_f = -g_8 \cdot i f_{imn} \bar\psi_m \psi_n \phi_i \quad ,
\end{equation}
with the antisymmetric structure constants of SU(3)
\begin{equation}
f_{imn} = \frac{1}{4i}{\rm Tr} ([\lambda_m ,\lambda_n]\lambda_i)
\end{equation}
and therefore also called F-type coupling. 
The octet coupling constant $g_8$ is a
universal coupling strength.
One can rewrite the structure of the above interaction in matrix form.
We normalize the SU(3) matrices as 
\begin{equation}
{\rm Tr} \lambda_i\lambda_j = 2\delta_{ij} \quad .
\end{equation}
and define the baryon and meson matrices as
\begin{equation}
\bar B = \frac{1}{\sqrt{2}} \lambda_m\bar\psi_m
\qquad
B = \frac{1}{\sqrt{2}} \lambda_n\psi_n
\qquad
M = \lambda_i\bar\phi_i
\end{equation}
This gives the compact notation
\begin{equation}
{\bf L}_f = - g_8 \cdot {\rm Tr} ([\bar B,B] M )
\end{equation}
The case for the symmetric coupling is analogous. 
Here the symmetric structure constants defined as 
\begin{equation}
d_{imn} =  \frac{1}{4}{\rm Tr} (\{ \lambda_m ,\lambda_n\} \lambda_i)
\end{equation}
give
\begin{equation}
{\bf L}_d =  -g_8 \cdot d_{imn} \bar\psi_m \psi_n \phi_i =
- g_8 \cdot {\rm Tr} (\{\bar B,B\} M ) \quad .
\end{equation}
This is also called the D-type coupling.

The baryon matrix is given explicitly as
\begin{equation}
B = \left( \begin{array}{ccc}
\frac{1}{\sqrt{2}}\Sigma^0+\frac{1}{\sqrt{6}}\Lambda^0 & \Sigma^+ & p \cr
\Sigma^- & -\frac{1}{\sqrt{2}}\Sigma^0+\frac{1}{\sqrt{6}}\Lambda^0 & n \cr
-\Xi^- & \Xi^0 & -\frac{2}{\sqrt{6}}\Lambda^0  \cr
\end{array} \right) \quad ,
\end{equation}
and analogous for the antibaryon matrix.
The general form of the Yukawa coupling is a mixture of 
D-type and F-type coupling
\begin{equation}
{\bf L}_{\rm SU(3)} = - g_8 \alpha{\rm Tr} ([\bar B,B] M )
+g_8 (1-\alpha) {\rm Tr} (\{\bar B,B\} M ) \quad .
\end{equation}
Here $\alpha$ denotes the F/(F+D) ratio.
For the vector mesons one gets the following coupling terms
\begin{eqnarray}
{\bf L}_{\rm SU(3)} &=&
- g_{NN\rho} \bar N \tau N \rho 
- g_{\Sigma\Lambda\rho}(\bar\Sigma\Lambda + \bar\Lambda\Sigma) \rho
- g_{\Sigma\Sigma\rho} \bar\Sigma \tau \Sigma \rho 
- g_{\Xi\Xi\rho} \bar \Xi \tau \Xi \rho \cr
&&
- g_{NN\omega_8} \bar N N \omega_8
- g_{\Lambda\Lambda\omega_8} \bar\Lambda\Lambda \omega_8
- g_{\Sigma\Sigma\omega_8} \bar\Sigma\Sigma \omega_8
- g_{\Xi\Xi\omega_8} \bar\Xi \Xi \omega_8 \cr
&&
- g_{\Lambda NK^*} (\bar N \tau \Lambda K^*  + \bar \Lambda \tau N \bar K^*) 
- g_{\Sigma NK^*} (\bar N \tau \Sigma K^* + \bar \Sigma \tau N \bar K^*) \cr
&&
- g_{\Xi\Lambda K^*} (\bar \Lambda \tau \Xi K^* + \bar \Xi \tau \Lambda \bar K^*)
- g_{\Xi\Sigma K^*} (\bar \Sigma \tau \Xi K^* + \bar \Xi \tau \Sigma \bar K^*)
\end{eqnarray}
and the meson-baryon coupling constants are now related to the F/(F+D)
ratio and the overall octet coupling strength by
$$
\begin{array}{rcl}
g_{NN\rho} = g_8 &
g_{\Sigma\Lambda\rho} = \frac{2}{\sqrt{3}} g_8 (1-\alpha_{ps}) &
g_{\Sigma\Sigma\rho} = 2g\alpha_{ps}
\\
g_{NN\omega_8} = \frac{1}{\sqrt{3}} g_8 (4\alpha_{ps}-1) &
g_{\Lambda\Lambda\omega_8} = - \frac{2}{\sqrt{3}} g_8 (1-\alpha_{ps}) &
g_{\Sigma\Sigma\omega_8} = \frac{2}{\sqrt{3}} g_8 (1-\alpha_{ps}) 
\\
g_{\Xi\Xi\omega_8} = -\frac{1}{\sqrt{3}} g_8 (1+2\alpha_{ps}) &
g_{\Lambda NK^*} = -\frac{1}{\sqrt{3}} g_8 (1+2\alpha_{ps}) &
g_{\Sigma NK^*} = g(1-2\alpha_{ps}) 
\\
g_{\Xi\Lambda K^*} = \frac{1}{\sqrt{3}} g_8 (4\alpha_{ps}-1) &
g_{\Xi\Sigma K^*} = -g_8 &
g_{\Xi\Xi\rho} = -g(1-2\alpha_{ps}) .
\end{array}
$$
As noted before, there is also a singlet state for the mesons which couples to
all baryons with the same strength
\begin{equation}
{\bf L}_1 = - g_1 (\bar N N + \bar \Lambda \Lambda + \bar \Sigma\Sigma
+ \bar\Xi\Xi ) \omega_1 \quad ,
\end{equation}
which means that 
\begin{equation}
g_{NN\omega_1} = g_{\Lambda\Lambda\omega_1} = g_{\Sigma\Sigma_1}
= g_{\Xi\Xi\omega_1} \quad .
\end{equation}
The physical states are not $\omega_8$ and $\omega_1$ but the $\omega$ meson
and the $\phi$ meson which are mixed states
\begin{eqnarray}
\omega &=& \omega_8 \cos \theta - \omega_1 \sin\theta \cr
\phi &=& \omega_8 \sin \theta + \omega_1 \cos\theta \quad .
\end{eqnarray}
For the vector meson it is known, that the $\phi$ meson is nearly a 
pure s\=s state as it decays mainly to kaons. Under this condition, the
mixing is called ideal and the mixing angle is 
$\tan \theta = 1/\sqrt{2}$, $\theta \approx 35.3^{\rm o}$. 

If the nucleon does not couple to the (purely strange) $\phi$ meson,
then 
\begin{equation}
g_1 = \sqrt{6} g_8 = \sqrt{6} g_{NN\rho} \quad .
\end{equation}
In SU(6) symmetry, which is a special case of SU(3) symmetry, only the F-type
coupling remains for the vector mesons. This is in accordance with
the vector dominance model \cite{Dov84}. In this case, the relations for the
vector mesons can be given as in table~\ref{tab:su6}.
These relations reflect simple quark counting rules.
So the $\Lambda$ couples to the $\omega$ meson only 2/3 as strong as the
nucleon, as it has only two light quarks. On the other side, the doubly strange
baryon $\Xi$ couples twice as strong to 
the (hidden) strange meson $\phi$ compared
to the $\Lambda$ which has only one strange quark.

The scalar coupling constants to the
$\sigma^*$-field are fixed by the potentials
\begin{equation}
U^{(\Xi)}_\Xi \approx U^{(\Xi)}_\Lambda \approx
2U^{(\Lambda)}_\Xi \approx 2U^{(\Lambda)}_\Lambda \approx 40 \mbox{ MeV}
\end{equation}
which is motivated from one-boson exchange models and the measured strong
$\Lambda\Lambda$ interaction \cite{Sch94}.
Note that the nucleons do not couple to these new fields.
This fixes all the parameters of model 2.

\jfigps{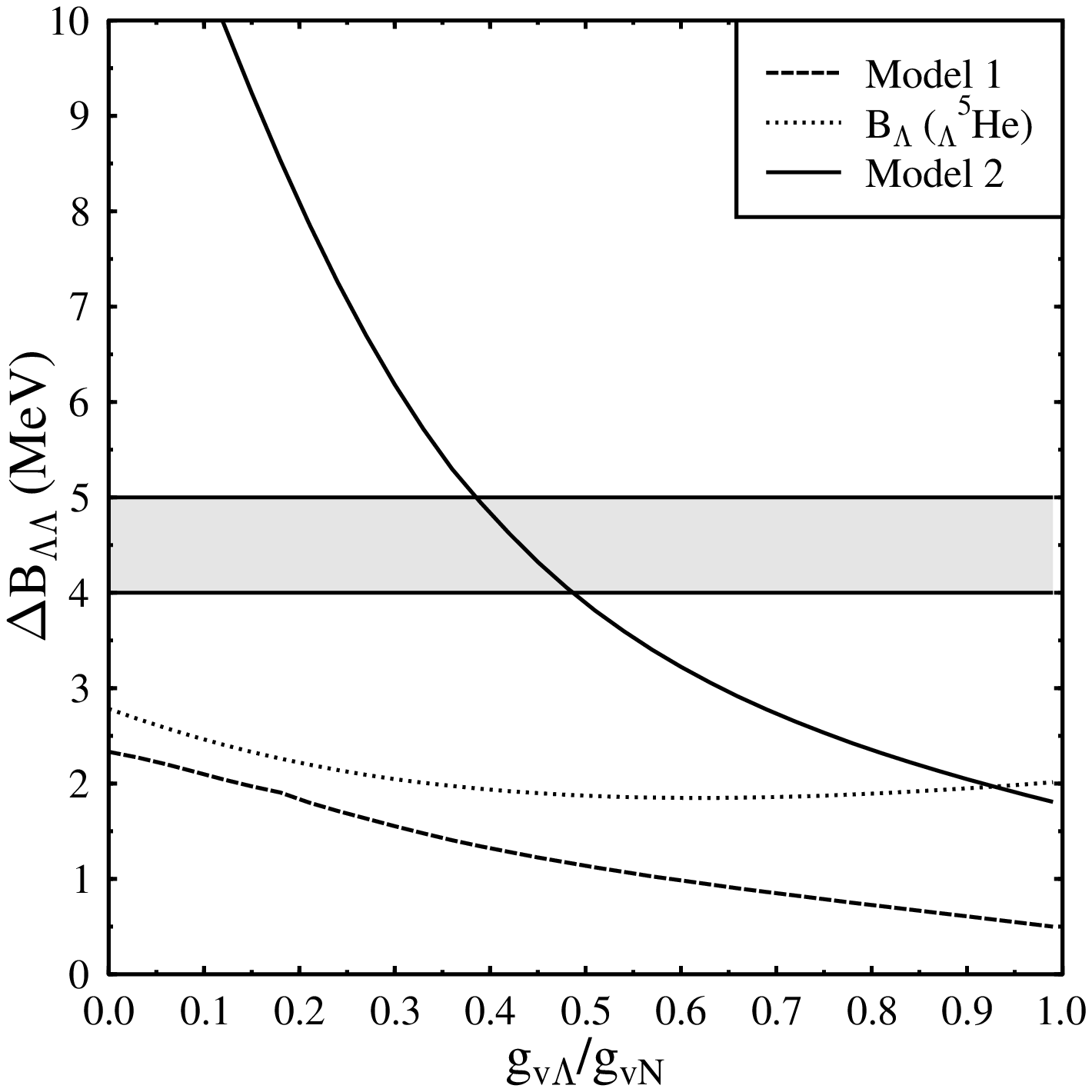}{The matrix element 
$\Delta$B$_{\Lambda\Lambda}$ 
as a function of the vector coupling constant in model 1 and 2.}

This extended model 2 is now closer to the experimental value of
the $\Lambda\Lambda$ interaction matrix element (see fig.~\ref{fig:pic4}).
Due to the additional attractive forces coming from the hidden strange meson
exchange, the $\Lambda\Lambda$ matrix element is now much stronger than in
model 1.


\subsection{Properties of Strange Hadronic Matter}


In the following we study the properties of metastable exotic multihypernuclear
objects (MEMO's). Metastability means that strong processes are forbidden,
so that the system can only decay weakly and lives therefore on the timescale
of the weak interaction of $10^{-10}$ s.
Inside a bound system some of the strong processes can be energetically
forbidden due to medium effects, i.e.\ the strong reaction is Pauli-blocked.
Only two processes 
\begin{eqnarray}
{\rm n}+\Xi^0&\longrightarrow&
\makebox[2cm][c]{$
\Lambda + \Lambda$}
\qquad  (\Delta E= 23   {\rm MeV}) \label{eq:ksi1}\\
{\rm p}+\Xi^-&\longrightarrow&
\makebox[2cm][c]{$
\Lambda + \Lambda$}
\qquad  (\Delta E= 28   {\rm MeV})\label{eq:ksi2}
\end{eqnarray}
have such low Q-values that this effect occurs in a bound system.

\jfigps{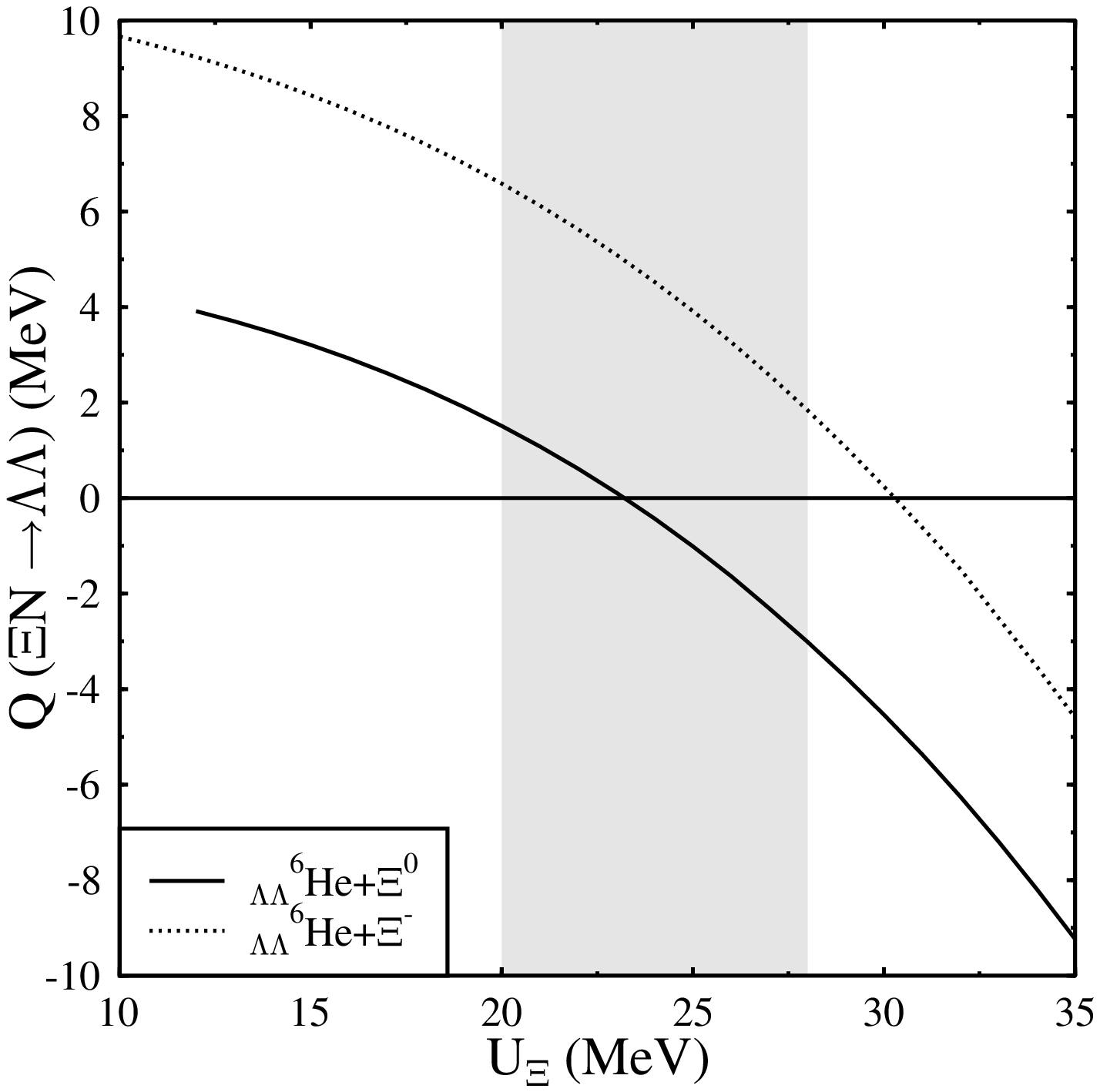}{The Q-value of the strong process $\Lambda\Lambda\to\Xi$N for
the system with $A=7$ and $S=-4$ as a function of the potential
depth of the $\Xi$ in nuclear matter. The process is not allowed
for negative values, hence the system can only decay weakly.}

Therefore we study systems composed of $\{ {\rm p,n,}\Lambda,\Xi^0,\Xi^- \}$
and search for metastable combinations by the following recipe:
\begin{itemize}
\item Start with a deeply bound nuclear core, e.g.\ $^4$He or $^{56}$Ni,
so that the neighboring nuclei are much less bound.
\item Fill up the $\Lambda$-levels so that the products of the above reactions
are Pauli-blocked.
\item Add as many $\Xi$'s as possible unless the strong reactions can open.
\end{itemize}
The effect of the Pauli-blocking is demonstrated in Fig.~\ref{fig:pic5}
for the lightest possible system $_{\Xi^0\Lambda\Lambda}{}^7$He.
The Q-value is given by
\begin{equation}
Q = m_\Xi + m_N - 2m_\Lambda - B_\Xi(1s) - B_N(1s)
\label{eq:qwert}
\end{equation}
and is plotted versus the potential depth of the $\Xi$ in nuclear matter.
The system is likely to be bound and metastable if this value gets negative.
This happens for $U_\Xi > 22$~MeV which is in accordance with experiment.

\jfigps{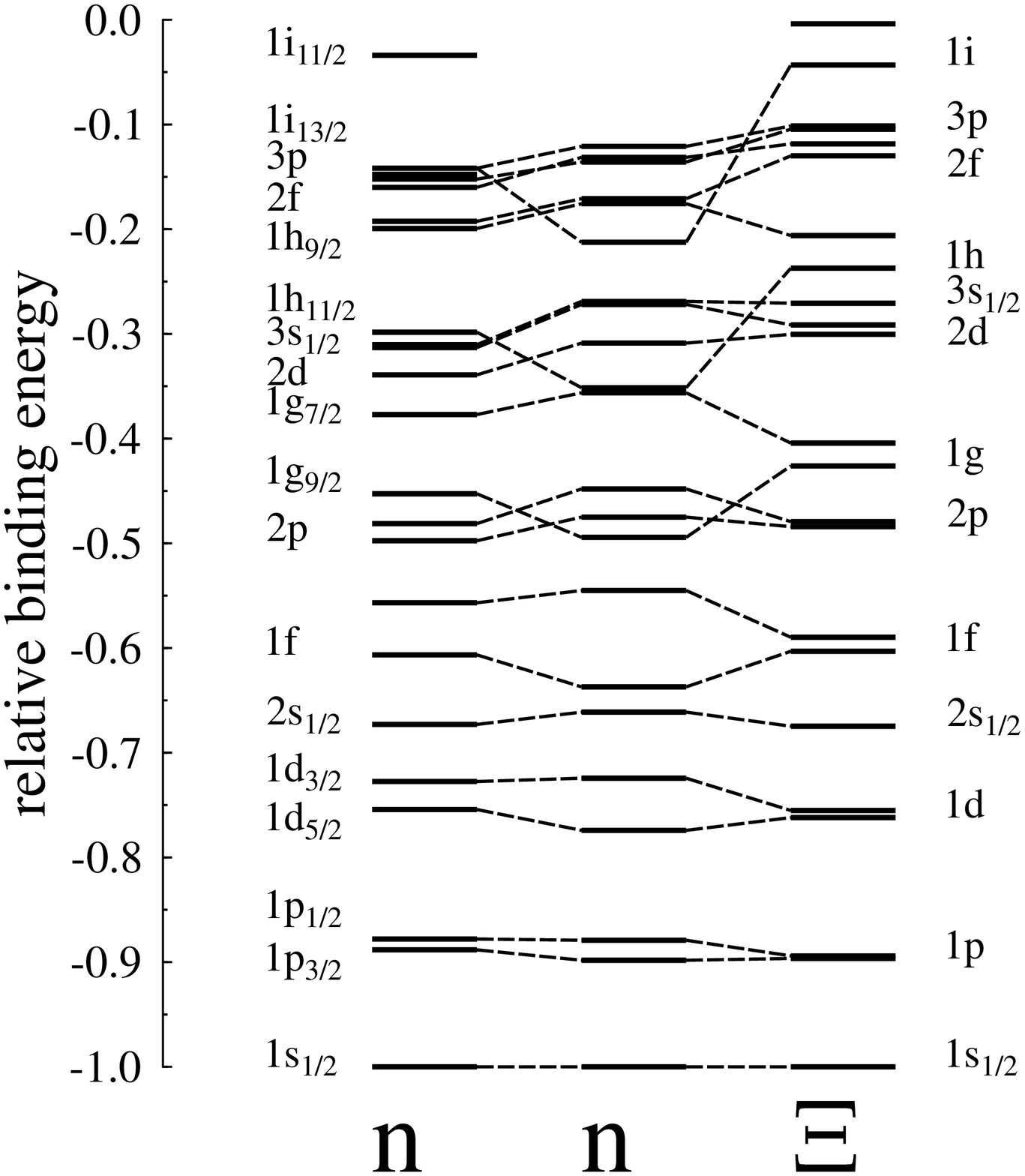}{The single particle energy of the system
$^{208}$Pb with 70 $\Lambda$, 18 $\Xi^0$, and 70 $\Xi^-$ (middle column: 
neutron levels, right column: $\Xi^0$ levels). The neutron levels
for $^{208}$Pb are shown in the left column for comparison.}

Heavier systems with a doubly magic nucleon core like $^{208}$Pb 
can be also filled up with
hyperons ($\Lambda$, $\Xi^0$, and $\Xi^-$) selfconsistently so that they are
metastable according to the recipe outlined. The presence of many hyperons will
change the sequence of single-particle energies and accordingly the magic
numbers appreciably. One gets a new shell ordering for a large strangeness
$|S|$. Fig.\ \ref{fig:pic6} shows the single particle energies of the system
$^{208}$Pb with 70 $\Lambda$, 18 $\Xi^0$, and 70 $\Xi^-$ which is bound by
$E_B/A = -12.9$ MeV. The neutron levels in the ordinary nucleus (left column)
and in the strangeness-rich system (middle column) are compared to each other.
The shell levels show some reordering especially for the higher lying levels
due to the presence of many hyperons.
For the protons, the change is more pronounced due to
Coulomb effects and the magic number is now 70. The single
particle levels of the $\Xi^0$ indicates a quite small spin-orbit splitting.
They are heavier than nucleons and in addition feel a smaller spin-orbit force
than nucleons due to the smaller coupling constants. The magic numbers here are
the ones of the Wood-Saxon potential without spin-orbit splitting: 2, 8, 18,
20, 34, 58, and 92.

\jfigps{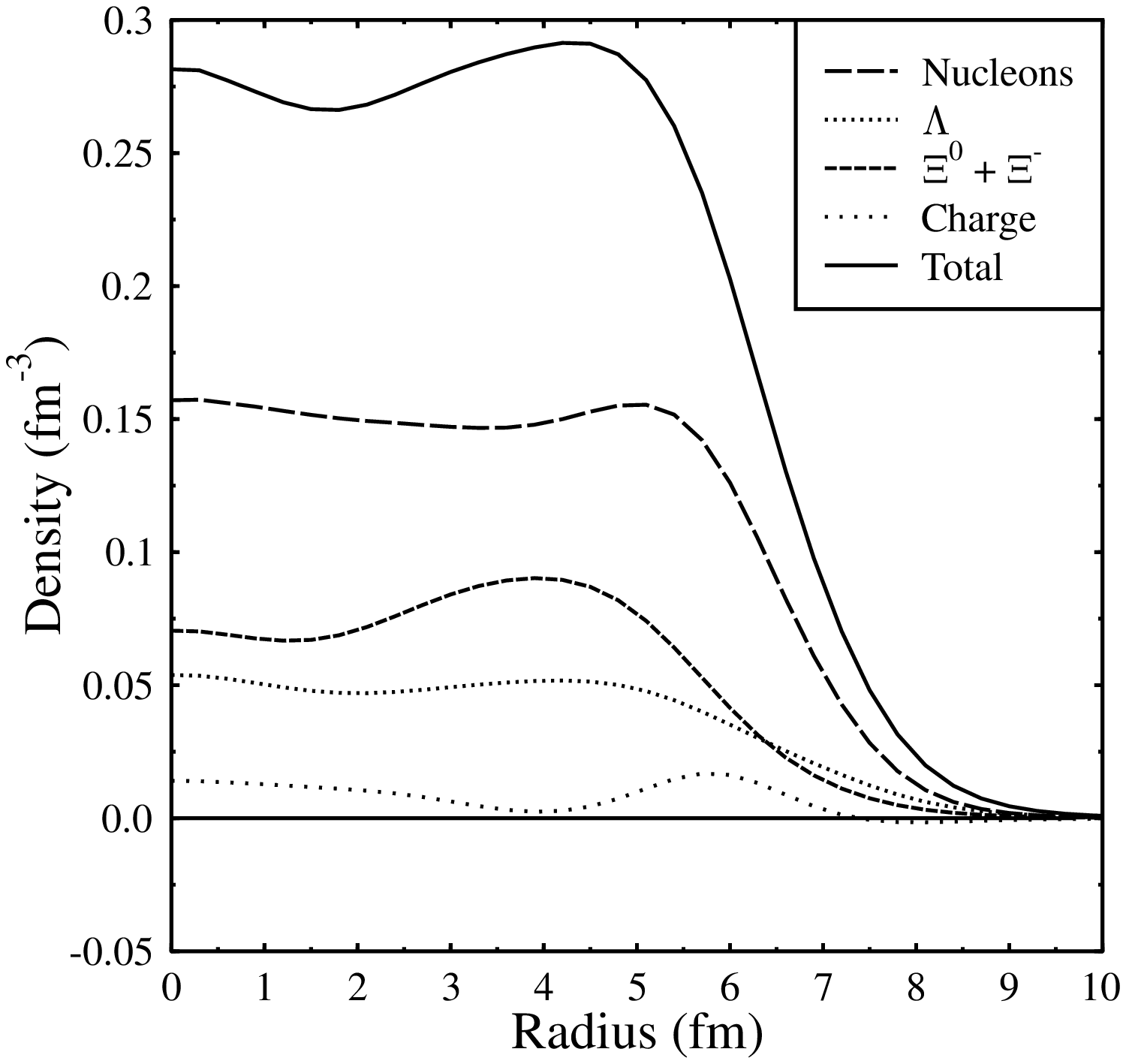}{The density distribution of the system
$^{208}$Pb with 70 $\Lambda$, 18 $\Xi^0$, and 70 $\Xi^-$.}

The density distribution of the system with hyperons is plotted in 
Fig.\ \ref{fig:pic7}.
The density distribution for the nucleons is still
around normal nuclear density. The total density is increased to about
2$\rho_0$ as the hyperon density has to be added up. Note, that the hyperons
are treated as distinguishable point particles. Effects which take into account
the substructure of the hadrons might alter this picture but will be not
discussed here. The $\Lambda$ density distributions shows a broad surface as
the $\Lambda$ single particle levels are occupied until the least bound level
which ensures the Pauli blocking mechanism for the $\Xi$'s. The $\Xi^-$'s are
less bound than the protons so that the charge density changes sign at the
surface and gets negative.

\jfigps{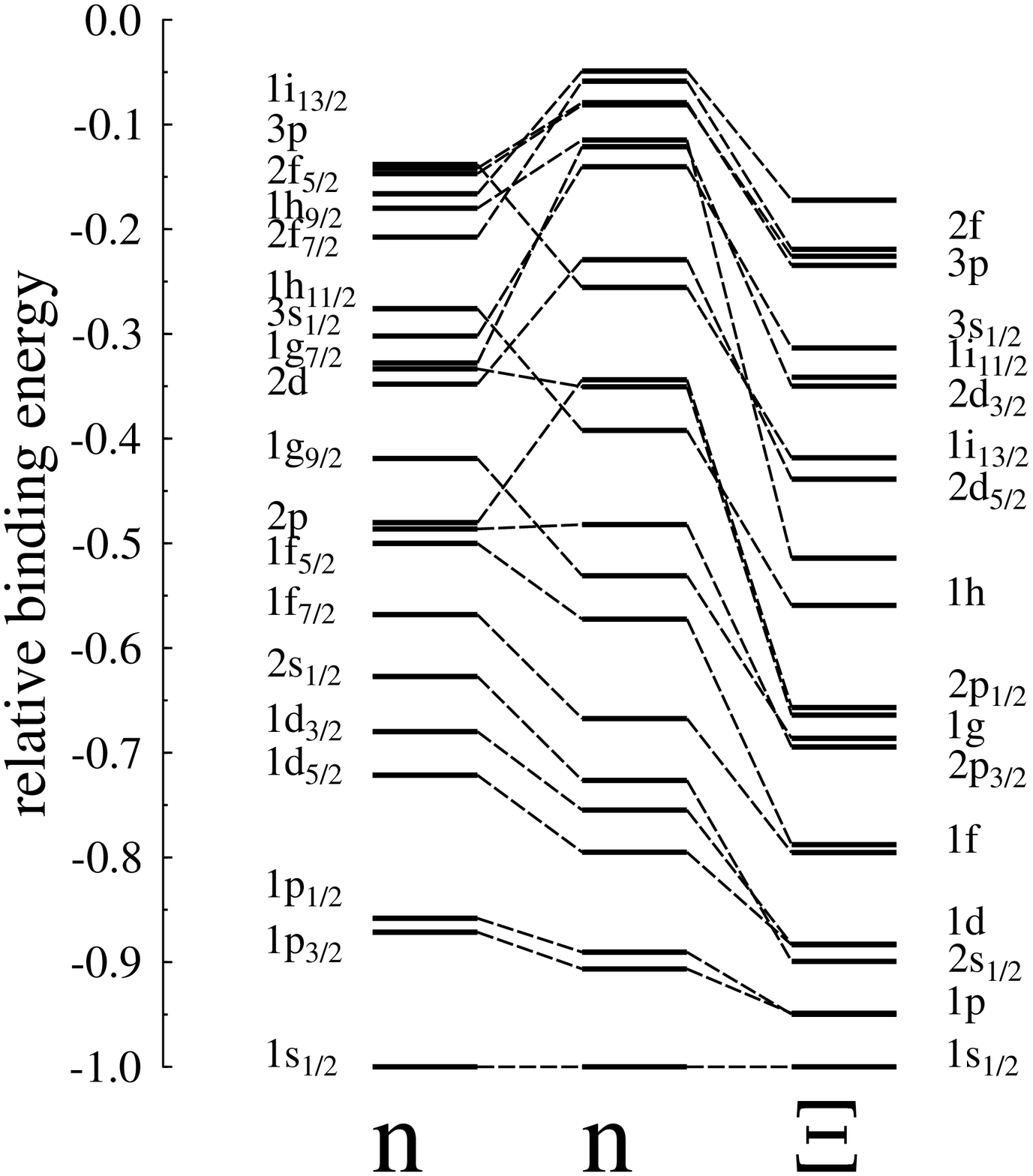}{The single particle energy of the system
$^{180}$Th with 92 $\Lambda$, 92 $\Xi^0$, and 70 $\Xi^-$
(middle column:  neutron levels, right column: $\Xi^0$ levels). 
The neutron levels
for $^{180}$Th are shown in the left column for comparison.}

An analogous example for model 2 with the strongly attractive hyperon-hyperon
interaction is the system $^{180}$Th with 92 $\Lambda$, 92 $\Xi^0$, and 70
$\Xi^-$. Now the binding energy is $E_B/A = -21.4$ MeV emerging from the
additional attractive forces between the hyperons. The nucleus $^{180}$Th
itself is unstable and emits protons immediately. The presence of the hyperons,
especially the negatively charged $\Xi^-$'s, shift the unbound proton levels
down, so that all the proton levels are bound. The magic numbers for nucleons
has changed then from 82 to 90 for both, neutrons and protons for large $|S|$.
The big level changes can be seen in 
Fig.\ \ref{fig:pic8}. 
In particular, the $n=1$ (nodeless) levels are more
bound than in ordinary nuclei. 
Also for the $\Xi$ levels, the nodeless 
1i level now appears below the 3s level contrary
to the ordinary oscillator shell ordering.

\jfigps{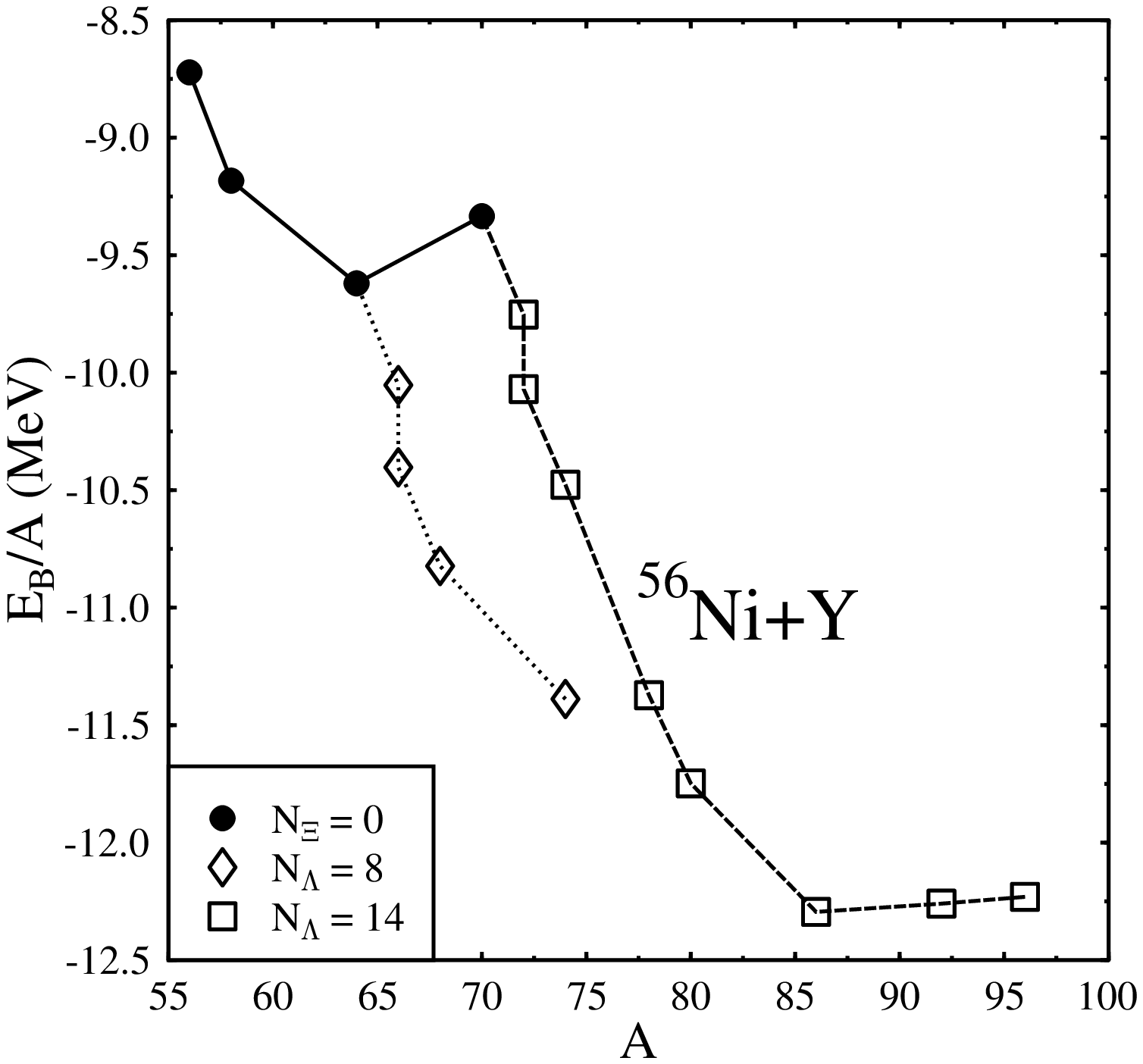}{The binding energy per baryon versus the baryon number $A$
for $^{56}$Ni with hyperons added in model 1.}

Now we discuss the
calculations for several sequences starting with a nuclear core and
subsequently filling up the levels with hyperons.
Fig.~\ref{fig:pic12} shows the case for the nuclear core $^{56}$Ni in model 1.
When the p-shell of the $\Lambda$'s is filled
(N$_\Lambda =8$) it is energetically favourable to add $\Xi$'s to the system.
If the next shell is also filled up (N$_\Lambda =14$), 
the maximum number of added $\Xi$'s
can be even higher enriching the system with a lot of strangeness.

\jfigps{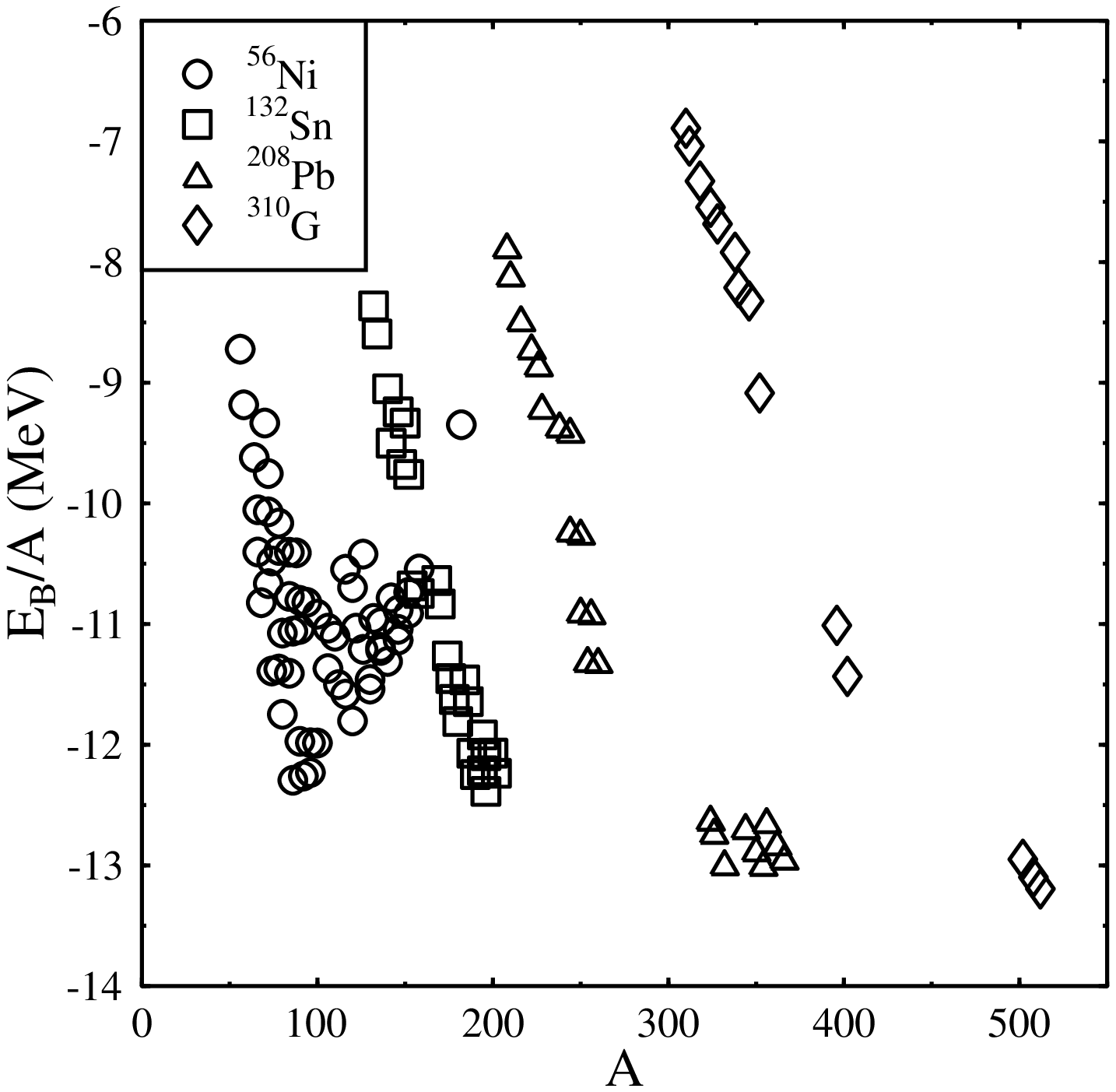}{The binding energy per baryon in model 1 for various sequences
starting with $^{56}$Ni, $^{132}$Sn, $^{208}$Pb and
the superheavy nuclear core $^{310}$G with 126 protons and 184 neutrons.}

Fig.\ \ref{fig:pic14} shows the sequences for the doubly magic cores 
$^{56}$Ni, $^{132}$Sn, $^{208}$Pb, and $^{310}$G ($Z=126$, $N=184$).
The glue-like effect of the hyperons result in an enhanced binding energy.
Adding hyperons stabilizes
the systems, i.e.\ enhances the binding energy due to the opening of a new
degree of freedom.
Mass numbers of $A\approx 500$ can be easily reached with a rather large
binding energy of $E_B/A \approx -13$ MeV. These systems are stabilized due to
the presence of the $\Xi^-$, which avoid the instability due to the Coulomb
repulsion. 
The Pb sequence has systems with a total charge of $Z=12$ (82
protons and 70 $\Xi^-$). For the Ni sequence, one gets also zero charged
systems with 28 protons and 28 $\Xi^-$. Adding further hyperons for the small
systems results in less binding, therefore the parabolic shape for the Ni
sequence. One sees a linear shape for the heavier cores by adding more and
more hyperons. The sequences of the heavier systems stop when
the $\Lambda$ levels are filled up and this happens before there are more
$\Xi^-$'s than protons so that one sees only a linear curve. 

\jfigps{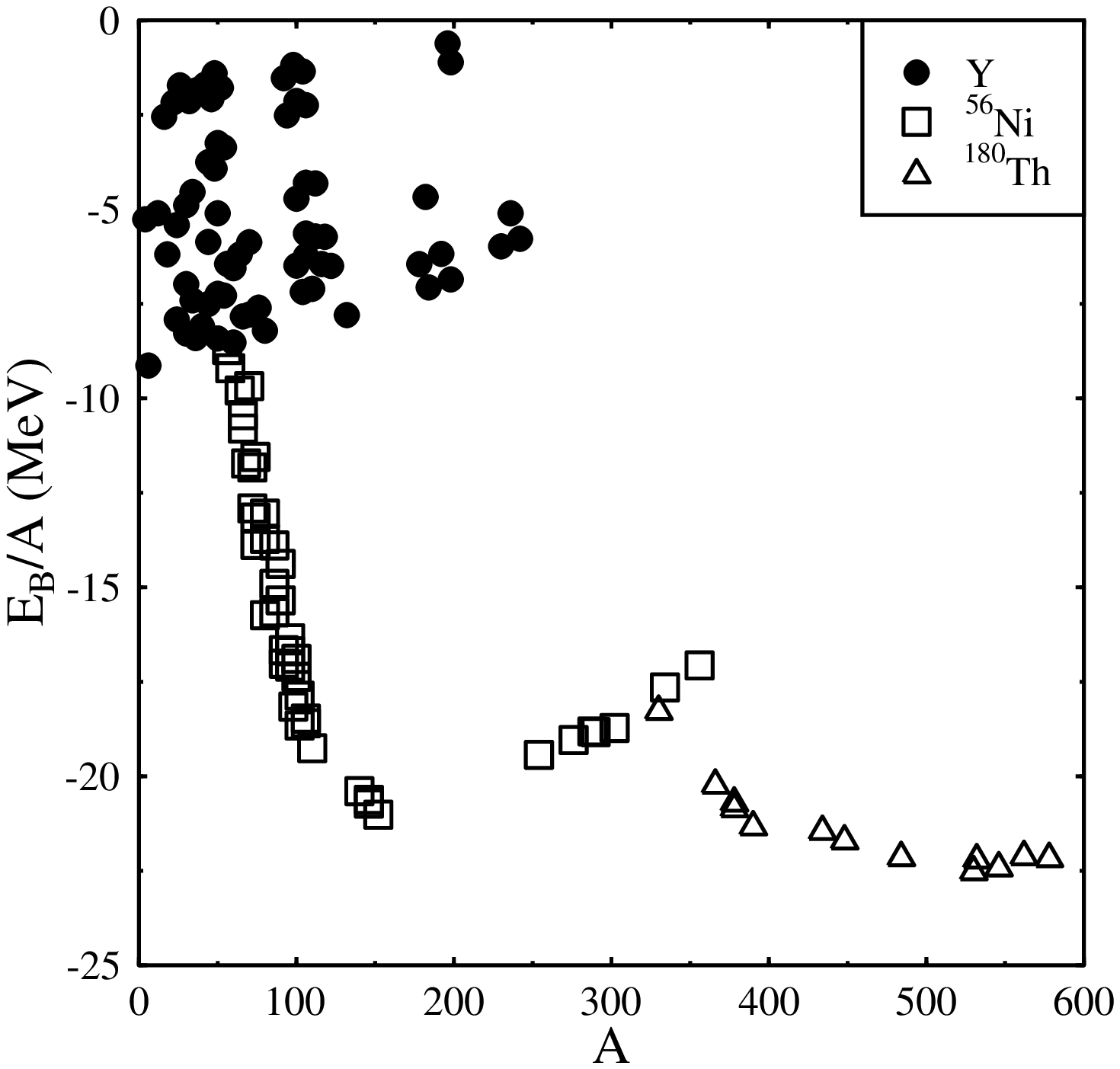}{The binding energy per baryon in model 2 for various sequences
starting with $^{56}$Ni, and $^{180}$Th.
The symbol Y denotes purely hyperonic systems.}

The analogous sequences for model 2 are depicted in Fig.\ \ref{fig:pic15}
for the nuclear cores $^{56}$Ni and $^{180}$Th. The other cores
($^{132}$Sn, $^{208}$Pb, and $^{310}$G) are not doubly magic anymore for large
$|S|$. The system $^{180}$Th is considerably stabilized with a large injection
of hyperons. About 150 hyperons are needed to achieve stability. 
The region of stability extends from $A=330$ to $A= 578$. 

\jfigps{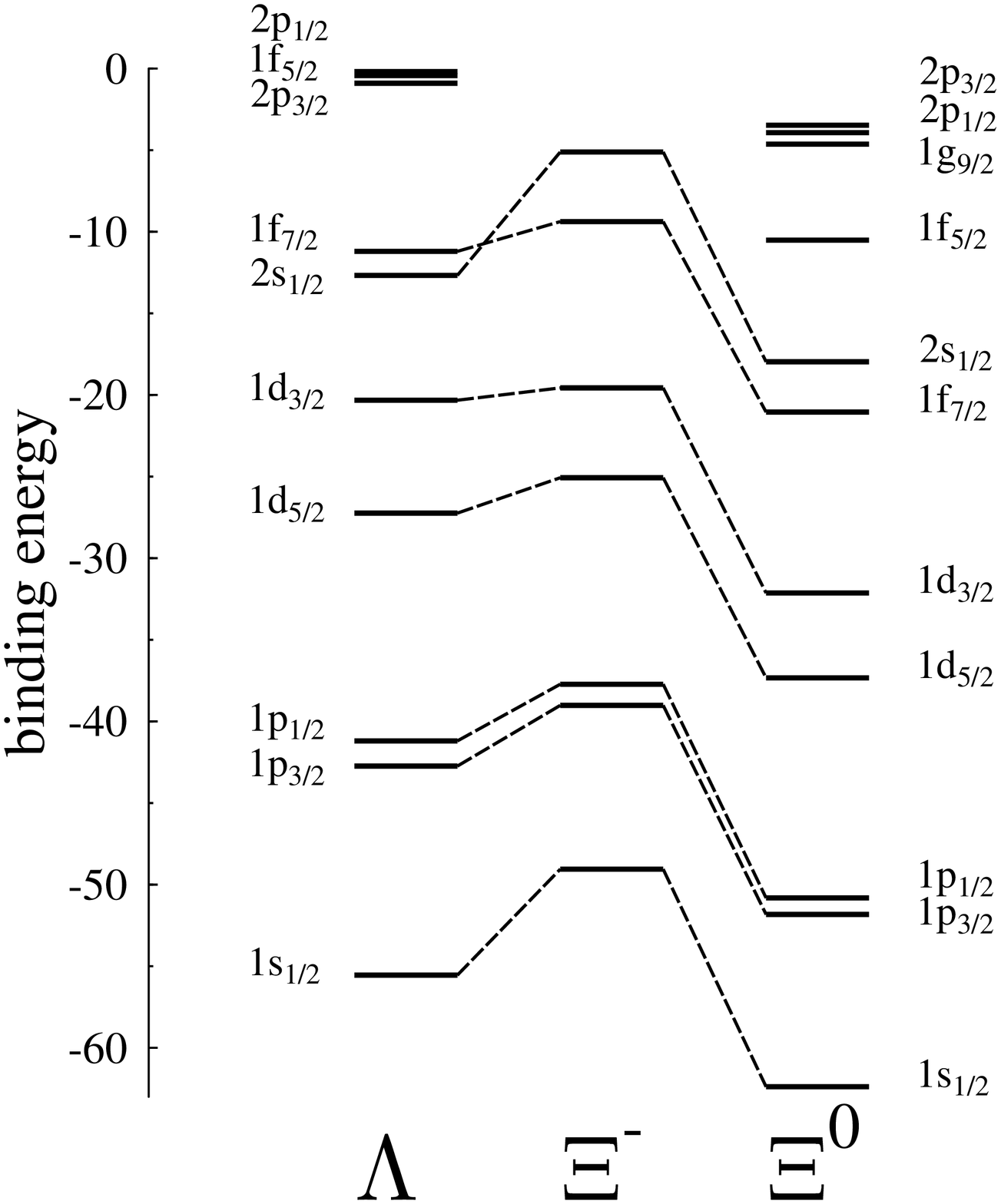}{The single particle energy of the purely 
hyperonic system $\{ 18\Lambda,34\Xi^0,28\Xi^- \}$ in model 2.}

Purely hyperonic
systems, composites of $\{\Lambda, \Xi^0,\Xi^-\}$ only, are stable in model 2
up to $A\approx 200$. They are denoted as Y which should not be confused with
the ordinary Ytterbium nucleus. Typical binding energies are $E_B/A=-5$ to
$-8.5$ MeV. The most stable system is the one for $A=60$ with
14 $\Lambda$, 28 $\Xi^0$, and 18 $\Xi^-$. The Coulomb repulsion prevents
systems larger than $A\approx 200$.
Interestingly, hyperonic matter is always highly negatively charged.
These systems are lying between $-0.5< Z/A < -0.2$.
Moreover, the strangeness fraction is also very high $f_s>1.6$ for this
type of matter. For isospin saturated systems of $\{ \Lambda,\Xi^0,\Xi^- \}$
one gets the values $f_s = 5/3$ and $Z/A = -1/3$, respectively.
The lightest bound system is likely to be $\{ 2\Lambda,2\Xi^0,2\Xi^- \}$
where every hyperon fills up their respective 1s-shell.
Fig.~\ref{fig:pic11l} shows as an example
the single particle energy for the hyperonic
system $\{ 18\Lambda,34\Xi^0,28\Xi^- \}$. The 1s-state has a binding energy
between 50 and 60 MeV for all three hyperon species. This reflects more or
less the assumed universal hyperon-hyperon interaction. The energy levels
of the $\Xi^0$ are a little bit deeper than the ones of the $\Lambda$ due
to the higher mass. The Coulomb repulsion shifts the states of the $\Xi^-$
to smaller binding energy compared to the others.

\jfigps{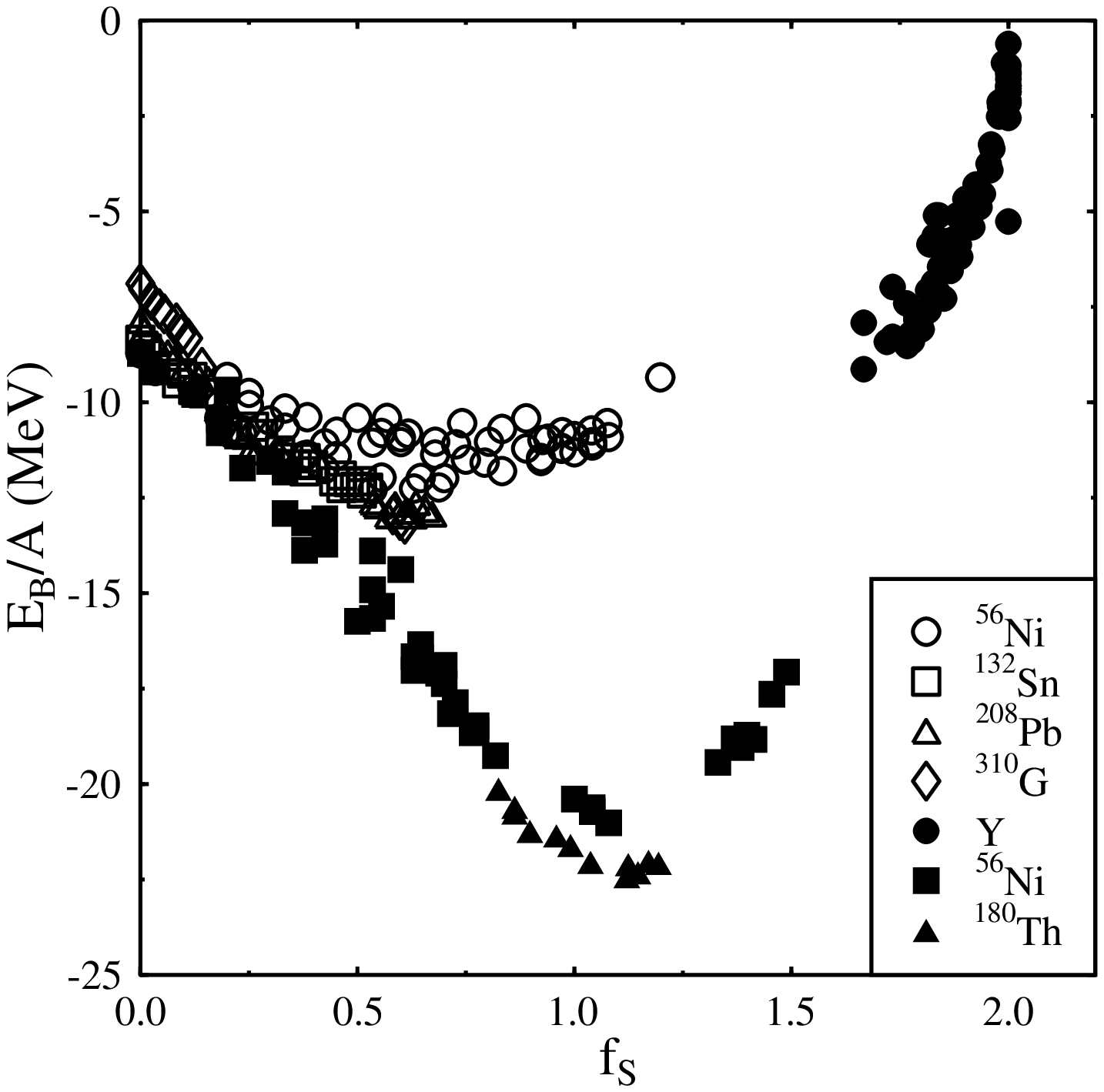}{Binding energy versus the strangeness fraction for SHM
in model 1 (nuclear cores $^{56}$Ni, $^{132}$Sn, $^{208}$Pb, $^{310}$G)
and model 2 ($^{56}$Ni,$^{180}$Th, Y denotes purely hyperonic systems).}

The binding energy of strange hadronic matter is plotted in
Fig.~\ref{fig:prl2} versus the strangeness fraction $f_s=|S|/A$ for both models
for various nuclear cores.
It is interesting to note that the minima of the curves are located around
$f_s \approx 0.6$ for model 1 and $f_s\approx 1.0$ for model 2 which
is in the same range as for strangelets.
The minima found in Fig.~\ref{fig:prl2} correspond to charge
to mass ratios around zero very
similar to the property of strange quark matter in its groundstate.
Nevertheless, the main difference between strangelets and strange hadronic 
matter, except, of course, of their internal structure, is that
the maximum binding energy of $E_B/A \approx -13$~MeV (model 1) and
$E_B/A \approx -23$~MeV (model 2) is by far to less to overcome
the mass difference between nucleons and hyperons of about 177~MeV.
Strange hadronic matter will thus decay weakly on a timescale
comparable to the lifetime of a $\Lambda $-hyperon.

\clearpage

\section{STRANGE MATTER IN
RELATIVISTIC HEAVY ION 
COLLISIONS}

Collision experiments of heavy ions at high bombarding energies
may be the only unrevealing chance and opportunity for both the
production and detection of small pieces of strange quark matter:
One major goal of these experiments is to unfold the temporary
creation and existence of the quark gluon plasma phase.
A direct proof of this fleetingly
small moment turns out to be a difficult and complicated task.
One therefore has to find signals which come directly and only 
from the transient existence of the QGP.
This will be different, if a (meta-)stable strange quark droplet will show up during
the break-up. This is quite similar
to the situation in the early universe \cite{Wit84},
where a remnant (or `ash') of the QGP
is created. In fact, it seems rather selfevident that a deconfined
quark matter state could probably act as a good starting point for
the agglomeration of quarks to strangelets, because the quarks are thought to
move quasifreely over longer distances similar as in the situation
inside strange quark matter.

In this section, we review the ideas that had led to the
search for strangelets at present heavy ion experiments at Brookhaven and at CERN:
In the first subsection, the important mechanism of strangeness separation
in the phase transition of a quark gluon plasma
back to hadronic degrees of freedom will be explained \cite{PRL87}.
The second subsection then addresses the exciting possibility of producing
strange quark matter droplets in heavy ion collisions. In particular
we will discuss why such droplets might cool and form long-lived cold
strangelets.
In the last subsection we will close our discussion in
critically emphasizing the detection
possibilities of strangelets and small MEMOs by their properties and lifetimes,
also in respect to the present experimental
undertaking at Brookhaven and at CERN.

\subsection{Strangeness separation during the QGP phase transition}
In nuclear collisions, strangeness
can only be produced in $s\bar s$ pairs due to the
conservation of hypercharge in strong interactions.
Thus, at first sight, there seems to be no chance to really succeed for
producing strangelets: The net strangeness of the plasma state
counting the difference of strange to antistrange quarks is zero from
the onset of the fireball's expansion. In this subsection we want to
give arguments,
that first if a {\em baryon-rich} and hot QGP is created in such collisions
and second if the strangeness degree of freedom is nearly saturated, the
strange quarks will separate from the antistrange quarks in a
nearly equilibrium
deconfinement $\leftrightarrow $ confinement transition and will predominantly
remain in the plasma phase \cite{PRL87}.

According to
fluid dynamic or microscopic models baryon stopping,
(baryon-)dense and hot matter formation and partial
thermalization has to be expected at AGS and CERN energies
for central collisions and heavy systems \cite{AK90}. One should be aware
of the fact, that the energy and baryon densities reached are so large,
that a phase transition to a quark gluon plasma may have occurred.

It was suggested and shown already some time ago
that an abundant number of strange and antistrange quarks are produced
in a hot QGP \cite{Mu82} by gluon fusion and that, accordingly, strangeness
saturates the phase space after a very short equilibration time which
may actually be shorter than the lifetime of the QGP phase \cite{Koc86,Ka86}.
In the meantime, an enhancement of strange particle production in nuclear collisions has
been observed in many experiments \cite{Ha92,WA85,NA36,NA35}. However,
it has also been learned that such an enhancement alone does not
make a reliable signature for the QGP. Strange particles, especially K mesons
and $\Lambda $ hyperons can be copiously produced in hadronic reactions
before the nuclear fireball reaches equilibrium \cite{Ma89}.
Yet, the enhancement of the $\Lambda $ hyperon over a wider rapidity range \cite{NA35}
and especially the observed and strongly enhanced yields of antihyperons
\cite{WA85,NA36,NA35}, which were also proposed
as a signature for QGP \cite{Koc86},
require particular microscopic modifications of the hadronic cascades:
Color rope formation \cite{So92}; multiple string breaking and
(!) decaying multi-quark droplets \cite{We93}. Such modifications
can already be considered as precursor phenomena associated with QGP formation.

In the following, the deconfinement phase transition is
assumed to be of first order, implying
that the relaxation times for chemical transmutations as well as the
hadronization time are small compared to the overall transition time.
The Gibbs criteria read
\begin{eqnarray}
\label{cg7}
T_{QGP} & = & T _{HG} \, \, \, ,\nonumber \\
P_{QGP} & = & P _{HG}  \, \, \, , \\
\mu^B_{QGP} & = & \mu^B_{HG} \, \, \, , \nonumber \\
\mu^s_{QGP} & = & \mu^s_{HG} \nonumber \, \, \, .
\end{eqnarray}
In particular the last condition is significant for the separation to be
discussed: In a Gibbs phase equilibrium the chemical potentials are continuous across
the phase boundary, whereas the corresponding densities are (or might be)
discontinuous.
This first implies that the abundance of especially the strange hadrons during
the equilibrium transition are governed by the same quarkchemical potentials,
$\mu_q$ for the light quarks and $\mu_s $ for the strange quarks,
e.g.
\begin{eqnarray}
\label{cg8}
\mu_{K,\bar{K}} & = & \pm \mu _q \mp \mu_s \, \, \, , \nonumber \\
\mu_{\Lambda,\bar{\Lambda }} & = & \pm 2 \mu _q \pm \mu_s \, \, \, , \nonumber \\
\mu_{\Xi ,\bar{\Xi }} & = & \pm  \mu _q \pm 2\mu_s \, \, \, ,  \\
\mu_{s,\bar{s}} & = & \pm \mu_s  \, \, \, . \nonumber
\end{eqnarray}
Second, the corresponding density to the strange
quark potential is the net strangeness content, i.e.
the strangeness ratio $f_s$ defined in (\ref{cg3}). In general
this ratios need {\em not} to be the same during transition:
$f_s^{QGP} (\mu_q, \mu_s, T) \neq
f_s^{HG} (\mu_q, \mu_s, T) $.

\begin{figure}[ht]
\vspace*{7cm}
\caption{
Schematic picture of the separation of strangeness during phase transition
in a baryon-rich system
\label{separ}
}
\end{figure}

Consider now the phase transition of the QGP to the hadron gas at some
critical temperature.
{\em How does the strange and antistrange
quarks hadronize during this transition?} There is no stringent reason why
these different quarks do hadronize in the same manner and time,
especially if one thinks of a baryon rich system.
At the beginning of the hadronization ($t=t_S$) there is only the QGP phase.
Because of an overall strangeness conservation the number of
strange quarks equals the number of antistrange quarks, the
strange chemical potential $\mu_s = 0$ and the net strangeness content
of the plasma phase is that of the system,
$f_s^{QGP} (t=t_S) \equiv 0$.
Now, quarks combine to hadronic
particles and leave the plasma phase, but hadrons may also decompose
and go back to the plasma state. Because of the progressing expansion
of the whole fireball, however, the hadron phase turns bigger while
the plasma phase decreases.
It is `simple' for the
antistrange quarks to materialize into kaons, because of the
(immense) surplus of massless quarks compared to their
antiquarks. The strange quarks can combine
to $\Lambda $-particles, but these are rather heavy and it is
energetically much easier for them to stay in the plasma, when
hadronization proceeds.
The possible strangeness content in the two different phases
is schematically drawn in Fig. \ref{separ}.
A large antistrangeness builds up in the hadron matter
while the QGP retains a large net strangeness excess. This separation
will occur only when the system carries a positive net baryon number.

\begin{figure}[ht]
\vspace*{8.5cm}
\caption{
Fraction of net strangeness to baryon number present in the QGP phase
as a function of the baryochemical potential $\mu_q $ and the
volume fraction $\chi $. Note that $f_s $ can exceed 0.5. The
path of an isentropic expansion is also shown.
\label{separ2}
}
\end{figure}

To be more specific, requiring that the total strangeness during the
phase transition in the
combined system of two phases vanishes, leads to the constraint
\begin{equation}
\label{cg9}
V_{QGP} (\rho _s - \rho _{\bar{s}}) \, + \,
V_{HG} (\rho _K + \rho _Y - \rho _{\bar{K}} - \rho _{\bar{Y}}) \, \equiv 0
\, \, \, .
\end{equation}
Together with the Gibbs criteria of pressure equilibrium, (\ref{cg7}),
the phase transition temperature and the strange quark potential are then
implicit functions of the quark chemical potential $\mu _q$, and, for example,
the volume fraction $\chi$ of the hadron phase
to the total system, i.e.
 $\chi = V^{HG} / (V^{HG} + V^{QGP})$
\cite{PRL87}.
In Fig. \ref{separ2} the fraction $f_s^{QGP}(\mu_q, \mu_s, T) $ of strange quarks to all quarks
present in the QGP is shown as a function of the quark potential
$\mu_q$ and the volume fraction $\chi$.
At the beginning of the phase transition ($V^{HG}=0$) the well
known result $\mu_s =0$ is recovered, $s$ and $\bar s$ quarks are produced
in pairs only. On the other hand, for vanishing QGP ($V^{QGP}=0$)
zero net strangeness leads to a nonzero value of the strange
chemical potential $\mu_s > 0$ \cite{PRL87}. A different strange particle
production shows up in the dominance of the associated production over
the direct pair production at finite baryon density.
During the coexistence of the two phases, an additional channel opens
up for the strangeness: besides the associated pair production in the
hadron gas it is possible to have, for example, associated production of
a K - meson in the sector of the hadron gas phase and the s-quark
staying in the QGP. This leads to a net strangeness content $f_s^{QGP} (t>t_S)$
larger than 0 in the quark phase and to a diminished hyperon abundance,
$f_s^{HG}<0$,
the hyperons being too massive, in the hadronic gas sector at finite net
baryon densities. At vanishing overall net baryon density, i.e.
at $\mu_q =0$, (and vanishing
net strangeness density) the transition is completely symmetric for strange
and antistrange particles, i.e. $\mu_s =0$ throughout the transition,
and separation can not occur as one would expect.

Of special importance is the fact that the accumulation of s quarks
in the plasma phase grows with decreasing plasma volume. This opens
up the possibility that s-quarks may be bound not only in hyperons and strange mesons:
They could form strange quark matter clusters or MEMOs
which might be metastable objects.
A baryon-rich environment may thus be
more feasible for strangelets than the situation in the hot quark
phase in the early universe with only a tiny small surplus.
Also the cooling of the plasma should in addition not only be due to
the expansion, but also due to
`prefreezeout' evaporation processes of hadronic
particles, because of the relatively small length scale in heavy ion
collisions compared to the astrophysical picture: Surface processes
like evaporation should become important.
This issue and the possible distillation of strangelets
will be discussed in the next subsection.

Finally we remark that the separation mechanism can be probed
by density interferometry with hyperons \cite{GrM88} or kaons
\cite{Gy92}. The hadrons with negative strangeness, the
$\Lambda $, $\bar{K}$ and $\bar{K}^0$, are expected to be produced
mainly at the last stage of the phase transition when the size of
the quark phase volume has become quite small. As the hyperons
or antikaons are quite heavy, their thermal velocity is relatively small
so that at freeze-out they probably still will be localized near the center of the
system. Such a close localization $r_0 \approx 2 - 3 $ fm can be seen by
measuring two particle correlations
with the Hanbury-Brown-Twiss method. In Fig. \ref{HBT}
the pair correlation function
of two $\Lambda $ particles is shown under the assumption that a
rather strong possible resonance channel
contributes at low momenta. This might
occur because of a possible strong attractive $\Lambda$-$\Lambda $-interaction
(see section 3).
Especially with such a channel small source sizes are relatively clean
to determine. (Of course, the observation of such a resonance would
give some exciting new {\em experimental} insight on the interaction among
two $\Lambda$s.) If the separation does not occur or if there is no QGP
the deduced source size should be nearly the same as for kaons
or pions and thus considerably larger.

\begin{figure}[ht]
\vspace*{8.5cm}
\caption{
The pair correlation function of emitted $\Lambda $ hyperons is shown for
different source sizes under the assumption of a relatively
strong resonance channel.
\label{HBT}
}
\end{figure}

\subsection{Strangelet distillation}

Besides the onset of strangeness separation
during the equilibrium phase transition
there exist another, somehow related argument
which could lead to strangelet formation in ultrarelativistic
heavy ion collisions \cite{PRD88,Liu84}.
Rapid kaon emission from the `surface' (or, in other words, the outer region)
of the fireball can result
in an even stronger enhancement of the $s$-quark abundance in the quark phase.
As important, this
prompt kaon (and, of course, pion) emission may in addition cool
the quark phase, so that there really might be a chance that the
quark plasma droplet condenses
into metastable or stable, rather cold
droplets of strange quark matter \cite{PRD88}.

In order to model the evolution of an initially hot fireball a two phase
{\em equilibrium} description between the hadron gas and the QGP was combined
with the {\em nonequilibrium} radiation by incorporating the rapid freeze-out
of hadrons from the hadron phase surrounding the QGP droplet during
the phase transition \cite{PRD91}. In particular we will then address the
question whether, how and why a hot QGP droplet can cool and will then
form (or better `become') a long-lived cold lump of strange quark matter.
Before we turn to the model we summarize the reasoning for the existence
and the possible consequences
of the evaporation processes.

Let us make some crude estimates about the qualitative
features of the surface radiation of a
{\em baryon-rich} fireball. One first might think of meson radiation
off the initial pure quark phase by various microscopic processes
\cite{Liu84,Da83,Ba83,Mu85}.
The most dominant particles to be evaporated are the pions.
For the strange mesons
the $K^{+}$, $K^{0}$ are more easily radiated
in a baryon-rich environment
off the hot surface
of the quark phase than $K^-$ and $\bar{K}^0$ \cite{PRD88}.
Similarly and probably even more important should be the meson
radiation while the system is in the phase coexistence region.
On one hand most of the total collision time is spent in this region
On the other hand, the possibility
of reabsorption of mesons is very unlikely, since the system approaches
freeze-out.
Meson evaporation in both stages of the expansion just described
carries away entropy, energy
{\em and} antistrangeness. Therefore, the residual expanding
fireball, which is in the mixed phase, loses entropy
and is charged up with a net strangeness larger than zero.
For a rough estimate a thermal black-body-type radiation formula
can be employed.
Accordingly the emission of $\overline{K}$s
containing an $\bar{q} $-quark are then suppressed from
baryon-rich matter by a factor $\sim $
$e^{-2\mu_q/T}$ as compared to the
$K$s.
Adding the resulting numbers over the different stages of the
fireball, a loss in entropy per baryon of about 15 units \cite{PRD88}
and a net strangeness
enrichment of the total system of about $f_s^{total} =0.75$ is found
\cite{Liu84,PRD88}!  The assumption of black body radiation seems
intuitively plausible
from detailed balance arguments \cite{Ba87} if one thinks that each hadron
moving towards the `surface' of the plasma phase will first coalescence
and then be completely absorbed.

The mechanism by which a QGP state is converted into hadrons is
a major uncertainty in the different descriptions.
The hadronization transition has often been
described by  geometric and statistical models,
where the matter is assumed to be in partial or complete equilibrium
during the whole expansion phase. A hydrodynamic expansion is often
assumed, which is substantially modified by the
phase transition \cite{Sub86,Ka86}.
Partial departure from (strange) chemical equilibrium is allowed in the flavour kinetic
model \cite{Koc86,Ka86,Ba88}. These rate calculations
suggested that chemical equilibrium is reached in the hadron phase,
if the system evolves from the deconfined phase: gluon fusion
yields fast equilibration of strangeness.
The subsequent re-hadronization affects that the strange
hadron yields closely approach the hadron equilibrium yields
{\em from above}.

A more realistic scenario must take
into account the competition of the collective expansion with
the particle radiation from the surface of the hadronic
fireball before `freeze out'.
In a second class of models \cite{Ho85,Ber88} it is assumed that the plasma
breaks up into droplets at some point during the hadronization transition.
In a `cascade' model
\cite{Ber88} the
emission and reabsorption of pions from the surface of the hot plasma
was studied.
For the present purpose, evaporation of other hadrons
must also be taken into account. Nucleons
and strange
particles (e.g. kaons and hyperons) are particularly important. One important
outcome of the cascade model \cite{Ber88}
was the observation that the total entropy
is approximately constant
during the hadronization.

\begin{figure}[ht]
\vspace*{\fill}
\centerline{\psfig{figure=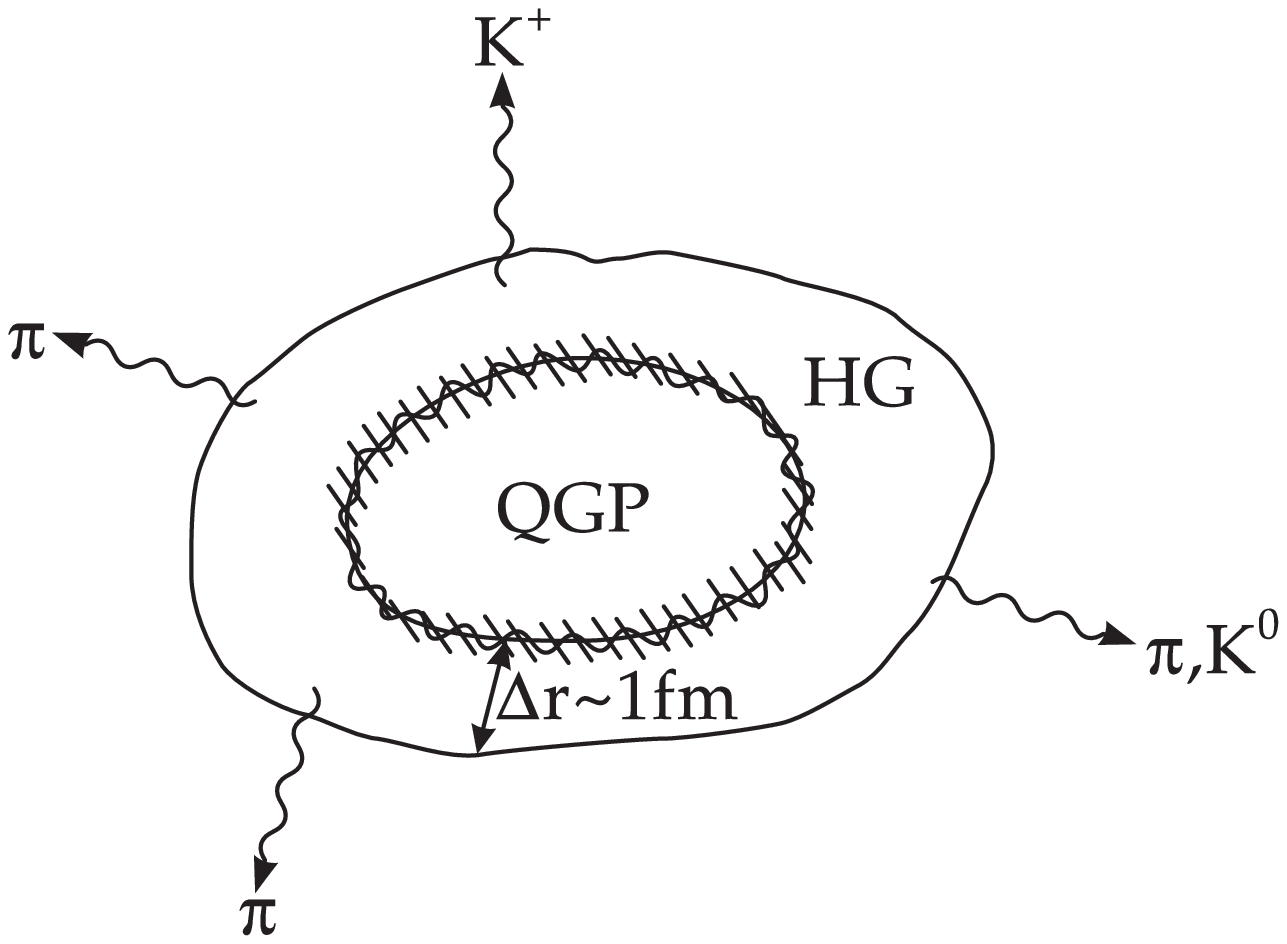,height=2.5in}}
\caption{
Hadron fluid surrounds the QGP
at the phase transition. Particles evaporate
from the hadronic region. New hadrons emerge
out of the plasma
by hadronization.
\label{schempic}
}
\end{figure}

Our following model \cite{PRD91} combines in a simple, yet plausible way the two different classes.
The expansion of the QGP
droplet during the phase transition will be described within a two-phase
equilibrium; in particular the
strangeness degree of freedom is thought to stay in chemical equilibrium
because the complete hadronic particle production is driven
by the plasma phase, as suggested by the rate calculations.
The competition of the collective expansion with
the particle radiation
is incorporated by rapid
freeze-out of hadrons from the outer layer of the hadron phase
surrounding the QGP droplet
at the Gibbs phase transition point. The expansion is here assumed to be
quasiisentrope, as suggested by the cascade model.
This scenario is visualized in Fig. \ref{schempic}.

The global properties of
the two phase system then change in time in accord with the
following differential equations
for the baryon number, the entropy and the net strangeness number
of the total system:
\begin{eqnarray}
\frac{d}{dt}A^{tot}  & = & -\Gamma \, A^{HG}  \nonumber \\
\frac{d}{dt}S^{tot}  & = & -\Gamma \, S^{HG}
\label{cg10}
\\
\frac{d}{dt}(N_s - N_{\overline{s}})^{tot}  & =  & -\Gamma \,
(N_s - N_{\overline{s}})^{HG} \, \, \, , \nonumber
\end{eqnarray}
where $\Gamma = \frac{1}{A^{HG}}
\left( \frac{\Delta A^{HG}}{\Delta t} \right) _{ev}$ is the effective
(`unique')
rate
of particle (of {\em converted hadron gas volume})
evaporated from the hadron phase.
These three equations may easily be combined in
a physically illustrative form:
\begin{eqnarray}
\frac{d}{dt}\left( {\frac{S}{A}}\right)^{tot} & = & -\Gamma \,
\lambda^{HG} \,
(1-\lambda^{HG})
\left( \left( \frac{S}{A}\right)^{HG} - \, \left( \frac{S}{A}\right) ^{QGP} \right)
\nonumber  \\[5mm]
\frac{d}{dt}\, f_s^{tot} & = & -\Gamma \,
\lambda^{HG}  \,
(1-\lambda^{HG})
\left( f_s^{HG} - \, f_s^{QGP} \right)  \, \, \, \, ,
\label{cg11}
\end{eqnarray}
where $\lambda ^{HG} = 1- \lambda ^{QGP} = A^{HG}/A^{tot} $ is
the ratio of baryon number contained in the hadron phase over
the total baryon number in the system,
$f_s$ defines the
total net strangeness content of the system,
and $\frac{S}{A}$ is the
entropy per baryon.
The three equations (\ref{cg10},\ref{cg11}) constitute
a set of {\em coupled} differential equations for
$A^{tot},(S/A)^{tot}$ and $f_s^{tot}$.
The latter two are given in terms of the
hadron and quark phase content by
\begin{eqnarray}
\label{cg12}
\left( \frac{S}{A} \right) ^{tot} & = & \left( \frac{S}{A}\right) ^{QGP}(\mu_q,\mu_s,T)
\,  (1- \lambda^{HG})  \,
+ \,   \left( \frac{S}{A}\right) ^{HG}(\mu_q,\mu_s,T)
\lambda^{HG}  \\[5mm]
f_s^{tot} & = & f_s^{QGP}(\mu_q,\mu_s,T) \, (1- \lambda^{HG}) \, +\,
f_s^{HG}(\mu_q,\mu_s,T)  \lambda^{HG} \, \, \, . \nonumber
\end{eqnarray}
The equation of state used consists of the Bag-model for the QGP and
a mixture of relativistic Bose-Einstein and Fermi-Dirac gases of the well
established nonstrange and strange hadrons up to M $\sim$ 2 GeV
for the hadron matter.

Although the baryon number $A$ and the strangeness
$f_s$ are conserved under strong interactions,
their value in the system
changes with time according to the above equations due to
the evaporation process.
This calculation requires solving
simultaneously
the equations of motion and the Gibbs phase equilibrium
conditions (\ref{cg7}),
which specify the intrinsic variables (e.g. the chemical potentials)
of the functions
$\left( \frac{S}{A} \right)^{QGP/Had}(\mu_q,\mu_s,T)$ and
$f_s^{QGP/Had}(\mu_q,\mu_s,T)$.
Still the effective rate $\Gamma $ and the baryon fraction
$\lambda^{HG}$
are not specified.
Intuitively, the plasma is surrounded by
the hadron phase, thus we take it as a shell of constant thickness
of $\Delta r \sim $ 1 fm (this is an ad hoc assumption, however,
as it turns out,the results
depend only weakly on this thickness).
This then specifies $\lambda^{HG}$.
Some unique
hadron gas volume of the outer layer is then evaporated in a small time interval
$\Delta t$. For the unknown evaporation time we use an averaged
rate of all particles at the considered temperature,
dictating the time-scale for the evolution of
the hadronic particles out of the hadron phase.
Altogether, the quasiadiabatic expansion during hadronization
and the evaporation of the hadrons in the most outer regions are
both incorporated and dictate the evolution of the (plasma) droplet
and its intrinsic variables $\mu_q,\mu_s$ and $T$.

For the following arguments it is important to study the expansion of an
initially hot QGP fireball with special attention to the evolution of the
strangeness during the phase transition. As outlined above, early non-equilibrium
particle radiation
off the pure hot and dense (baryon-rich)
QGP fireball is expected to be important in the initial phase of the
reaction.
Pions and kaons (containing an
$\bar{s} $-quark)
(and only a minor number of antikaons)
are emitted from the surface of the plasma.
Thus the QGP is enriched with finite
net strangeness even {\em before} the phase transition point is reached.
The net strangeness enrichment resulting from the early black body
radiation off the pure QGP-droplet
has been estimated to be
in the range $f_s^{init}$ $\stackrel{<}{\sim } $ 0.5
\cite{Liu84,PRD88}.
These values have been used in the present model as input for
the initial condition $f_s(t_0)$ when hadronization starts.
In a complete isentropic expansion with a total (initial) net strangeness
content but {\em without} further particle evaporation
it was (already) shown \cite{PRD88,He93} that for not so high initial entropies
the evolution can end with a small and (meta-)stable strangelet, if the
employed QGP equation of state
will allow for such exotic configurations to exist at small temperatures.
We will see in the following how things will work.

\begin{figure}[ht]
\vspace*{16cm}
\caption{
(a) Baryon number, strangeness content and temperature of the
quark glob during complete hadronization as a function of time
for a very large bag constant
$B^{\frac{1}{4}}=235 $ MeV.
The initial values are an initial baryon content of $A_B(t_0)=100$,
an entropy per baryon ratio of $\frac{S}{A}(t_0)=25$ and an initial net
strangeness fraction of
$f_s(t_0)=0.25 $.
Note the strong increase of the strangeness content with time.
(b) The same situation as in (a), however, for a small bag constant
$B^{\frac{1}{4}}=145 $ MeV, when a strangelet is {\em distilled}.
One observes a strong decrease in the evolving temperature.
\label{distill}}
\end{figure}

Fig. \ref{distill} gives an impression
how the hadronization proceeds for a large bag constant
($B^{1/4}=235$ MeV -- no strangelet in the groundstate)
and a small bag constant ($B^{1/4}=145$ MeV).
The initial parameters are a small net
strangeness content of $f_s(t_0) = 0.25$ and a moderate
entropy per baryon ratio of
$\frac{S}{A}(t_0) = 25$ (which is expected at CERN SPS energies):
For the large bag constant the system hadronizes completely in
$t \sim $ 8 $\frac{fm}{c}$, which is
customarily expected and thus not surprising.
The quark droplet remains unstable until the
strange quarks have clustered 
into $\Lambda $-particles and other strange hadrons to carry away 
the strangeness 
and the plasma has completely vanished into standard particles.
Yet, a strong increase of the net strangeness of the system
is found in both situations, which is basically a confirmation of
the strangeness separation mechanism and will be analyzed in some more
detail below. The plasma drop reaches a strangeness fraction of
$f_s \stackrel{\sim }{>} 1.5$ when the volume becomes small.
Indeed, for the {\em small} bag constant, however,
a cold
strangelet emerges from the expansion and evaporation process
with an approximate baryon number of $A \sim$  22, a radius of
$R \sim  2.5  fm$, and a net strangeness fraction of
$f_s \stackrel{>}{\sim }  1.5$, i.e. a charge to baryon
ratio
$Z/A
\sim  - 0.25$!
This would be an interesting object also from its global
properties: It would comprise
a nucleus of positive baryon number, but negative charge.

\begin{figure}[ht]
\vspace*{\fill}
\centerline{\psfig{figure=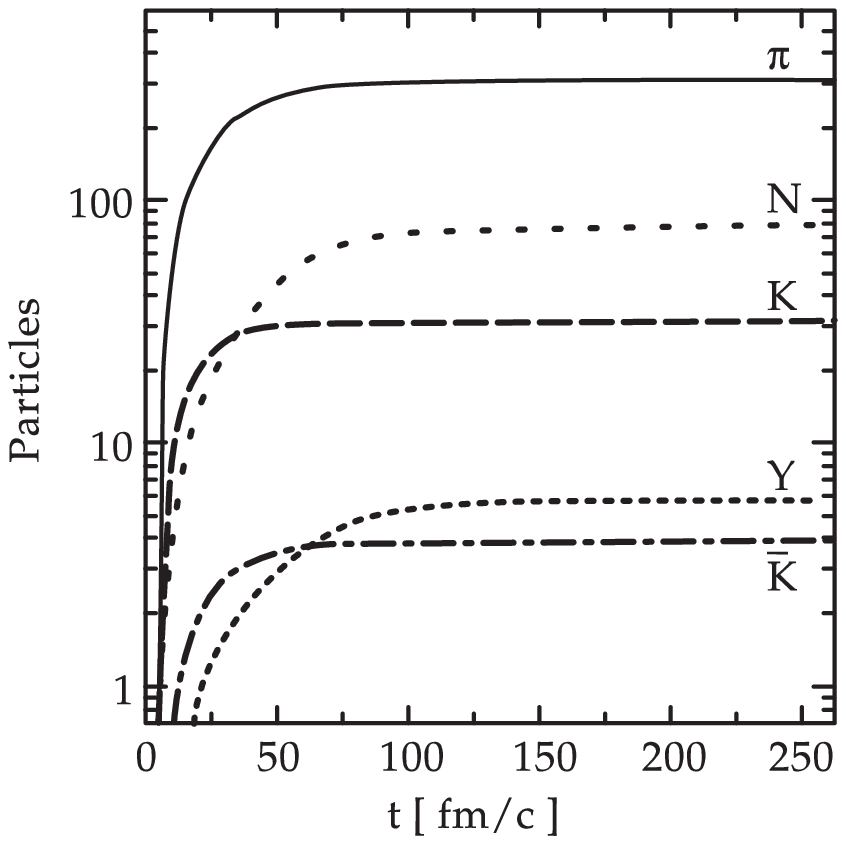,height=3.0in}}
\caption{
The number of emitted particles are shown versus time in a situation,
when a strangelet is distilled. The conditions are the same as in Fig.
\ref{distill},
however, no initial net strangeness is assumed.
\label{particles}
}
\end{figure}

The particle yields of the most
frequently emitted species are shown in Fig. \ref{particles} for the case,
where a strangelet condenses after $\sim$ 100 $\frac{fm}{c}$,
although the total net strangeness was zero at the beginning of the
hadronization.
Their time dependence is more or less as expected:
Roughly 300 pions and 32 kaons are produced
prior
to most of the other hadron species,
followed by the nucleons ($\sim $ 77) and at the late stage by some
antikaons ($\sim $ 3) and hyperons ($ \sim $ 6).
The total $K^+/\pi^+$-ratio is then approximately 0.16. The surplus of
17 in baryon number and of 23 in strangeness are contained
in the missing strangelet, which has to have a strangeness fraction
of $\sim$ 1.4!
The temporal behavior of the emitted particles
underlines the important role played by the kaons
(and pions) during hadronization. The early leaving kaons do enrich the
remaining system with net strangeness.
The pions carry about
half of the available entropy off the system.
For this reason one might expect that the nonequilibrium aspects
can become even more efficient at still higher entropies, when
more kaons and pions are produced.

One observes that in both cases depicted in Fig. \ref{distill}
the net strangeness increases.
However, in the first case, at high critical temperatures,
the temperature of the system increases slightly, although one might have
expected some cooling by the emission of particles, whereas in second case,
strangelet formation goes hand in hand with
strong cooling.
As a compelling consequence
the temperature (and entropy)
decreases drastically
throughout
the hadronization and condensation process. Thus
the pressure goes to zero and an
absolutely stable, cold strangelet is created.
The energy per
baryon of the remaining strange quark droplet converges to its zero temperature
value.

Why does in both (model) cases the net strangeness
content increase?
Consider therefore the physical content of eqs. (\ref{cg11}).
Obviously, $\Gamma $, $\lambda^{HG}$ and $(1-\lambda^{HG})$ are larger
than zero.
Therefore,
the total net strangeness fraction
$f_s^{tot}$ of the system
will always increase or decrease, respectively, with time,
if the fraction $f_s^{HG}$
is smaller or larger, respectively, than $f_s^{QGP}$.
Of course, this is just the question whether or not the
separation of strangeness does occur.
The abundant presence of the kaons enforces $f_s^{HG}<0$ early on!
Also later on
$f_s^{HG}$ remains smaller than $f_s^{QGP}$, and
therefore
$f_s^{tot}$ {\em increases} as function of time
($\frac{d}{dt} \, f_s^{tot} >0$).

Why is there cooling in the one and reheating in the other case?
Here a rather similar argument holds for the total entropy.
The total entropy of the remaining system
$(S/A)^{tot}$ will decrease ($\frac{d}{dt}(S/A)^{tot} <0$), i.e.
the system will be cooled, if and only if
the specific entropy per baryon in the hadron phase
exceeds that in the quark phase,
$(S/A)^{HG} > (S/A)^{QGP}$.
The system will reheat, i.e.
$(S/A)^{tot}$ will increase ($\frac{d}{dt}(S/A)^{tot} >0$), if
and only if
$(S/A)^{HG} < (S/A)^{QGP}$.
It was pointed out
already in ref. \cite{PRD88}
that the strangelet formation can only go hand in hand with
strong {\em cooling} rather than reheating. This is the case
when the bag constant $B^{\frac{1}{4}}$ is small and allows for
the existence of (meta-)stable strange quark matter states.
Although this intimate connection between cooling of the
plasma phase and the existence of strange quark matter is intriguing,
it might be valid only within the simple parametrization of the
quark gluon plasma phase within a bag model description.
Ultimately, whether
$(S/A)^{HG}$ is larger or smaller than $(S/A)^{QGP}$ at finite, nonvanishing
chemical potentials might theoretically
only be proven rigorously by lattice gauge calculations
in the future, as also the principle existence of (meta-)stable strange quark
matter.

In the situation when a strangelet gets distilled out
of the original QGP phase,
one observes \cite{PRD91} that the values
of the two chemical potentials approach `rather' quickly
their asymptotic numbers, even when the temperature of the
strangelet is still not so small, $T \sim 40 - 60$ MeV:
or metastable):
\begin{itemize}
\item
$\mu_q  \sim 200 - 275 $ MeV \, , \,
$\mu_s  \sim 300 - 325 $ MeV \, \, ($B^{\frac{1}{4}}=145$ MeV); \vspace*{5mm}
\item
$\mu_q  \sim 200 - 250 $ MeV \, , \,
$\mu_s  \sim 350 - 400 $ MeV \, \, ($B^{\frac{1}{4}}=160$ MeV).
\end{itemize}
The evaporation of the particles forces these potentials to
become quasistationary; a detailed balance for the emission of
nucleons and {\em antikaons} has adjusted itself. The hyperon
emission is still strongly suppressed compared to the nucleons,
although the strange chemical potential is already larger than the
light quark potential.
(For higher bag parameters, the hyperons become
more and more important.
It is exactly when strangelets are getting barely stable to unstable
when the hyperon emission becomes as or even more efficient in
the late stages of the plasma phase than the one for the nucleons.)
A `simple' but basic idea for the
possible formation of strange
quark matter droplets is due to behavior of the nucleons.
As discussed in the section 2, the total energy per baryon
(at finite temperature this has to be substituted by the free energy)
can easily be lowered by
assembling the non-strange quarks into pure
nucleonic degrees of freedom as long as the strangelet is not at its
minimum value in $f_s$. This happens rather fast.
For finite temperatures
the surplus in thermal energy
provides additional nucleon evaporation, so that
this minimum point will be `overshot' and the
strangelet turns out to have a strangeness fraction larger than 1.
To further illustrate how then stable strangelets cool and survive,
we note first that with the above numbers
an effective nucleon binding energy
$I_n = m_n-3\mu_q \, \sim 120-350$ MeV is found in the late
stages of the process.
The energy per baryon of a strangelet at
small temperature and zero pressure is approximately given by
\begin{equation}
\label{cg14}
\frac{E}{A}=
\left( \frac{E}{A} \right) _{T=0} + \gamma T^2 \, \, , \,   \, \,
\gamma \sim \frac{1}{20} \frac{1}{MeV}    \nonumber
\end{equation}
Any late emission of a nucleon
will now decrease the energy per baryon of the blob:
$$ \Delta \left( \frac{E}{A} \right) =
 \frac{I_n}{A(T)} \, \Delta A \, , \, \, \, \, \,
\Delta A <0 \, .  $$
In addition, the differential of eq. (\ref{cg14}) is
$\Delta (E/A) = 2 \gamma T \Delta T$.
Combining these two simplified expressions and integrating from the initial
temperature $T_i$ to $T=0$ yields roughly
the final baryon number of the strangelet:
\begin{equation}
A^{final}\, = \, A (T_i) \, \exp \left(
\frac{-\gamma T_i^2}{I_n} \right) \, \, \, .  \nonumber
\end{equation}
For $I_n \sim 250 $ MeV and $T_i \sim 20$ MeV, a total of $\sim 92$
percent of the initial baryon number would remain in the droplet.
Hence, evaporation of baryons from quark matter will
be suppressed at very low temperatures (and will cool the droplet),
because there is simply not enough thermal energy available to power the complete
evaporation. The strangelet becomes cold and stable.

We have so far demonstrated that
under certain circumstances,
the hadronization can result in the distillation of a rather cold
strange quark matter droplet. This debris of the quark gluon plasma
can, of course, only survive, if the energy per baryon of the
strange quark object is smaller than the energy per baryon of the
corresponding hadronic system at the same $f_s$. For the bag model
the strangelet is stable for $B^{\frac{1}{4}}=145$ MeV or metastable
for $B^{\frac{1}{4}}=160-180$ MeV.
For realistic parameters modeling the initial situation of the
QGP fireball either in the stable or metastable situation
one might expect a reasonable strangelet with a baryon number
$\sim $ 10 -- 30 \cite{PRD91}.
Negatively charged strangelets result for all cases calculated in \cite{PRD91}:
\begin{itemize}
\item
$f_s(t\rightarrow `\infty ') \sim  1.1 - 1.5 \, , \,
Z/A \sim (- 0.05) - (-0.25) $ \quad \mbox{-- stable case} \, ; \vspace*{5mm}
\item
$f_s(t\rightarrow `\infty ') \sim  1.5 - 2.0 \, , \,
Z/A \sim (- 0.25) - (-0.5) $\quad \mbox {-- metastable case} \, .
\end{itemize}
One should emphasize the fact
that the charge of
the distilled strangelet (with positive baryon number) is found to be negative!
This seems counterintuitive as the most stable configurations should actually
be, though small, positively charged. This has to do with the final
nucleon emission which is still energetically possible if the strangelet
carries still a positive charge.
Its charge then can only be changed by weak processes.
(Please note our previous discussion above.
We will come back to this point also in the next section when we address
the important question of
what are the most likeliest candidates to be seen in the dedicated experiments.)

As long as finite size effects like surface tension or curvature
contributions are neglected, the equations of motion (\ref{cg10}) are
scale invariant. All extensive properties, the rate $\Gamma $ and the time $t$
do scale for example with the initial baryon content $A_B(t_0)$.
Thus, in order to distill a strangelet with a baryon number much larger than 1,
the initial quark gluon plasma droplet must be fairly large.
(Not only) for this reason the heavy ion experiments using the heaviest ions,
like the available Au- and Pb-beams, should be most favourable for strangelet search.

The distillery works even for larger initial
entropies S/A=50 or 100 \cite{PRD91,Sp95}.
A high initial entropy does not necessarily prohibit
strangelet formation.
Abundant kaon production enriches the plasma rapidly with net strangeness
at high entropies.
This might offer
to look for strangelet production at the highest bombarding
energies available in the present and future for very heavy systems, e.g. at the
CERN SPS ($E_{LAB}$ $ \sim  200 \frac{GeV}{N} $) or even at
RHIC ($E_{LAB}$ $ \sim  20 \frac{TeV}{N}  $) facilities.
One might even think that the distillation indeed could also work for
very small initial chemical potentials, i.e. $\mu_q/T \ll 1$, as will
be expected in the central rapidity region at RHIC energies (although
there still might be a noticeable, nonvanishing net baryon excess) and
LHC energies \cite{Sp95}. There might be fluctuations
at some, but no particular rapidity region where a small but net
excess in baryon number as well as in strangeness number statistically
occurs. Within the present model this can lead to a distillation
of very small strangelets of a size $A_B \leq 5$ \cite{Sp95}.
There is a lot of wishful thinking in this scenario, as it assumes that at the onset
of the transition the system is in chemical equilibrium and due to the
fluctuations is described by very small, but finite chemical potentials.
It is amusing to note some analogy with the cosmological scenario proposed
by Witten \cite{Wit84}. The tiny net baryon excess in the early
universe hiding in the `high-temperature' phase is tremendously small, even
compared to the numbers expected at the LHC energies due to fluctuations.
The distillation of strange quark matter Witten visualizes as a
shrinking, leaking `balloon', where only neutrinos are allowed to escape,
leaving the net baryon density back and trapped inside the balloon.
This picture assumes neutrino losses are the main way for the high-temperature
phase to lose energy and baryon diffusion to be negligible,
while in fact neutrino losses and surface evaporation
might appear comparable. In the present case the mesons take the part of the neutrinos
and it is the question how efficient their emission is compared to the
evaporation of baryons, so that a cooling really might take place.

We conclude that there are two essential processes which can favour the formation of
strangelets from a baryon-rich QGP formed
in ultrarelativistic heavy ion collisions.
The first (as a prerequisite)
is the $s$-$\bar{s}$--separation mech\-a\-nism
discussed in the section before.
This leaves us with a quark phase in the coexistence region
of hadronic and quark matter, which is charged up with strangeness. The
$s$-quarks, created in pairs with $\bar{s}$-quarks in the early quark-gluon
fireball, remain in the quark phase during the phase transition, the
$\bar{s}$-quarks materialize mainly into kaons.
The second is the evaporation of the hadronic gas
with its antistrangeness excess, which charges the remaining
system with net strangeness. Besides the expansion additional
pion and nucleon evaporation should help to
allow for a possible, yet necessary cooling.
All these processes are statistical in nature.
These calculations are strictly valid only for large systems.
Large fluctuations around these average predictions
have to expected, in particular for the strangeness production
per event and for the evaporation processes. In a real heavy
ion collision an idealized situation as put forward in the present
description is probably never reached. The results here should be
seen with this in mind. They should illustrate that if (meta-)stable strangelets
do (or can) exist in nature, there might really be a chance for their
production at the present and future relativistic heavy ion experiments.

\subsection{Detectability in heavy ion experiments}

An important prediction of the exploited model in the previous subsection
is the
{\em negative} charge to mass ratio,
$Z/A \sim  - 0.1$ for absolutely
stable strangelets, and $Z/A \sim  - 0.45$ for metastable droplets.
Still, not restricting too much to this model, the produced
strangelets or MEMOs might also be slightly positive \cite{PRD88}.
It then could be detected by its unusual
charge-to-mass ratio
($\sim +0.15 < Z/A < -0.5$).

On the other hand,
simple coalescence estimates give production probabilities
of strange clusters of the order of $10^{3-A_B-|S|}$, where $S$ denotes the
strangeness and $A_B$ the baryon number of the cluster
\cite{Chi79,mat90}.
Small clusters with $A_B+|S|\le r+3$, where $r$ is the sensitivity of the
apparatus (presently $r \le 12$), are most favoured for detection.
Therefore, if strangelets or MEMOs are formed due to this scenario, baryon numbers
of $A_B \le 12$ are expected.

It is important to note that these objects
are a {\em new form of matter},
not a specific new particle. The strange droplets produced in
these reactions do not
come in the form of a single type of particle.
Many different sizes of droplets may be produced, spanning
a range in mass, charge, and strangeness content. 
The experimental task of finding the new form of matter 
is therefore challenging.
Here any detected particle having
an unusual charge to mass ratio is a potential {\em hypermatter}
candidate.

To identify a particle or cluster, 
its charge and mass need to be measured. To 
determine that the particle is a new form of strange matter, its 
strangeness content must also be revealed.
The experimental approach is first to
find `objects' having a peculiar or new charge/mass ratio.
(The strangeness might be seen by interaction
with a secondary nucleus : multiple production of
$\Lambda$s, $\Sigma$s, $\Xi$s and $\bar{K}$s
in such a secondary reaction would signal its existence.)
The key idea here is that the charge/mass ratio will be unlike that of any
normal nuclear isotope (the $^8${\em He} with a $Z/A=0.25$ would be
the isotope candidate with the smallest ratio).
Strangelets (or MEMOs) would have a charge $\sim 0$,
being slightly positively or negatively charged. In particular
in the range $-0.25 < Z/A < +0.25$ there exists no quasistable form of
nuclei or antinuclei.
Such a range will has been covered by the E864\cite{Sa91} experiment
at Brookhaven. E878, the successor of E858\cite{Cr91},
using a focusing spectrometer at zero degree,
is seizing a much smaller selected range, which
in respect to cover still the full range of interest
can be
steadily adjusted.
A similar technique like in E878 is also been employed
by the Newmass collaboration\cite{Pr93} (NA52 experiment)
at CERN at much higher energies, and at the present time
also with the heavy {\em Pb}-beam.

The question remains, on which time-scale weak decay
or flavour changing modes will appear. How would a strangelet or MEMO
then look like when passing through the detector?
Employing TOF-techniques to reveal the velocity and thus the charge to mass ratio,
the experimental setup sets a natural time scale $\sim 10^{-8}$ sec.
So, an important question we finally have to address are the lifetimes of these
objects.
In the following,
we subsequently discuss the properties of both forms of strange matter
and the possible long- and short-lived candidates referring to \cite{scha97}.

Whether or not strangelets exist depends crucially on the value of
the bag constant which is not known for such strange and big systems.
For a bag constant of $B^{1/4}=145$ MeV, the original value of the MIT bag
model fit, strangelets are absolutely stable, for bag constants up to 
$B^{1/4}\approx 180$ MeV strangelets are metastable, 
i.e.\ they can decay by weak interactions.
So anything between absolutely stable and unbound is possible.
Nevertheless, for the following arguments one needs only three basic
assumptions: 
\begin{enumerate}
\item 
Strange quark matter is at least metastable.
\item 
There exists a local minimum for the total energy per baryon 
of strange quark matter at a finite strangeness fraction $f_s=|S|/A$.
\item 
The relativistic shell model can be used for strangelets.
\end{enumerate}
With these assumptions we demonstrate
that there exists a valley of stability at low mass numbers
and that these strangelets are
highly negatively charged contrary to former findings.

\begin{figure}[ht]
\centerline{\psfig{figure=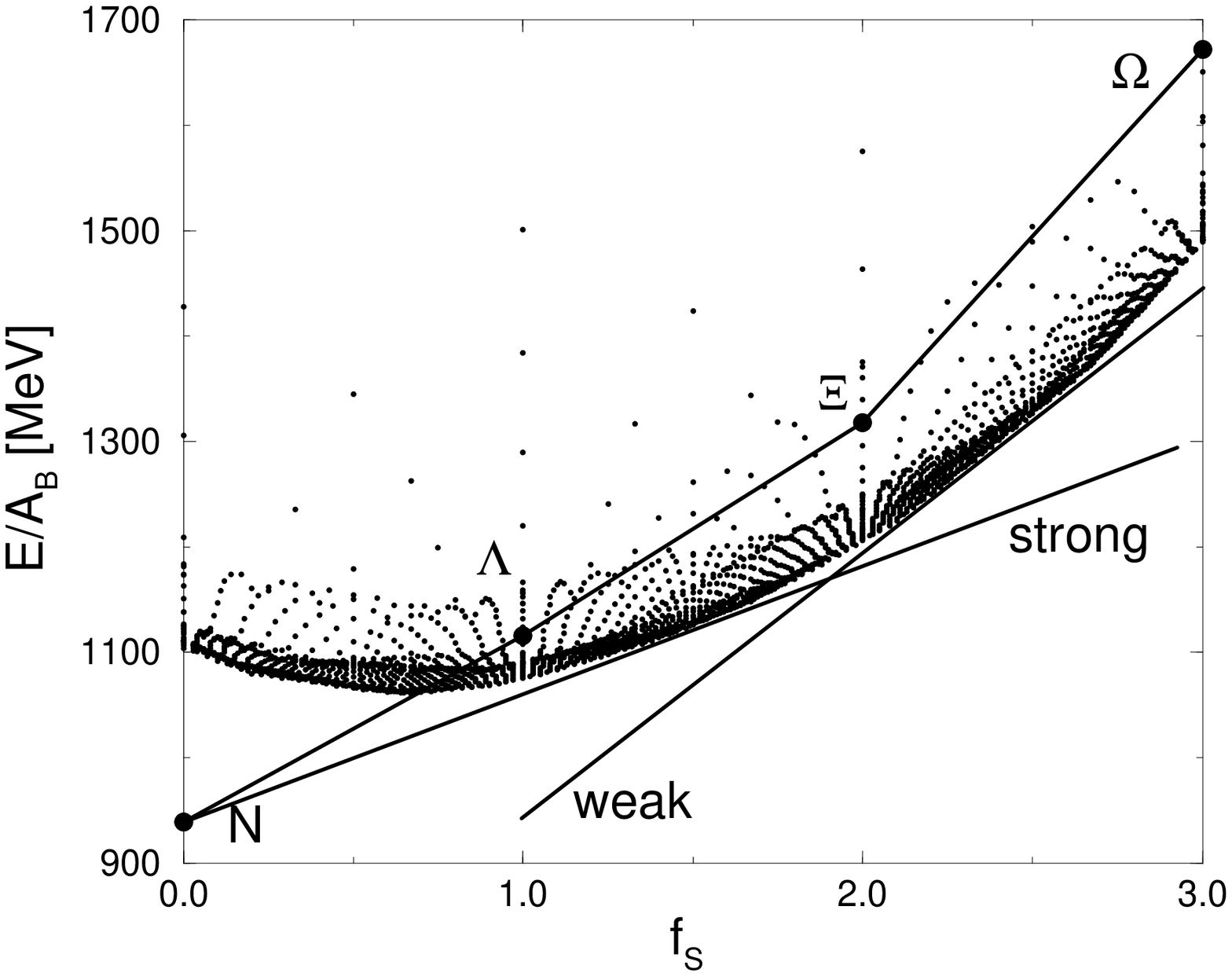,height=10cm}}
\caption{The energy per baryon $E/A_B$ of isospin symmetric strangelets
with $A_B\le40$ for a bag constant of $B^{1/4}=170$ MeV versus
the strangeness fraction $f_s$. The solid line
connects the masses of nucleon, $\Lambda$, $\Xi$ and $\Omega$ and stands
for free baryon matter.}
\label{fig:qsbags}
\end{figure}

The MIT bag model is used here as a guideline only.
Fig.\ \ref{fig:qsbags} shows the energy per baryon number of isospin symmetric 
strangelets as a function of $f_s$ for $A\leq 40$ for a bag parameter of
$B^{1/4}=170$ MeV. Now there are three different processes which will shift a
strangelet emerging from a heavy ion collisions to a very high strangeness
fraction. First, the strangelets sitting above the line drawn between the
nucleon and the hyperon masses will decay to a mixture of nucleons and
hyperons by strong interactions completely as this is energetically
favored. Second, the strangelets located between that line and the tangent
construction starting at the nucleon mass (denoted as strong) can still decay
{\em strongly} on a timescale of $10^{-20}$ sec
by emitting nucleons and hyperons. They will be shifted to a higher
strangeness fraction until they reach the tangent point at $f_s\approx 1.4$
(confer our discussion at the end of the previous subsection).
Third, weak nucleon decay can occur for the strangelets between the former
tangent and the other tangent (denoted as weak) 
starting at the nucleon mass and $f_s=1$ (as weak
interaction change one unit of strangeness) \cite{Chi79}.
The timescale for a weak nucleon decay has been estimated to lie
between $10^{-7}$--$10^{-9}$ sec \cite{henning}. More conservatively, one could
also argue that it might be comparable to that of a $\Lambda $-particle.
For a strangelet with $f_s>1$ 
the weak nucleon decay will enhance the strangeness fraction as 
\begin{equation}
\Delta f_s = \frac{|S|-1}{A-1}- \frac{|S|}{A} = \frac{f_s -1}{A-1}
\quad .
\end{equation}
Hence, strangelets surviving strong and weak
nucleon decay (and which thus can be detected with present experimental setups)
can be sitting at a very high strangeness fraction of
$f_s\approx 2.2$ which is the weak tangent point in Fig.\ \ref{fig:qsbags}.
For isospin symmetric systems, this large strangeness fraction corresponds to a
charge fraction of 
\begin{equation}
 \frac{Z}{A} = \frac{1}{2} \left(1- f_s\right) = -0.55
\end{equation}
which indicates highly charged strange quark matter. This is contrary to the
conventional picture that strangelets have a slightly positive 
charge-to-mass ratio which is the case for strange matter sitting in the
minimum of the curve plotted in Fig.\ \ref{fig:qsbags}.
But as pointed out before,
the combined effect of strong and weak hadronic decay will shift strangelets
emerging from a heavy ion collision to much higher values of $f_s$ and
therefore to highly negatively charged objects!

This simplified 
picture is only valid in bulk matter. For finite systems, which we
are interested in, shell effects will be important (see already our discussion
in section 2).
Already in Fig.\
\ref{fig:qsbags} one sees that shell effects are at the order of 100 MeV per
baryon number! Hence, we expect that strangelets with a closed shell
can be very deeply bound. These 'magic' strangelets are 
most likely to be stable against
strong and weak hadronic decay modes as their decay products have a much higher
total mass. 
The single particle levels inside a cavity (as for the MIT bag model) or for
ordinary nuclei or hypernuclei show the same order of levels for the lowest
eigenstates. First, there is a 1s$_{1/2}$ shell, then the 1p$_{3/2}$ 
and the 1p$_{1/2}$ shells follow. 
Due to relativistic effects, the spin-orbit splitting is quite sizable
for nucleons. As the quarks are much lighter and relativistic effects are even
more pronounced, the spin-orbit splitting for quarks is at the order of 100
MeV for very light bags, i.e.\ on a similar scale as the splitting between the
s and p shell. 
One can put 6 quarks in the s-shell due to the color degree of freedom, then
12 quarks in the 1p$_{3/2}$ shell and again 6 quarks in the 1p$_{1/2}$ shell.
The smallest and most pronounced 
magic numbers for quarks are then 6, 18, and 24 (the next one would be already
at 42).

Studying isospin asymmetric systems reveals another important effect.
The weak nucleon decay by emitting a proton carries away positive
charge. Nevertheless, the neutron does not carry away negative charge if it is
not accompanied by a $\pi^-$. But this decay is suppressed by the mass of the
pion and the phase space of the three body final state. Therefore, a strangelet
stable against weak nucleon decay is most likely to be negatively charged.

Let us look now for strangelets which have closed shells for all three quark
species with a negative charge 
and a high strangeness fraction as these are the most likely
candidates. The first magic strangelet is the quark alpha
with 6 quarks of each quark species at $A=6$ which has zero charge
\cite{Mi88}.
The magic strangelets with a high strangeness fraction and a negative
charge are then at $A=10$, $Z=-4$ (with 6 up, 6 down and 18 strange quarks),
$A=12$, $Z=-6$ (with 6 up, 6 down and 24 strange quarks),
$A=14$, $Z=-8$ (with 6 up, 18 down and 18 strange quarks), and
$A=16$, $Z=-10$ (with 6 up, 18 down and 24 strange quarks).
One sees a correlation, that 
adding two units of baryon number decreases the charge by two. 
These strangelets
have a rather high and negative charge fraction of $Z/A
\approx -0.5$ very similar 
to an antideuteron but with a much higher mass and charge! 
These strangelets constitute a valley of stability which is due to 
pronounced shell effects. 

This picture holds, i.e.\ these candidates remain, 
also within an explicit calculation using the MIT bag model
with shell mode filling \cite{scha97}. 
We calculated the masses of strangelets with 
all possible combinations of up, down and strange quarks up to a baryon number
of $A=30$. Then we look for possible strong decays as the emission of 
baryons (p,n,$\Lambda,\Sigma^-,\Sigma^+,\Xi^-,\Xi^0,\Omega^-$) and mesons
(pions and kaons) by calculating the mass difference between the strangelet and
its possible decay products. For the strong interactions, we also allow for
multiple hadron emission, like the strong decay of a strangelet via a neutron and
a pion, and the complete evaporation to hadrons.
For example, the strong proton decay $Q'\to Q+{\rm p}$ is checked by
\begin{equation}
M(A,S,Z) < M(A-1,S,Z-1) + m_p
\end{equation}
where $M(A,S,Z)$ stands for the mass of the strangelet for a given baryon
number, strangeness and charge.
Afterwards we check for weak hadronic decay, the single emission of baryons and
mesons within the same procedure simply by changing one unit of strangeness in
the final products.
The weak proton decay $Q'\to Q+{\rm p}$ is now checked by
\begin{equation}
M(A,S,Z) < M(A-1,S\pm 1,Z-1) + m_p
\end{equation}
where we allow for both strangeness changing processes of $\Delta S=\pm 1$.
This calculation has been done for several bag parameters. We choose a strange
quark mass of $m_s=150$ MeV if not otherwise stated. The value of $B^{1/4}=145$
MeV and $m_s=280$ MeV is taken from the original MIT bag model 
fit to the hadron masses.

\begin{figure}[th]
\centerline{\psfig{figure=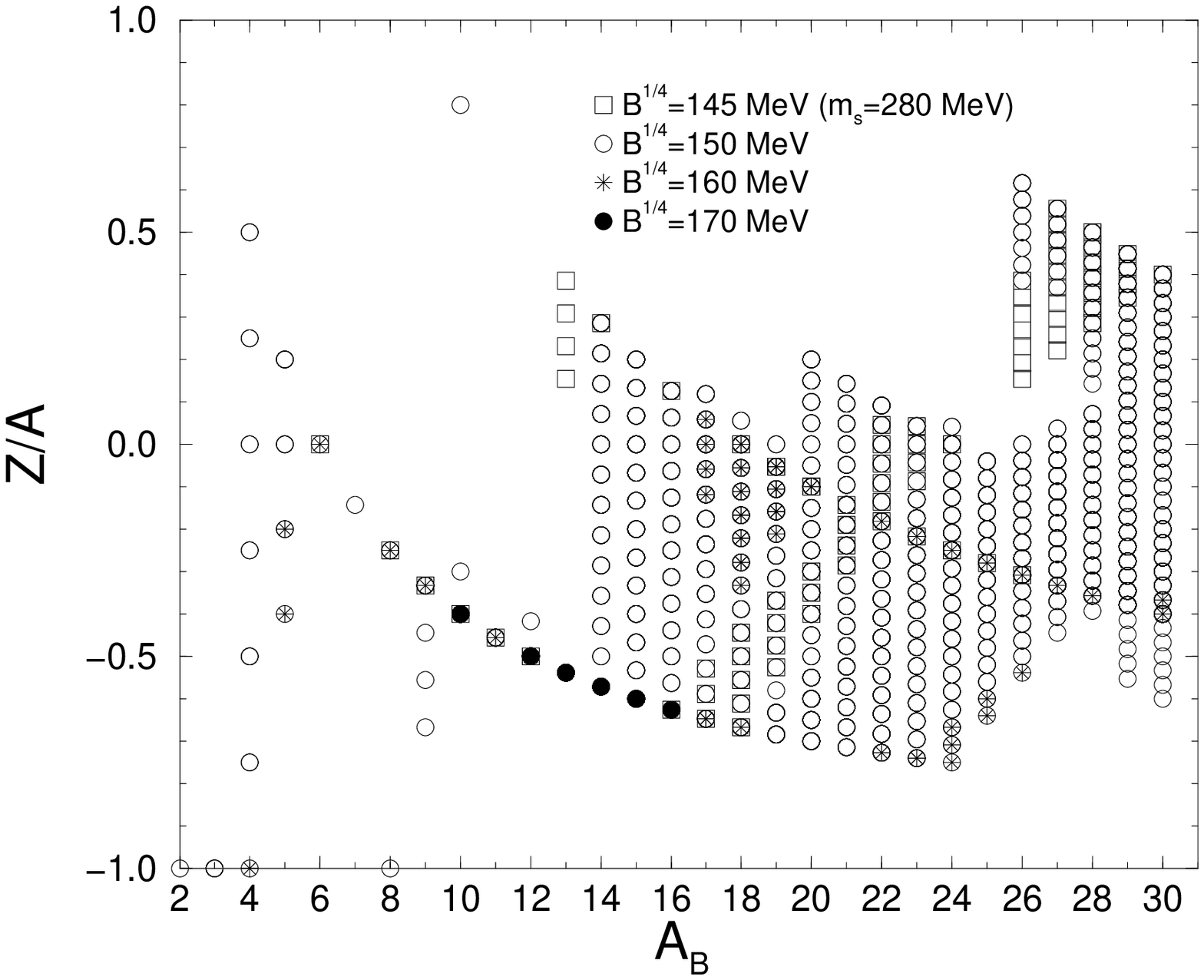,height=10cm}}
\caption{The charge fraction $Z/A$ 
for long-lived strangelets, which are stable against 
strong and weak hadronic decay, for different choices of the bag parameter.
The case for the original MIT bag model 
parameters ($B^{1/4}=145$ MeV, $m_s=280$
MeV) is also plotted.}
\label{fig:stab30}
\end{figure}

The candidates which are stable against strong and weak hadronic decay 
are 
plotted in Fig.\ \ref{fig:stab30} in a scatter plot as a function of their
baryon number and charge fraction.
In all the parametrizations shown, we find the candidates at
$A=10$ with $Z=-4$,
at $A=12$ with $Z=-6$, and at $A=16$ with $Z=-10$.
We do not find any candidates for a bag parameter of $B^{1/4}=180$ MeV 
or higher as strange quark matter starts to get unstable.

As expected and outlined before, 
the main long-lived strangelets stable against strong and weak hadronic
decay are lying
in the valley of stability and are highly negatively charged.
This finding is contrary to the common belief that strangelets have a small
positive 
charge and will have serious impact on present heavy ion searches for strange
matter.
These long-lived candidates are still unstable against
weak semileptonic decay (emission of
electrons and antineutrinos) and thus will live on a timescale of
maybe $10^{-4}$ sec \cite{henning}.

One can also ask about the much richer spectrum of short-lived candidates
which do decay by weak hadronic processes and thus probably
do live only
as short as the hyperons ($\tau\approx 10^{-10}$ s).
MEMOs can decay weakly on the timescale of the free hyperon weak decay
and are thus also belonging to the short-lived candidates.
MEMOs have quite distinct properties, they can be negatively charged while
carrying a positive baryon number due to the negatively charged hyperons, the
$\Sigma^-$ and the $\Xi^-$ \cite{scha97}.
Light candidates are the combinations
$\{2{\rm n},2\Lambda,2\Xi^-\}$,
$\{2{\rm p},2\Lambda,2\Xi^0\}$, $\{2\Lambda,2\Xi^0,2\Xi^-\}$,
$^{~~~~6}_{\Xi^0\Xi^0}$He and
$^{~~~~7}_{\Lambda\Lambda\Xi^0}$He
discussed in section 3.

\begin{figure}[th]
\centerline{\psfig{figure=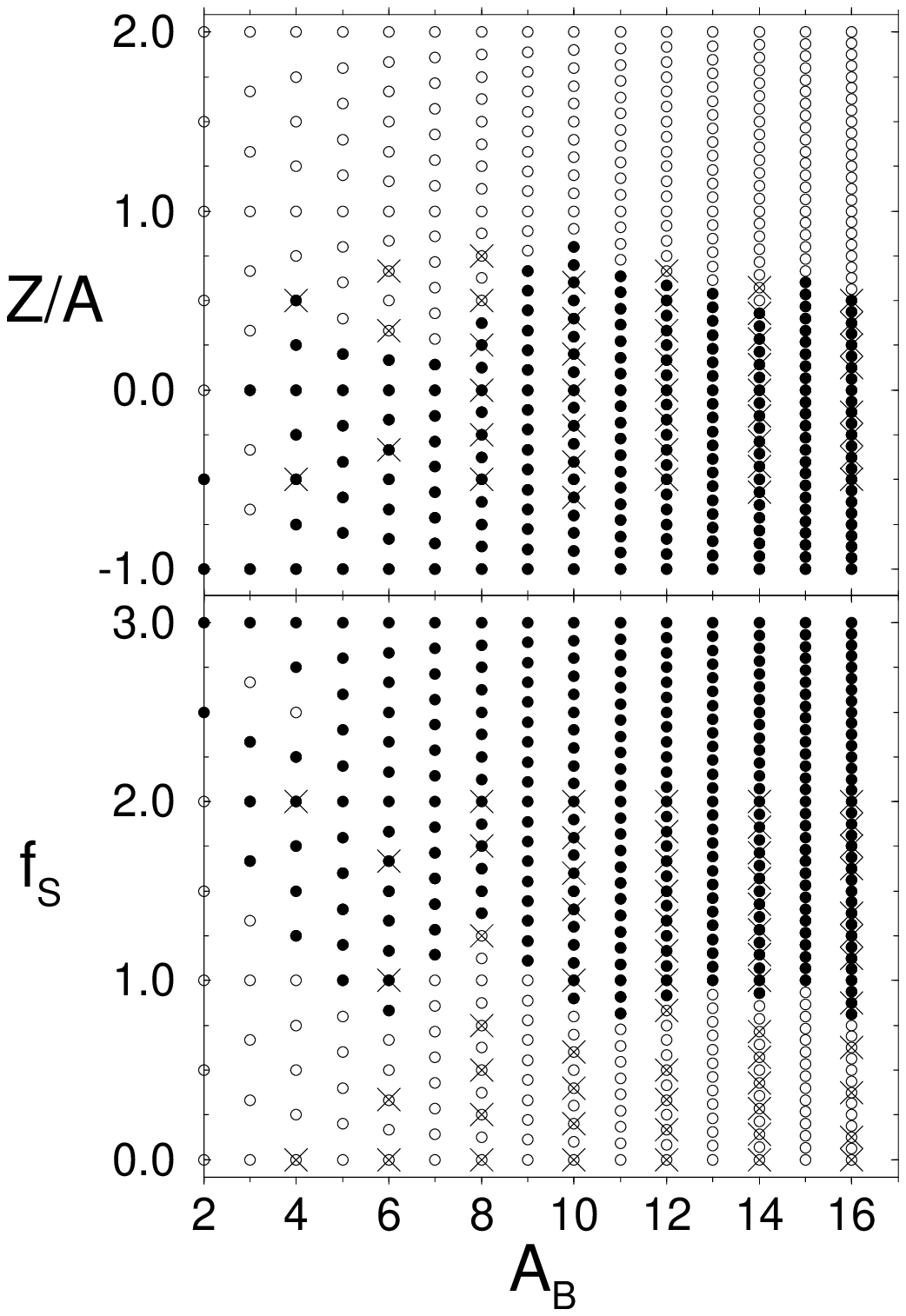,height=10cm}}
\caption{The strangeness per baryon $f_s$ (lower part) and the charge
fraction $Z/A$ (upper part) as a function of the baryon
number $A_B$ for short-lived strangelets (dots) and
unstable strangelets (open circles) for a bag constant of
$B^{1/4}=160$ MeV. The hadronic counterparts, MEMOs, are shown by crosses.}
\label{fig:B160}
\end{figure}

MEMOs thus compete with short-lived strangelets as they are of similar strangeness content.
We calculated light MEMOs up to a closed p-shell and checked for
metastability (strong decay). We analyzed the strangelet candidates without the
weak hadronic decay, i.e.\ allowing for the strong decay only.
The short-lived candidates for MEMOs and 
strangelets for a bag constant of $B^{1/4}=160$ MeV 
are shown in Fig.\
\ref{fig:B160} in a scatter plot as a function of strangeness fraction $f_s$,
charge fraction $Z/A$ and baryon number $A$. 

As can be seen, 
there are many more short-lived candidates than long-lived.
Light MEMOs can have very unusual charge fractions 
between $Z/A=\pm 0.6$ indicating a rich structure of strange hadronic matter.
Strangelet candidates also cover a wide range of charge
fraction but are mainly located at negative charge. This comes from the strong
decay which shifts strangelets to higher strangeness fraction and to negative
charge. 
There are MEMOs and strangelets with the same 
strangeness content and baryon number.
Here, the energetically least favourable object
can decay into the other via strong interactions.
A strangelet created
in a quark gluon plasma can then possibly decay into a MEMO.
Or vice versa, MEMOs can coalesce from the hot and hyperon-rich
zone of a relativistic heavy ion collision first and then they form
a strangelet. 

Presently, there are only experiments designed to look for long-lived
composites with a lifetime of $\tau> 50$ ns.
Designing an experiment for short-lived composites is challenging but 
planned for future colliders and can reveal
the possibly rich structure of strange matter.

\clearpage
\section{STRANGE MATTER IN NEUTRON STARS}

Originally (strange) quark matter in bulk was thought to exist
only in the interior of neutron stars where the pressure is
high enough so that the neutron matter melts into its quark substructure
\cite{Col75,Ba76,Fr78}.
At least in the cores of neutron stars (where the density rises up to the order of $10$ 
times normal nuclear density)
it is not very likely that matter consists of individual hadrons. 
For an overview of the structure of neutron stars see
\cite{SMProc,WeberGlendenning1996,Glen97}.

On the other hand it is also known that the pure `neutron' matter
is not really a nuclear matter state made solely out of neutrons, but
at least at higher densities consists also of a considerable amount of
protons as well as hyperons. The extrapolation of strange
hadronic matter to higher densities in fact has influence on the
constituents of `neutron' matter.
The gross structure of a neutron star like its mass M and radius R is
influenced by the composition of its stellar material. 
This holds especially in the case of the existence of strangeness bearing 
``exotic'' components like hyperons or strange quark matter
which may significantly change the
characteristic mass-radius (MR) relation of the star.

In this final section we therefore address briefly on how
the two different forms of strange matter might influence the properties
of neutron stars. This section, however, can only give an impression
of its implication on neutron star physics. It is based primarily on
some own work on this interesting issue and can thus provide only
an idea of possible effects. The interested reader should therefore
also consult the cited literature for a much wider range of topics.

\subsection{Strange quark matter stars}

In section 2 we have described the equation of state of cold strange quark
matter by means of a simple noninteracting Fermi gas of up, down and
strange quarks. In the following we adopt a somewhat modified model
which was developed quite recently, the effective mass bag model \cite{Sche97}:

In condensed matter as well as in nuclear physics medium effects play an 
important role. One of the most prominent medium effects are effective
masses generated by the (average) interaction of the particles with the system.
For example in the case of a gluon gas at high temperature
the consideration of an effective
mass for the gluon within the ideal gas approximation
leads to an excellent description of the equation of state found in lattice 
calculations \cite{GS,PKPS}.
If one considers now analogously the case of a quark gas at
zero temperature the situation is as follows:
The quarks are considered now as dressed
quasi-particles which acquire an effective
mass by the interaction with the other quarks of the dense system. 
The effective quark masses are obtained from the zero momentum limit of the 
quark 
dispersion relation, which follows from the
so called hard dense loop (HDL) approximation
of the one-loop quark self-energy at finite chemical potential 
\cite{EffeMass,Sche97}.
They are given by
\begin{equation} \label{defmq}
{m_q^*}^2(\mu)=\frac{g^2\mu ^2}{6\pi ^2}
\end{equation}
for the light quarks (i.e. $u$,$d$-quarks) and
\begin{equation} \label{defms}
m_s^*(\mu)=\frac{m_s}{2}+\sqrt {\frac{m_s^2}{4}+\frac{g^2\mu ^2}{6\pi ^2}}
\end{equation}
for massive quarks with current quark mass $m_s$ ($s$-quarks).
$g$ denotes the QCD coupling constant and is treated as a
further phenomenological
parameter as in the MIT model.
As can be readily expected from the formulas (\ref{defmq}) and (\ref{defms}),
the effective masses
increase with the coupling constant $g$ and the quark chemical potential $\mu$.
These effective masses are used in the ideal Fermi gas EOS at temperature $T=0$. 
Similar to the expressions stated in section 2, the pressure,
the particle density $\rho$ and energy density $\epsilon$
of this quasiparticle Fermi gas take the following form

\begin{eqnarray} \label{defpgeneral}
p(\mu) + B^*(\mu) & = & \frac{d}{2 \pi^2} \int_{k=0}^{k_F} dk \, k^2 (\mu - \omega^*(k))  \\
 & = & \frac{d}{24 \pi^2} \left[ \mu \, k_F \left( \mu^2 - \frac{5}{2} {m^*}^2 \right)
   +\frac{3}{2} {m^*}^4 \ln {\left( \frac{k_F + \mu}{m^*} \right)} \right]
\, \, \, ,   \nonumber
\end{eqnarray}

\begin{equation} \label{defrhogeneral}
\rho(\mu)= \frac{d}{6 \pi^2} k_F^3 \, \, \, ,
\end{equation}

\begin{equation} \label{defepsgeneral}
\epsilon(\mu) - B^*(\mu)=\frac{d}{16 \pi^2} \left[ \mu \, k_F \left( 2 \mu^2 - {m^*}^2 \right)
   - {m^*}^4 \ln {\left( \frac{k_F + \mu}{m^*}\right)} \right]  \, \, \, .
\end{equation}

Here $d$ denotes the degree of degeneracy (e.g. $d=6 n_f$ for $n_f$ flavors). Up to the 
additional function $B^*$, which can be regarded as a $\mu $-dependent bag constant,
these are the
ideal Fermi gas formulas at temperature $T=0$ for quasiparticles of mass
$m^*$ and chemical potential $\mu$.
Due to the $\mu$-dependence of $m^*(\mu)$ the introduction of the
function $B^*(\mu)$ is necessary in order
to maintain thermodynamic self-consistency on the one hand and to receive the
final expressions (as given above) for the particle and energy density on the other hand,
having the exact form as a noninteracting, but massive Fermi gas \cite{Sche97}.
The MIT bag constant $B_0$ still enters as $B^*(\mu =0)$ and is thus
a further phenomenological parameter of the model.
When inspecting this improved equation of state for strange matter numerically,
the overall energy per baryon number $E/A$ increases for increasing
coupling constant g \cite{Sche97}. This behavior one would intuitively
expect as the masses of the quarks do increase with the coupling constant.
In return, strange matter thus becomes less energetically
favorable for realistic QCD coupling constants $\alpha_s = \frac{g^2}{4\pi }$.
(One should be aware, however, that the value of the `free' bag parameter is
still far from being settled. For example,
the results of Peshier et al. \cite{PKPS}
suggest, that, by including effective medium masses in describing the hot gluon plasma
and compare it to lattice calculations, the employed free bag parameter
$B_0^{1/4}\approx 0.7 T_c$ might still be considerably smaller
than even the already very low
original MIT bag constant of $B_0^{1/4}=$145 MeV. In this sense, all the
speculative conclusions of section 2 are all still valid.)

We now show how a  neutron star would look like
under the assumption that it {\em entirely} consists
of cold electrically charge neutral SQM in equilibrium
with respect to the weak 
interactions.

This requires the inclusion of electrons into the model.
The thermodynamics of electrons at $T=0$
can be described by a relativistic Fermi gas

\begin{eqnarray}
\rho_e & = & \frac{\mu_e^3}{3 \pi^2} \\
\epsilon_e & = & \frac{\mu_e^4}{4 \pi^2} \\
p_e & = & \frac{\mu_e^4}{12 \pi^2}.
\end{eqnarray}
In return, one has four chemical potentials ($\mu_u$, $\mu_d$, $\mu_s$, $\mu_e$)
which are related by the chemical equilibrium between the quark flavors and the leptons. 
The basic weak reactions are given by

\begin{eqnarray}
d & \longrightarrow & u+e^- +\bar{\nu}_{e^-} \\
s & \longrightarrow & u+e^- +\bar{\nu}_{e^-}.
\end{eqnarray}
The equilibration of flavors is provided by

\begin{equation}
s + u \longrightarrow d + u.
\end{equation}
Hence, the four chemical potentials are reduced to two independent ones

\begin{equation}
\mu \equiv \mu_s = \mu_d \qquad \mbox{and} \qquad \mu_u = \mu - \mu_e.
\end{equation}
Finally, the overall condition of electrically charge neutrality
(`neutron' stars are locally charge neutral)

\begin{equation} 2/3 \rho_u -1/3 (\rho_d + \rho_s) -\rho_e = 0 \end{equation}
just leaves one independent chemical potential, say $\mu$. Therefore, the EOS
can be written as a function of $\mu $ only:

\begin{eqnarray}
\rho_B& = & (\rho_u+\rho_d+\rho_s)/3, \\
\epsilon & = & \epsilon_u + \epsilon_d +\epsilon_s +\epsilon_e +B_0 \label{rhostar}, \\
p & = & p_u +p_d +p_s +p_e -B_0. \label{pstar}
\end{eqnarray}
In the following we assume
a cold, static, spherical star. It is described by the solutions of the
famous Oppenheimer-Volkoff-Tolman (OVT) equations of hydrostatic equilibrium:
\begin{eqnarray}
\frac{dp(r)}{dr} &=& - \frac{\epsilon (r) m(r)}{r^2}
\left( 1+\frac{p(r)}{\epsilon (r)} \right)
\left( 1+\frac{4\pi p(r)r^3}{m(r)} \right)
\left( 1-\frac{2m(r)}{r} \right)^{-1} \, \, , \nonumber \\[2mm]
\frac{dm(r)}{dr} &=& 4\pi r^2\epsilon (r) \, \, .
\end{eqnarray}
They follow from general relativity \cite{OVT} which one has to apply due to highly 
concentrated matter and therefore curved space-time.
The solutions of the OVT equations 
are the pressure $p(r)$ and
mass m(r) inside a sphere of radius r.
The total radius $R$ of the star is determined by the condition $p(R)=0$
while the total gravitating mass of the star is given by

\begin{equation} M=\int_{r=0}^R 4 \pi r^2 \epsilon(r) dr. \end{equation}
The OVT equation can be solved specifying the central energy density $\epsilon_c=\epsilon(r=0)$ 
and the EOS in the form $p=p(\epsilon)$.

\begin{figure}[ht]
\vspace*{\fill}
\centerline{\rotate[r]{\psfig{figure=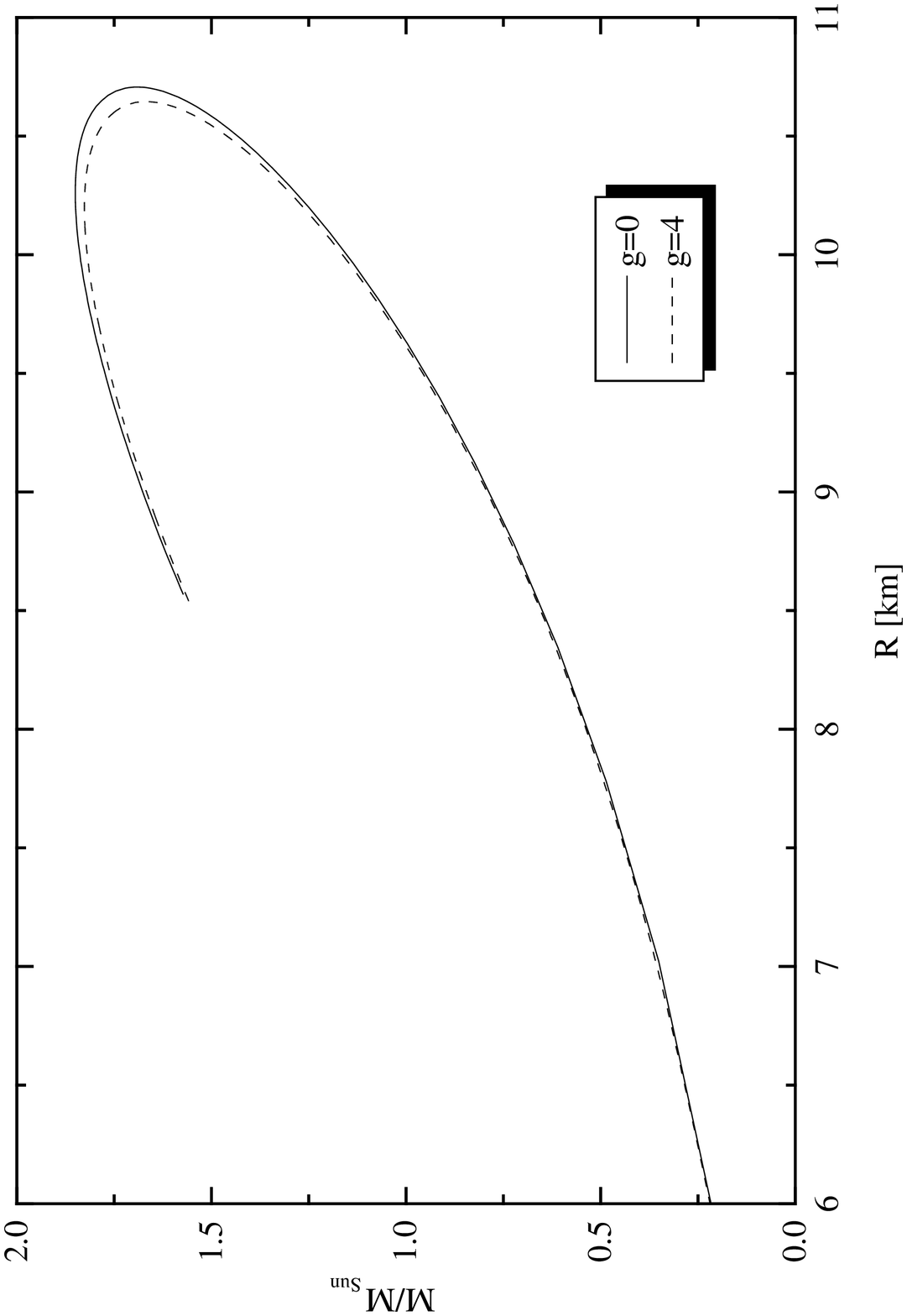,height=13cm}}}
\caption{Mass $M/M_\odot$ (in units of solar mass)
versus radius $R$ of a SQM star, $m_s=150 MeV$, $B_0^{1/4}=145MeV$.
\label{mr}}
\end{figure}

In Fig. \ref{mr} the MR relation of such a hypothetical
star is depicted.
The numerical calculation of $p=p(\epsilon)$ from 
(\ref{rhostar}) and (\ref{pstar}) shows that there is no
noteworthy influence by a change of the coupling constant $g$.
The EOS gets slightly softer for high values of the coupling constant (e.g. $g=4$) but there is 
no remarkable impact on $R$ and $M$ (see Fig. \ref{mr}).
The maximum mass of such a star lies in the range of 1.6-2 times the
solar mass and is thus compatible with the predictions of normal
(nonexotic) neutron stars.
One should also note one special feature inherent to the description
of pure quark matter stars. As they are inherent self-bound as a possible
consequence of QCD (which just corresponds to the speculation about
the existence of stable quark matter), for smaller masses
$M\stackrel{<}{\approx }M_{sun}$, the
quark matter stars do become also smaller. This is different, however, for
normal neutron stars, where their size typically increases for lower masses
due to the very diffuse (i.e. low density) surface.

\begin{figure}[ht]
\vspace*{\fill}
\centerline{\rotate[r]{\psfig{figure=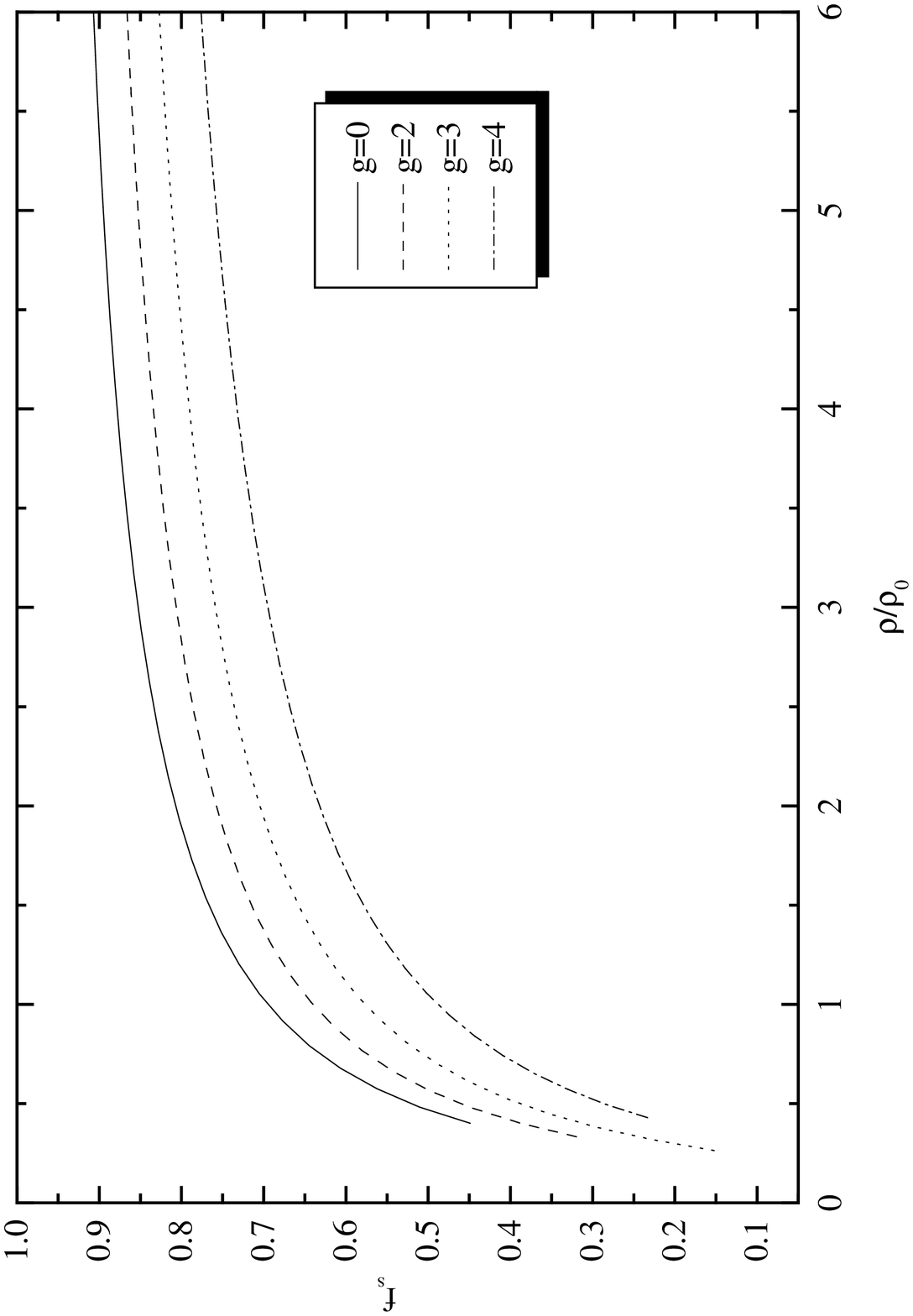,width=10cm}}}
\caption{Strangeness fraction $f_s$ versus
$\rho/\rho_0$ in a SQM star, $m_s=150 MeV$, $B_0^{1/4}=145MeV$
and $\rho_0=0.17 fm^{-3}.$
\label{fsrho} }
\end{figure}

Furthermore, Fig.\ \ref{fsrho} shows the change of
the strangeness fraction versus the baryon density for different values of $g$.
Throughout the interior the quark matter is indeed strange, with a net strangeness
content of $f_s \approx 0.7 - 0.9$. Therefore the quark matter star is in fact
a strange quark matter star.

Note, however, that the illustrative results
are only valid under the assumption of a pure SQM star. Although there
is no change in mass and radius of the star, there is, nevertheless, an increase of the energy
$E/A$ at a given radius inside the star due to medium effects.
Owing to the increase of $E/A$ in the entire
star a phase transition to hadronic matter will take place at a smaller
radius in the interior of the star \cite{WeberGlendenning1996,Web92}.
The `neutron' star would then have the form of a
so called hybrid star,
i.e. a star which is made of baryonic matter in the outer region, but with
a quark matter core in the deep interior.
It would be interesting to see how the
presented improved equation of state will affect the picture of such a hybrid star.
Such a study has recently been carried out \cite{Sche98}.
The deconfinement phase transition from hadronic matter to the SQM phase 
is constructed according to a construction
proposed by Glendenning \cite{Glen92,Glen97}.
One here requires the weaker condition of global charge neutrality
instead of assuming charge neutrality in either phase.
The latter assumption would have the
drastic consequence of strictly excluding a
spatial extended region of possible mixed phase inside the
star due to a resulting {\em constant} pressure in the mixed phase.
In the mixed phase, however,
the fraction of the quark matter phase on the one side, and
the fraction of the hadronic matter phase on the other side, might
both be (oppositely) charged if
only their volume proportion
$\chi$ is always chosen to fulfill the condition of global charge neutrality
\begin{equation}
\chi \, q_{QP} + (1-\chi) \, q_{HP} = 0.
\end{equation}
Here $q_{QP}$ and $q_{HP}$ denotes the charge density of each phase.
(This argument of Glendenning is very similar in its spirit to that
one given in section 3 concerning the idea of strangeness separation
during the mixed phase in the QGP phase transition.)
Since the pressure
inside a star must fall monotonically from
its interior to its surface, a constant pressure mixed phase could not exist 
over a finite
region inside the star.
However, with the above description (which is the thermodynamic correct one)
one finds that the
pressure varies smoothly and continuously with the proportion
of phases in equilibrium
\cite{Glen92,Glen97} leading to a
mixed phase of finite radial extent inside the star.

\begin{figure}[ht]
\vspace*{9cm}
\caption{Radii and inner structure of a hybrid star of mass $M=1.4 M_\odot$
($m_s=150 MeV$, $B_0^{1/4}=165MeV$)
\cite{Sche98a}.
\label{Rba}}
\end{figure}

In Fig. \ref{Rba} the schematic view of possible hybrid stars of
(an assumed and fixed) canonical mass
of $M=1.4\,M_\odot$ is shown for different and increasing coupling constant
g.
The bag constant was chosen so that a pure quark matter star cannot exist
even for $g=0$. The hadronic phase is described within the
Walecka model including (some) hyperon degrees of freedom.
(Details will not be given here and we refer to \cite{Sche98}.)
One finds
that already a small coupling constant of $g=1.5$
($\alpha_s\approx0.18$) is able to
halve the radius of the pure SQM core (denoted as QP) of initially
$R \approx 6\,km$ (with neglected medium effects, Fig.\,\ref{Rba}a) to $R \approx
3\,km$ (Fig.\,\ref{Rba}c).
Already at $g=2$ ($\alpha_s\approx0.32$) the pure SQM core has completely
vanished (Fig.\,\ref{Rba}d), while the
pure hadron phase (HP) has grown only moderately.
SQM is now only present in the mixed phase (MP) which dominates the
interior structure of the star.
For $g\approx 3.5$ a phase transition to SQM and thus the occurrence
of the mixed phase is completely suppressed.

\subsection{Strange hadronic matter inside neutron stars}

As shown in the previous section, strange quark matter can appear in the core
of a neutron star. Here we will discuss the scenario that strange hadrons, as
hyperons and kaons, can
appear in the interior of a neutron star. 
Indeed, it was shown by Glendenning that
hyperons \cite{Glen85} appear at a moderate density of about
$2\div 3$ times normal nuclear matter density $\rho_0=0.15$ fm$^{-3}$.
These new species influence the properties of the equation of state
of matter and the global properties of neutron stars.
There are so many hyperons in the neutron star that the whole object is more
appropriately dubbed a giant hypernucleus.
Hyperons considerably soften the EOS and reduce the maximum mass
of a neutron star.

The appearance of another form of hadrons with strangeness, kaon condensates,
was discussed also in many papers
\cite{Kap86,Brown,Prakash,Sch96,Fujii96}. 
Chiral perturbation theory (ChPT) gives a rather robust
prediction of the onset of antikaon condensation at 
$\rho\approx (3-4)\rho_0$ \cite{Brown} taking into account 
in-medium modifications of the antikaon energy with density. 
Antikaon condensation will soften the EOS and reduce the maximum mass
of a neutron star similar to the case of hyperons. 
This will allow for the existence of low-mass black
holes and implications have been discussed in \cite{Bethe}.
In \cite{Maru94,Sch94b} it was
criticized that effects nonlinear in density were not taken into
account which will shift the appearance of a kaon condensed phase to
higher density. Moreover, hyperons were not considered in this approach.  

Rather recently, antikaons and hyperons were considered on the same
footing, where it was found that hyperons shift
the onset of antikaon condensation to higher density
\cite{Prakash} or that it is very unlikely that
it appears at all due to the rather strong hyperon-hyperon
interactions \cite{Sch96}. 

In the following we will discuss the appearance of hyperons and kaons 
in neutron star matter.

The equation of state for neutron star matter is derived by standard methods
(see e.g.\ ref.\ \cite{Glen85} for the RMF approach without
the hidden strange meson fields).
The equations of motion for the meson fields in uniform matter at rest 
are given by
\begin{eqnarray}
m_\sigma^2 \sigma + \frac{\partial}{\partial \sigma} U(\sigma ) &=&
\sum_{\rm B} g_{\sigma B} \rho^{(B)}_S \cr
m_{\sigma^*}^2 \sigma^* &=&
\sum_{\rm B} g_{\sigma^* B} \rho^{(B)}_S \cr
m_\omega^2 V_0 + d V_0^3 &=&
\sum_{\rm B} g_{\omega B} \rho^{(B)}_V \cr
m_\rho^2 R_{0,0} &=&
\sum_{\rm B} g_{\rho B} \tau_0^B \rho^{(B)}_V \cr
m_\phi^2 \phi_0 &=&
\sum_{\rm B} g_{\phi B} \rho^{(B)}_V
\quad .
\end{eqnarray}
where $\rho_S$ and $\rho_V$ denote
the scalar and vector densities, respectively.
The equations can be solved for a given total baryon density $\rho_B$ and
charge density $\rho_c$
including the contributions from the free electrons and muons
\begin{eqnarray}
\rho_B &=& \sum_{\rm B} \rho^{(B)}_V \cr
\rho_c &=& \sum_{\rm B} q_B \rho^{(B)}_V + \sum_{l=e,\mu} q_l \rho_l = 0
\end{eqnarray}
where $q_i$ stands for the electric charge of a species $i$.
In $\beta$-equilibrium the chemical potentials of the particles
are related to each other by
\begin{equation}
\mu_i = b_i \cdot \mu_B + q_i \cdot \mu_e
\end{equation}
where $b_i$ is the baryon number of a species $i$.
This means that all reactions which
conserve charge and baryon number are allowed, as e.g.\
\begin{equation}
{\rm n} + {\rm n} \to \Lambda + {\rm n} \quad , \qquad
\Lambda + \Lambda \to \Xi^- + {\rm p} \quad, \qquad \dots
\end{equation}
Since we consider neutron stars on a long time scale, the strangeness quantum
number is not constrained and the net strangeness is determined by the
condition 
of $\beta$-equilibrium.
The above equations fix the fields and the equilibrium composition of
neutron star matter.

Fig. \ref{fig:fig-1} shows the composition of neutron star matter
for the parameter set TM1 with
hyperons including the hyperon-hyperon interactions.
Up to the maximum density considered here all effective masses
remain positive and no instability occurs. 
The proton fraction
has a plateau at $(2-4)\rho_0$ and exceeds 11\%{} which allows for the
direct URCA process and a rapid cooling of a neutron star.
Hyperons, first $\Lambda$'s and $\Sigma^-$'s, appear at $2\rho_0$, then
$\Xi^-$'s are populated already at $3\rho_0$. The number of electrons
and muons has a maximum here and decreases at higher densities, i.e.\
the electrochemical potential decreases at high densities. 
The fractions of all baryons show a tendency towards
saturation, they asymptotically reach similar values 
corresponding to spin-isospin and hypercharge-saturated matter.
Hence, a neutron star is more likely a giant hypernucleus \cite{Glen85}!

\jfigpsb{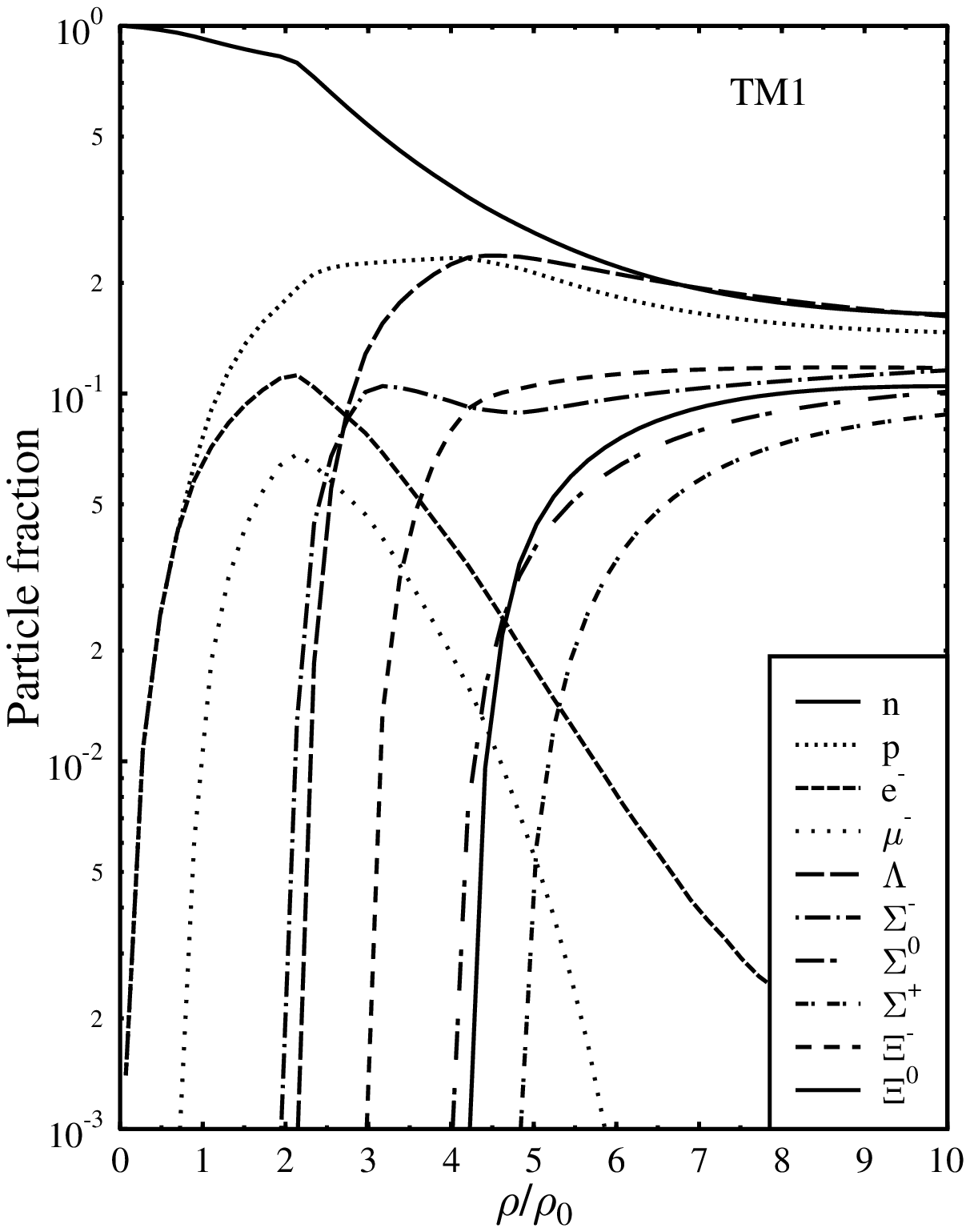}{The composition of neutron star matter with hyperons
which appear abundantly in the dense interior (from Ref.\ \cite{Sch96}).}

In the following
we adopt the meson-exchange picture for the KN-interaction simply
because we use it also for parametrizing the baryon interactions.
We start from the following Lagrangian
\begin{equation}
{\bf L}^{\rm RMF}_{KN} = D^*_\mu \bar K D^\mu K - m_K^2 \bar K K
- g_{\sigma K} m_K \bar{K}K \sigma
- g_{\sigma^* K} m_K \bar{K}K \sigma^*
\label{RMFLagr}
\end{equation}
with the covariant derivative
\begin{equation}
D_\mu = \partial_\mu +
ig_{\omega K} V_\mu + ig_{\rho K} \vec{\tau}\vec{R}_\mu + ig_{\phi K}\phi_\mu
\quad .
\end{equation}
The scalar fields essentially decrease the kaon mass, while
the vector fields will increase (decrease) the energy of the kaon
(antikaon) in the dense medium.
The scalar coupling constants are fixed by the s-wave KN-scattering lengths.
The coupling constants to the vector mesons are chosen from
SU(3)-relations.
The onset of s-wave kaon condensation is now determined by the condition
$- \mu_e = \mu_{K^-} \equiv \omega_{K^-} (k=0)$.

The density dependence of the K and $\bar K$ effective energies is displayed
in Fig. \ref{fig:fig-2}.
The energy of the kaon is first increasing in accordance with 
the low density theorem.
The energy of the antikaon is decreasing steadily at low densities.
With the appearance of hyperons the situation changes dramatically.
The potential induced by the $\phi$-field cancels the contribution
coming from the $\omega$-meson. Hence, at a certain density the energies of
the kaons and antikaons become equal to the kaon (antikaon) effective mass, 
i.e.\ the curves for kaons and antikaons are crossing at a 
sufficiently high density. At higher densities
the energy of the kaon gets even lower than that of the antikaon!
Since the electrochemical potential never reaches values above 160 MeV here 
antikaon condensation does not occur at all.
The possibility of antikaon condensation was checked for several parameter
sets and found that at least 100 MeV are missing
for the onset of kaon condensation \cite{Sch96}.
This is in contrast to previous
calculations disregarding hyperons \cite{Brown} but in line with 
the findings in \cite{Prakash} where it was seen that hyperons shift
the critical density for kaon condensation to higher density.

\jfigpsb{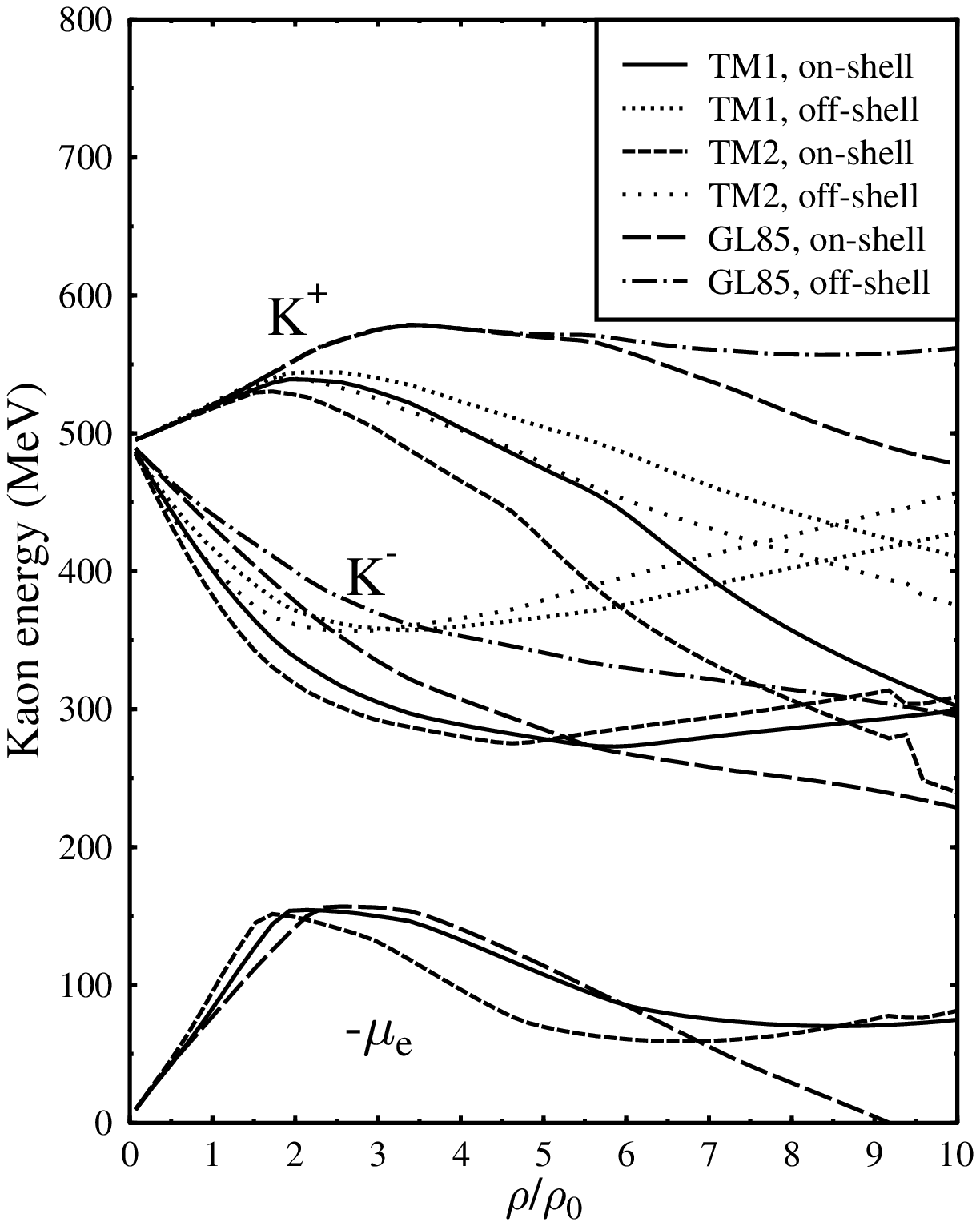}{The effective energy of the kaon and the antikaon
and the electrochemical potential.
Kaon condensation does not occur over the whole density region
considered (from Ref.\ \cite{Sch96}).}

These illustrations clearly demonstrate the `richness'
of possible scenarios for the interior structure of neutron stars.
In return it represents  a very interesting and still open and lively field
to pin down the possible structure of neutron stars by its observed properties
and phenomena.


\section*{Acknowledgments:}

J. S.-B. thanks the Alexander von Humboldt Stiftung
for its support with a Feodor Lynen scholarship.
Thanks go to our friends and colleagues A. Diener, C. Dover, A. Gal,
P. Koch-Steinheimer, I. Mishustin,
D. Rischke, K. Schertler, M. Thoma and especially to H. St\"ocker.
Without their help and collaboration this work would not have been possible.
This work is supported in part by 
the Director, Office of Energy Research,
Office of High Energy and Nuclear Physics, Nuclear Physics Division of the
U.S. Department of Energy under Contract No.\ DE-AC03-76SF00098.


\begin{thebibliography}{99}
\bibitem{Com1}
In later papers, like e.g.
in reference \cite{Far84}, the fine
structure constant has been redefined to $\alpha _c = g_c^2/4\pi$,
so that the original MIT parameter would there have to be taken as
$\alpha _c = 2.2$.
\end{thebibliography}
\end{document}